\documentclass[11pt,a4paper]{article}
\usepackage{jheppub}
\usepackage{xcolor}
\usepackage{amsmath,amssymb,amsfonts}
\usepackage{physics}
\usepackage{graphicx}
\usepackage{svg}
\graphicspath{{./images}}
\usepackage{subcaption}
\usepackage{array}
\usepackage{tikz}
\usepackage{url}
\usepackage{multirow}
\usepackage{comment}
\usepackage{tabularx}
\newcolumntype{Y}{>{\centering\arraybackslash}X}
\usepackage{makecell}

\usepackage{hyperref}

\usepackage{hyperref}

\begin{document}

\definecolor{lime}{HTML}{A6CE39}
\DeclareRobustCommand{\orcidicon}{\hspace{-1mm}
	\begin{tikzpicture}
	\draw[lime, fill=lime] (0,0) 
	circle [radius=0.12] 
	node[white] {{\fontfamily{qag}\selectfont \tiny \,ID}};
	\draw[white, fill=white] (-0.0525,0.095) 
	circle [radius=0.007];
	\end{tikzpicture}
	\hspace{-3mm}
}

\foreach \x in {A, ..., Z}{\expandafter\xdef\csname orcid\x\endcsname{\noexpand\href{https://orcid.org/\csname orcidauthor\x\endcsname}
		{\noexpand\orcidicon}}
}

\newcommand{\orcidauthorA}{0009-0000-2346-2273}
\newcommand{\orcidauthorB}{0000-0001-6639-0951}

%%%%%%%%%%%%%%
\title{Refractive neutrino masses in the solar DM halo:\\ Can the dark-LMA solution be revived?}
\preprint{TIFR/TH/25-20}

\author[a]{Susobhan Chattopadhyay\orcidA{}}
\author[a]{Amol Dighe\orcidB{}}

\affiliation[a]{Tata Institute of Fundamental Research, Homi Bhabha Road, Colaba, Mumbai 400005, India}
%%%%%%%%%%%%%%%

\emailAdd{susobhan.chattopadhyay@tifr.res.in}
\emailAdd{amol@theory.tifr.res.in}

\begin{abstract}
{Neutrinos can acquire ``refractive masses" as a consequence of their interactions with ultralight dark matter (DM). We explore a model with two additional sterile neutrinos and an ultralight scalar field which acts as DM and interacts with all five neutrinos. We show that the effective $5 \times 5$ Hamiltonian for neutrino propagation can be diagonalized by a unitary matrix $\mathbb{P}$ parametrized by 6 mixing angles and 1 complex phase. When active-sterile mixing angles are small, we identify a parametrization for $\mathbb{P}$ that reduces neutrino propagation inside the Sun to a two-flavor problem for a uniform DM background. In the presence of a DM halo inside the Sun, however, the propagation shows additional features in the region of halo dominance. We derive approximate analytic expressions for the electron neutrino survival probability in the presence of the DM halo. We show that this probability has a strong dependence on the neutrino production region even for a fixed energy, and numerically calculate the effects of averaging over these production regions. Comparisons with the re-interpreted solar data, in the light of possible active-sterile neutrino conversions, would allow putting bounds on the halo parameters. Finally, we examine the possibility of reviving the dark-LMA solution in this context, where the survival probability spectrum can have attractive features aligned with the measurements at Super-Kamiokande.}
\end{abstract}
\maketitle

\setcounter{page}{0}
\pagenumbering{arabic}

\section{Introduction}

Despite the phenomenal success of the Standard Model (SM) of particle physics in describing the Nature, there are still important questions that have remained unanswered. Two of such most pressing questions are the origin of the neutrino masses and the nature of dark matter (DM). 

Experiments have confirmed the oscillations among three active flavors of neutrinos through a combination of solar, atmospheric, reactor, and accelerator-based observations. Results from these experiments indicate the presence of at least two non-vanishing mass-squared differences, commonly known as the solar and the atmospheric mass-squared differences, along with three mixing angles \cite{ParticleDataGroup:2024cfk,Esteban:2024eli,Capozzi:2025wyn,Capozzi:2021fjo}. Although we know the differences between the squares of neutrino masses to a considerable precision, the absolute magnitude of neutrino masses is yet unknown. Several experiments like KATRIN \cite{Lokhov:2022zfn} are designed to achieve this goal of measuring the absolute masses of neutrinos. Over the years, theorists have proposed several mechanisms for generating neutrino mass \cite{deGouvea:2016qpx,Cai:2017jrq}. Another open question is the possible existence of sterile neutrinos. Although collider experiments such as the Large Electron-Positron collider (LEP) have placed stringent constraints disfavoring more than three active flavors of neutrinos \cite{ParticleDataGroup:2024cfk}, the existence of sterile neutrinos is still a subject of active theoretical and experimental investigation \cite{Dasgupta:2021ies}. Anomalies at short baseline experiments like LSND (Liquid Scintillator Neutrino Detector) \cite{Sung:2001ps} and MiniBooNE \cite{MiniBooNE:2021bgc,MiniBooNE:2022emn} point towards eV-scale sterile neutrinos, while super-light sterile neutrinos with masses $\sim 10^{-5}\ \text{eV}^2$ have also been proposed \cite{deHolanda:2003tx,deHolanda:2010am,BhupalDev:2012jvh,Liao:2014ola,Divari:2016jos} to explain the survival probability spectrum of solar neutrinos.

There is now convincing evidence for the presence of DM from many independent observations. Cosmological surveys indicate that DM constitutes $\sim 27\%$ of the energy budget of the Universe \cite{Planck:2018vyg}. Observations have demonstrated that DM halos affect not only the rotation velocities of galaxies but also play a major role in the formation of their structure and evolution. A plethora of DM models populate the existing literature \cite{Cirelli:2024ssz}, with their masses spanning a large range from ultralight particles around $10^{-21}\ \text{eV}$ to macroscopic objects reaching up to $10^{37}\ \text{kg}$. Ultra-light DM (ULDM) models, also popularly referred to as fuzzy/wave-like DM models, have recently gained significant attention owing to their detection prospects across a wide variety of experimental approaches. These encompass table-top experiments using atomic interferometry and nuclear clocks, to cosmological and astrophysical probes including stellar cooling observations, pulsar timing arrays, and cosmic microwave background measurements. From a theoretical perspective, ULDM naturally emerges in several well-motivated extensions beyond the SM aimed at solving the strong CP problem \cite{Preskill:1982cy,Abbott:1982af,Dine:1982ah,Duffy:2009ig,DiLuzio:2020wdo} or the electroweak hierarchy problem \cite{Graham:2015cka,Davidi:2017gir,Davidi:2018sii,Banerjee:2022wzk,Chattopadhyay:2024rha}. However, establishing the particle nature of DM, as well as its possible interactions with the SM particles, continues to be elusive.

When neutrinos propagate through a medium of ULDM, they may experience an effective potential generated due to scattering via the exchange of these light mediators \cite{Choi:2019zxy,Choi:2020ydp,Smirnov:2021zgn,Chun:2021ief,Sen:2023uga}. Neutrino-ULDM interactions and their related phenomenology have been studied in considerable detail in the literature \cite{Berlin:2016woy,Brdar:2017kbt,Capozzi:2018bps,Dev:2020kgz,Losada:2021bxx,Huang:2022wmz,Dev:2022bae,Davoudiasl:2023uiq,Lopes:2023vxn,Martinez-Mirave:2024dmw,Goertz:2024gzw,Sahu:2025vyy}. Recently, an intriguing possibility has been propounded \cite{Sen:2023uga} wherein the three active neutrinos have exactly zero mass in vacuum, while they acquire an effective mass due to the potential induced by neutrino-ULDM interactions. These ``refractive neutrino masses" can account for all the observed neutrino oscillation phenomena. Unlike the standard vacuum masses, refractive masses depend on the DM density along the path of neutrinos. When the energy of the neutrino is much greater than a `resonance energy' $E_R$, the refractive masses are almost independent of energy \cite{Sen:2023uga}. On the other hand, when the neutrino energy is much smaller in comparison to $E_R$, these masses are extremely small. These distinctive features of refractive masses help to reconcile constraints from neutrino oscillation experiments with the cosmological bounds on the sum of neutrino masses \cite{DESI:2025zgx} by choosing $10\ \text{eV} \leq E_R \leq 0.1\ \text{MeV}$ \cite{Sen:2023uga,Sen:2024pgb}. The proposed mechanism consists of an ultralight scalar $\phi$ that acts as ULDM and two additional sterile neutrinos $\chi_{1,2}$ that act as mediators to produce the neutrino-ULDM potential. The solar neutrino data, in particular, gives stringent constraints on the parameters of the model, namely, $m_\chi$ and $m_\phi$ \cite{Sen:2023uga}.

Some other developments in the context of refractive masses are the studies of the implications of this mechanism for Supernova neutrinos and the diffuse supernova neutrino background (DSNB) \cite{Ge:2024ftz,Perez-Gonzalez:2025qjh,Pompa:2025lbf}. If the DM background maintains coherence over relevant time scales, it may give rise to temporal fluctuations in neutrino masses and mixing angles, which can be tested against available neutrino oscillation data \cite{Berlin:2016woy,Brdar:2017kbt,Capozzi:2018bps,Dev:2020kgz,Losada:2021bxx,Huang:2022wmz,Dev:2022bae,Davoudiasl:2023uiq,Lopes:2023vxn,Martinez-Mirave:2024dmw,Goertz:2024gzw,Sahu:2025vyy,Sen:2024pgb}. Recently \cite{Cheek:2025kks}, it was pointed out that for $m_\phi \ll 10^{-14}$ eV, fast coherent oscillations in the ULDM field configuration may alter the oscillation probabilities in a way that is in tension with data from experiments like KamLAND \cite{KamLAND:2013rgu}. Moreover, for $m_\phi \gg 10^{-14}$eV, spatial fluctuations $\sim \mathcal{O}(1)$ may attenuate the neutrino oscillation amplitudes. However, as noted in \cite{Pompa:2025lbf}, the ability of ULDM to retain coherence across such long cosmological timescales is uncertain, particularly in the presence of virialization processes within galactic DM halos.

In this work, we analyze in detail the propagation of solar neutrinos in the refractive mass scenario. We develop a parametrization of the $5\times5$ unitary matrix, $\mathbb{P}$, that diagonalizes the Hamiltonian describing neutrino propagation. Our parametrization naturally incorporates the experimentally observed active-neutrino mixing matrix $U_{PMNS}$ as well as the mass-squared differences. It allows us to work in terms of 6 mixing angles, instead of 10, since the other 4 are guaranteed to stay close to zero during adiabatic propagation through the Sun. An appropriate ordering of rotations with these six angles reduces the $5\times 5$ analysis of the solar neutrino oscillations to a two-flavor oscillation problem. In the process, we also identify a larger class of structures of the neutrino-ULDM coupling matrix\footnote{In \cite{Sen:2023uga}, the analysis was carried out with a nearly Tri-bimaximal (TBM) mixing matrix.} which can satisfy all experimental constraints. We verify our analytical understanding of this parametrization by diagonalizing the mass matrix numerically to find the effective mixing angles and mass-squared values at different distances from the center of the Sun.

ULDM halos can form inside/around the Sun due to the gravitational capture and self-interactions of the scalar field $\phi$ on timescales comparable to the age of the Sun \cite{Budker:2023sex, Banerjee:2019xuy, Banerjee:2019epw}. We explore the implications of such a halo on the propagation of refractive neutrinos. We find that, though the reduction to a two-neutrino scenario is no longer valid in the presence of a halo, our parametrization remains robust, i.e., only the 6 chosen mixing angles are sufficient to analyze the problem.
If the propagation is adiabatic, the electron neutrino survival probability is affected by the halo only if neutrinos are produced deep inside the Sun in the ``region of halo domination" (RHD) or its periphery. In such cases, even very small and light DM halos, with radius $\sim 0.1 R_\odot$ and masses of $\sim 10^{-8} M_\odot$, can significantly affect the probability spectrum. If neutrinos are produced outside the RHD or its periphery, the aforementioned two-neutrino analysis remains valid and the solar neutrino solution is very close to the standard Mikheyev-Smirnov-Wolfestein (MSW) solution \cite{Mikheev:1986wj,Wolfenstein:1977ue}.

Solar neutrinos are produced via various nuclear reaction channels occurring within the Sun. Neutrinos in each channel are produced at different distances from the center of the Sun and contribute to the neutrino flux in specific energy ranges. As a result of a solar DM halo, the survival probability of $\nu_e$ of a given energy also depends on its production point. It therefore becomes instructive to examine the flux-averaged probability spectrum in order to compare our theoretical predictions with the observations made by neutrino detectors at the Earth. For this, we construct a general formula for the flux-averaged survival probability. We compare the features of upturn and jumps in the flux-averaged probability spectrum in the presence of the halo with the flux-averaged MSW probability spectrum.

While the above analysis obtains the solution to the solar neutrino problem as a small deviation from the standard MSW solution, we find that another qualitatively distinct solution, the so-called ``dark side of the Large Mixing Angle" (dark-LMA) solution \cite{deGouvea:2000pqg}, becomes more viable in the presence of the solar DM halo. In addition, it gives a flat probability spectrum in the $5-10$ MeV range as indicated by the Super-Kamiokande (SK) elastic scattering data. We identify several features in the spectrum which are specific to this scenario, and hence can be used to distinguish it from the standard MSW scenario with future data.

The paper is organized as follows. In section \ref{sec:the_model}, we outline the procedure of diagonalizing the $5\times5$ Hamiltonian in the basis of $\{\nu_e,\ \nu_\mu,\ \nu_\tau,\ \chi_1,\ \chi_2\}$ and we show how this $5\times5$ solar neutrino problem can be reduced to an effective two-neutrino problem. In section \ref{sec:dark_matter_effects}, we analyze the effects of a DM halo on the solar neutrino survival probability spectrum, with and without flux-averaging. In section \ref{sec:Dark_LMA}, we calculate the solar neutrino survival probability with the dark-LMA solution and point out its distinctive features. Finally, in section \ref{sec:conclusions}, we summarize our results.

\section{Refractive Neutrino Masses}\label{sec:the_model}
We work in the context of the model \cite{Sen:2023uga} where the SM is extended with two\footnote{It was argued in \cite{Sen:2023uga}, that this is the minimal choice possible for the number of additional sterile neutrino species required to generate two non-vanishing mass-squared differences.} additional light sterile Majorana neutrinos, $\chi_1$ and $\chi_2$, and an ultra-light complex scalar $\phi$. The $\phi$ in our model can constitute the DM. The relevant terms in the Lagrangian are

\begin{equation}
    \mathcal{L} \subset \sum_{\alpha=e,\mu,\tau} \sum_{k=1,2} \left(g_{\alpha k}\,\overline{(\chi_{k,L})^C}\,\nu_{\alpha,L}\,\phi^\star + m_{\chi_k}\,\overline{(\chi_{k,L})^C}\,\chi_{k,L} + \text{h.c.}\right) + V_\phi~,
\end{equation}
where $V_\phi$ denotes the self-potential of $\phi$. The propagation of neutrinos in the background of this ultra-light boson $\phi$ generates an effective potential for $\nu$ through the elastic forward-scattering $\nu\phi\to\nu\phi$, mediated by $\chi_1$ and $\chi_2$. The effective potential matrix, $V_{\alpha \beta}$, generated by the processes $\nu_\alpha \phi \to \nu_\beta \phi$ and $\nu_\alpha \bar{\phi} \to \nu_\beta \bar{\phi}$, is given by (see \cite{Choi:2019zxy,Choi:2020ydp,Smirnov:2021zgn})
\begin{equation}
    V_{\alpha \beta} = \sum_k g_{\alpha k}g_{\beta k}^\star \left[\frac{\bar{n}_\phi(2 E m_\phi - m_{\chi_k}^2)}{(2 E m_\phi - m_{\chi_k}^2)^2+(m_{\chi_k}\Gamma_{\chi_k})^2} + \frac{n_\phi}{2Em_\phi + m_{\chi_k}^2} \right], 
\end{equation}
where $E$ is the energy of neutrino, $m_\phi$ is the mass of $\phi$, $n_\phi(\bar{n}_{\phi})$ is the number density of $\phi(\bar{\phi})$, and $\Gamma_{\chi_k} \equiv \sum_k g_{\alpha k}^2m_{\chi_k}/8\pi$ is the total decay rate of $\chi_k$.
It can be seen from the above equation that the potential $V_{\alpha\beta}$ asymptotically approaches 
\begin{equation}
    V_{\alpha \beta} = \frac{1}{2E}\sum_k g_{\alpha_k}g_{\beta_k}^\star \frac{n_\phi+\bar{n}_\phi}{m_\phi}
\end{equation}
in the limit $E/E_R \gg 1$, where $E_R \simeq m_\chi^2/2m_\phi$ is the energy at which the potential has a resonant behavior. The effective quantities $m^2_{\alpha\beta}\equiv 2E\,V_{\alpha\beta}$ are constants in this asymptotic limit and may be treated as elements of the effective mass-squared matrix of the neutrinos. Thus, the potential due to background DM can induce ``refractive masses" for neutrinos.

\subsection{Diagonalizing the $5\times5$ mass matrix in the DM background} \label{subsec:texture}

Let us now work in the basis of $\{\nu_e,\ \nu_\mu,\ \nu_\tau,\ \chi_1,\ \chi_2\}$. In terms of the parameters of the Lagrangian, the $5 \times 5$ effective mass matrix can be written as
\begin{equation}\label{eq:mass_mat}
    \mathbb{M} = \begin{pmatrix}
        0_{3\times3} & \ \ & [g]_{\alpha k}\,\langle\phi\rangle^\star_{\text{coh}}\\
        [g^T]_{k \alpha}\,\langle\phi\rangle^\star_{\text{coh}} & \ \ & [m_{\chi_k}]^D
    \end{pmatrix},
\end{equation}
where $\left[g\right]$ is the matrix with elements $g_{\alpha\beta}$ and $[m_{\chi_k}]^D$ is a diagonal matrix with elements $m_{\chi_1}$ and $m_{\chi_2}$. Here $\langle\phi\rangle_{\text{coh}}$ is the expectation value of the DM field $\phi$ in the coherent state of particles, given by $\langle\phi\rangle_{\text{coh}} = Fe^{i\Phi}$ with $F=\sqrt{\frac{n_\phi+\bar{n}_\phi}{m_\phi}}$ \cite{Sen:2023uga}. This subsection outlines the procedure to diagonalize the mass matrix and identify its structure which can generate the observed solar and atmospheric mass-squared differences, $\Delta m_\text{sol}^2$ and $\Delta m_\text{atm}^2$, as well as the observed mixing, $U_{\text{PMNS}}$, in the active neutrino sector.

From eq. (\ref{eq:mass_mat}), we obtain the effective Hamiltonian $\mathbb{H}$ as
\begin{equation}
    \mathbb{H} = \frac{1}{2E}\mathbb{M}\mathbb{M^\dagger} = \frac{1}{2E} \begin{pmatrix}
        F^2\, [g] \cdot [g^\dagger] & \ \ \ \  & F\, e^{-i\Phi} [g]\cdot [m_{\chi_k}]^D \\
        F\, e^{i\Phi} [m_{\chi_k}]^D \cdot [g^\dagger]  & \ \ \ \ & F^2\,[g^T] \cdot [g^\star] + [m_{\chi_k}^2]^D
        \end{pmatrix}. \label{eq:hamltonian_H}
\end{equation}
From the above Hamiltonian, the time-evolutions of the states $\lvert\nu_\alpha\rangle$ and $\lvert\chi_k\rangle$ are
\begin{eqnarray}
    2E\left(i\frac{d}{d t}\lvert\nu_\alpha\rangle\right) &=& \sum_{\beta=e,\mu,\tau}\sum_{j=1,2} F^2 g_{\alpha j}g_{\beta j}^\star\lvert\nu_\beta\rangle + \sum_{j=1,2} F g_{\alpha j} m_{\chi_j}e^{-i\Phi}\lvert\chi_j\rangle\,,\\
    2E\left(i\frac{d}{d t}\lvert\chi_k\rangle\right) &=& \sum_{\beta=e,\mu,\tau} Fg_{\beta k}^\star m_{\chi_k}e^{i\Phi}\lvert\nu_\beta\rangle + \sum_{j=1,2}\left(\sum_{\beta=e,\mu,\tau}F^2g_{\beta k}^\star g_{\beta j} + m_{\chi_j}^2\delta_k^j\right)\lvert\chi_j\rangle. \quad
\end{eqnarray}
In order to get rid of the time-dependent complex phase $\Phi(t)$, we make a field-redefinition $\lvert \chi_k\rangle \to e^{-i\Phi} \lvert\chi_k\rangle$ to obtain the ``tilde" basis,
\begin{equation}
\lvert \widetilde{\nu}_\alpha \rangle = \left\{\lvert\nu_e\rangle,\lvert\nu_\mu\rangle,\lvert\nu_\tau\rangle,\lvert\widetilde{\chi}_1\rangle,\lvert\widetilde{\chi}_2\rangle\right\} = \left\{\lvert\nu_e\rangle,\lvert\nu_\mu\rangle,\lvert\nu_\tau\rangle,e^{-i\Phi}\lvert\chi_1\rangle,e^{-i\Phi}\lvert\chi_2\rangle\right\}.\label{eq:tilde_basis}
\end{equation}
In this basis, the effective Hamiltonian becomes
\begin{equation}
    \widetilde{\mathbb{H}} = \frac{1}{2E} \begin{pmatrix}
        F^2\, [g] \cdot [g^\dagger] & \ \ \ \ \ \ \ \  & F\, [g]\cdot [m_{\chi_k}]^D \\
        F\, [m_{\chi_k}]^D \cdot [g^\dagger] & \ \ \ \ \ \ \ \  & F^2\, [g^T] \cdot [g^\star] + [m_{\chi_k}^2+ 2E\dot{\Phi}]^D
        \end{pmatrix}. \label{eq:H_tilde_Fg}
\end{equation}
It is always possible to perform a Singular Value Decomposition (SVD) of the off-diagonal $3\times2$ block $[g]$ as
\begin{equation}
    [U^\dagger] \cdot F\, [g] \cdot [V] = \begin{pmatrix}
        0 & 0\\
        m_{a_1} & 0\\
        0 & m_{a_2}
    \end{pmatrix},\label{eq:SVD}
\end{equation}
using some $3\times3$ and $2\times 2$ unitary matrices, $[U]$ and $[V]$, respectively. 

We now provide a neat procedure for the diagonalizing $\widetilde{\mathbb{H}}$ when $m_{\chi_1} = m_{\chi_2} \equiv m_\chi$ and $[V]$ is an orthogonal matrix. In this case\footnote{Another case, where such a neat diagonalization can be carried out, is when $m_{\chi_1} \neq m_{\chi_2}$ but $[V] = \mathbb{I}_{2\times2}$.}, using $[U]$ and $[V]$ above, one can take the first step towards the block-diagonalization of $\widetilde{\mathbb{H}}$ as
\begin{eqnarray}
     \mathbb{H'} &=& \begin{pmatrix}
        [I]_{3\times3} & 0\\
        0 & [V^\dagger]_{2\times2}
    \end{pmatrix} \cdot \begin{pmatrix}
        [U^\dagger]_{3\times3} & 0\\
        0 & [I]_{2\times2}
    \end{pmatrix} \cdot \widetilde{\mathbb{H}} \cdot \begin{pmatrix}
        [U]_{3\times3} & 0\\
        0 & [I]_{2\times2}
    \end{pmatrix} \cdot \begin{pmatrix}
        [I]_{3\times3} & 0\\
        0 & [V]_{2\times2}
    \end{pmatrix}\nonumber\\
    &=& \frac{1}{2E}\begin{pmatrix}
    0 & 0 & 0 & 0 & 0\\
    0 & m_{a_1}^2 & 0 & m_{a_1} m_\chi & 0\\
    0 & 0 & m_{a_2}^2 & 0 & m_{a_2} m_\chi\\
    0 & m_{a_1} m_\chi & 0 & m_{a_1}^2+ m_{\chi}^2 + 2E\dot{\Phi} & 0\\
    0 & 0 & m_{a_2} m_\chi  & 0 & m_{a_2}^2+ m_{\chi}^2 + 2E\dot{\Phi}
    \end{pmatrix}.
\end{eqnarray}
Thus, using SVD, we have reduced the Hamiltonian to a system in which one of the neutrino species is completely decoupled and the rest are decoupled as pairs of two neutrinos, with the decoupled sub-blocks
\begin{eqnarray}
    \begin{pmatrix}
    H'_{22} & H'_{24}\\
    H'_{42} & H'_{44}
\end{pmatrix} = \begin{pmatrix}
    m_{a_1}^2 & m_{a_1} m_\chi\\
    m_{a_1} m_\chi & m_{a_1}^2+ m_{\chi}^2 + 2E\dot{\Phi}
\end{pmatrix}~,\label{eq:H24subblock}\\
\begin{pmatrix}
    H'_{33} & H'_{35}\\
    H'_{53} & H'_{55}
\end{pmatrix} = \begin{pmatrix}
    m_{a_2}^2 & m_{a_2} m_\chi\\
    m_{a_2} m_\chi & m_{a_2}^2+ m_{\chi}^2 + 2E\dot{\Phi}
\end{pmatrix}~,\label{eq:H35subblock}
\end{eqnarray}
where $H'_{ij} = \left[\mathbb{H'}\right]_{ij}$. We diagonalize these two sub-blocks with the rotation matrices $R_{22'}(\alpha_{22'})$ and $R_{33'}(\alpha_{33'})$, respectively, with the mixing angles given by
 \begin{equation}\label{eq:vacuum_mix_angles}
     \alpha_{22'} = \frac{1}{2} \text{tan}^{-1}\left(\frac{2m_{a_1}m_\chi}{m_\chi^2 + 2 E \dot{\Phi}}\right)\ \  \text{and}\ \ \alpha_{33'} = \frac{1}{2} \text{tan}^{-1}\left(\frac{2m_{a_2}m_\chi}{m_\chi^2 + 2 E \dot{\Phi}}\right)~.
 \end{equation}
The final mass eigenstates are labeled as $\{\nu_1,\nu_2,\nu_3,\nu_{2'},\nu_{3'}\}$. 

Thus, we have diagonalized $\widetilde{\mathbb{H}}$ as
\begin{equation}
    \mathbb{H}_D =  \mathbb{P}^\dagger \cdot \widetilde{\mathbb{H}}  \cdot \mathbb{P}~,
\end{equation}
where $\mathbb{H}_D$ is a diagonal matrix and
\begin{equation}
    \mathbb{P} =   \begin{pmatrix}
        [U]_{3\times3} & 0\\
        0 & [I]_{2\times2}
    \end{pmatrix} \cdot \begin{pmatrix}
        [I]_{3\times3} & 0\\
        0 & [V]_{2\times2}
    \end{pmatrix} \cdot R_{22'}(\alpha_{22'})\cdot R_{33'}(\alpha_{33'})~.\label{eq:P_defn}
\end{equation} 
When $\alpha_{22',33'} \ll 1$, the matrix $[U]_{3\times3}$ approximately diagonalizes the active $3\times3$ sub-matrix of $\widetilde{\mathbb{H}}$. Hence, we may identify $U_{\text{PMNS}} \approx [U]_{3\times3}$. We have thus parametrized $\mathbb{P}$ as 
\begin{equation}
\mathbb{P} = R_{23}(\theta_{23})\cdot R_{13}(\theta_{13},\delta_{\text{CP}}) \cdot R_{12}({\theta_{12}}) \cdot R_{2'3'}(\theta_{2'3'})\cdot R_{22'}(\alpha_{22'})\cdot R_{33'}(\alpha_{33'}), \label{eq:parameterization}   
\end{equation}
where we define $R_{2'3'}(\theta_{2'3'}) \equiv \begin{pmatrix}
        [I]_{3\times3} & 0\\
        0 & [V]_{2\times2}
    \end{pmatrix}$. 
    
We denote the final diagonalized mass-squared matrix, 
$\mathbb{M}_D^2 = 2 E\cdot\mathbb{H}_D$, by
    \begin{equation}
        \mathbb{M}_D^2 = \text{diag}\left\{m_1^2,\,m_2^2,\,m_3^2,\,m_{2'}^2,\,m_{3'}^2\right\}\,,
    \end{equation}
where the lowest eigenvalue is $m_1^2= 0$. We express the remaining two pairs of eigenvalues in terms of their averages $\overline{m^2}$ and splittings $\Delta m^2$ as
\begin{equation}
    m_{2}^2 = \overline{m^2}_{\!22'}-\frac{\Delta m_{22'}^2}{2},\ \ m_{2'}^2 = \overline{m^2}_{\!22'}+\frac{\Delta m_{22'}^2}{2}\, ,
\end{equation}
\begin{equation}
    m_{3}^2 = \overline{m^2}_{\!33'}-\frac{\Delta m_{33'}^2}{2},\ \ m_{3'}^2 = \overline{m^2}_{\!33'}+\frac{\Delta m_{33'}^2}{2}\, .
\end{equation}
Using eqs. (\ref{eq:H24subblock}) and (\ref{eq:H35subblock}), the quantities $\overline{m^2}$ and $\Delta m^2$ may be expressed as
\begin{equation}\label{eq:avg_mass}
    \overline{m^2}_{\!22'} = m_{a_{1}}^2+\frac{m_\chi^2+2E\dot{\Phi}}{2},\ \ \ \ \overline{m^2}_{\!33'} = m_{a_{2}}^2+\frac{m_\chi^2+2E\dot{\Phi}}{2}~,    
\end{equation}
and 
\begin{equation}\label{eq:active_sterile_mass_diff}
\Delta m_{22'}^2 = \sqrt{4 m_{a_{1
}}^2 m_\chi^2+\left(m_\chi^2+2E\dot{\Phi}\right)^2},\ \ \ \Delta m_{33'}^2 = \sqrt{4 m_{a_{2}}^2 m_\chi^2+\left(m_\chi^2+2E\dot{\Phi}\right)^2}~.    
\end{equation}
 We confine ourselves to normal mass ordering. In this scenario, the measured values of $\Delta m^2_\text{sol}$ and $\Delta m^2_\text{atm}$ may be identified with
 \begin{equation}\label{eq:mass_sq_matching}
     \Delta m^2_\text{sol} \approx \Delta m^2_{21} = \left( \overline{m^2}_{\!22'}-\frac{\Delta m_{22'}^2}{2} \right)~,\ \ \ \ \Delta m^2_\text{atm} \approx \Delta m^2_{31} = \left( \overline{m^2}_{\!33'}-\frac{\Delta m_{33'}^2}{2}\right)~.
 \end{equation}
 Henceforth, we denote $m_{\text{sol}} = \sqrt{\Delta m^2_{\text{sol}}}$ and $m_{\text{atm}} = \sqrt{\Delta m^2_{\text{atm}}}$. 

When $\alpha_{22'} \ll 1$, eq. (\ref{eq:vacuum_mix_angles}) gives $m_\chi^2+2E\dot{\Phi} \gg  2 m_{a_{1}} m_\chi$, and hence $\Delta m_{22'}^2 \approx m_\chi^2+2E\dot{\Phi}$. Then from eq. (\ref{eq:avg_mass}) and eq. (\ref{eq:active_sterile_mass_diff}), we get  $\overline{m^2}_{\!22'} - \Delta m_{22'}^2 \approx m_{a_{1}}^2$, leading to $m_{a_{1}}^2 \approx \Delta m_{\text{sol}}^2$. On the other hand, when $\Delta m_{22'}^2 \ll \overline{m^2}_{\!22'}$, eq. (\ref{eq:active_sterile_mass_diff}) gives $m_\chi^2+2E\dot{\Phi} \ll  \overline{m^2}_{\!22'}$, which in turn from eq. (\ref{eq:avg_mass}) implies $m_{a_{1}}^2 \approx \overline{m^2}_{\!22'} \approx \Delta m_{\text{sol}}^2$. Thus, in either of these regimes, we get $m_{a_1} \approx m_{\text{sol}}$. Similar arguments lead to $m_{a_2} \approx m_{\text{atm}}$.

Note that the orthogonal matrix $[V]_{2\times2}$ does not appear in the above identification. Therefore, the present measurements of active neutrino oscillations give no hint about its value. On the other hand, the couplings between neutrinos and DM, $g_{\alpha k}$, depend on $[V]_{2\times2}$ as
\begin{equation}
    [g] = \frac{1}{F}\, [U] \cdot \begin{pmatrix}
        0 & 0\\
        m_{a_1} & 0\\
        0 & m_{a_2}
    \end{pmatrix} \cdot [V^T] \approx \frac{1}{F}\, U_{\text{PMNS}} \cdot \begin{pmatrix}
        0 & 0\\
        m_\text{sol} & 0\\
        0 & m_\text{atm}
    \end{pmatrix} \cdot [V^T]~.
\end{equation}
The angle of rotation defining $[V]_{2\times2}$ controls the structure of the coupling matrix $[g]_{3\times2}$. Given a $U_{\text{PMNS}}$ and $\Delta m^2$'s, our procedure allows for a one-parameter family of elements of $[g]_{3\times2}$. 
We thus offer a larger class of structures of $[g]_{3\times2}$ that can satisfy all the experimental constraints, than has been considered earlier\footnote{For example, the $[g]$ considered in \cite{Sen:2023uga} corresponds to the special case $U_{\text{PMNS}} = U_{\text{TBM}}$ and $[V] = \mathbb{I}_{2\times2}$.}. Our choice of the order of rotations, as given in eq. (\ref{eq:P_defn}), has the additional advantage of reducing the solar neutrino propagation into an effective two-neutrino problem, as will be discussed later in section \ref{subsec:two_flavor_red}.

\iffalse
Using eq.(\ref{eq:SVD}), $[g]_{\alpha k}$ is given by
\begin{equation}
    [g] = [U] \cdot \begin{pmatrix}
        0 & 0\\
        m_{a_1} & 0\\
        0 & m_{a_2}
    \end{pmatrix} \cdot [V^T] \simeq U_{\text{PMNS}} \cdot \begin{pmatrix}
        0 & 0\\
        m_\text{sol} & 0\\
        0 & m_\text{atm}
    \end{pmatrix} \cdot [V^T], 
\end{equation}
where we have assumed that $m_{a_1} \simeq m_\text{sol}$ and $m_{a_2} \simeq m_\text{atm}$, which should be the case in order to avoid active-sterile oscillations as discussed in the next section \ref{sec:matter_effects}. The authors in \cite{Sen:2023uga} assumed a nearly tribimaximal form(TBM) of the neutrino mass matrix. Although close to the $U_\text{PMNS}$ observed so far, TBM fails to account for a non-zero $\theta_{13}$ and $\delta_{CP}$. As already discussed, our approach on the other hand does not have any such limitations. Indeed, this is only a particular choice of parameterization, but it has an additional advantage of reducing the solar neutrino propagation into an effective two-neutrino problem, as discussed in subsection \ref{subsec:two_flavor_red}.\fi

\subsection{Diagonalizing the $5\times 5$ Hamiltonian inside the Sun}\label{sec:matter_effects}\label{subsec:two_flavor_red}

We now apply the diagonalization procedure as described in the last section to analyze the solar neutrino oscillations. Since the active-sterile oscillations in this case are severely restricted by present experimental data \cite{deGouvea:2009fp}, we should have either very small active-sterile mixing angles ($\alpha_{22',33'} \lesssim 10^{-2}$) or very small active-sterile mass squared differences ($\Delta m_{22',33'}^2 \lesssim 10^{-12}\ \text{eV}^2$). These are indeed the same conditions mentioned in section \ref{sec:the_model} where our diagonalization procedure works well. 

When $\phi$ is a real scalar field, $\langle\phi\rangle_{\text{coh}} = F e^{i \Phi}$ is real, implying $\Phi = 0$. When $\alpha_{22',33'} \lesssim 10^{-2}$, eq. (\ref{eq:vacuum_mix_angles}) would imply a large value of $m_\chi \sim \text{eV}$. This would lead to a large decay rate $\Gamma_{\chi_k}$, thus invalidating the scenario. On the other hand, the scenario of $\Delta m_{22',33'}^2 \lesssim 10^{-12}\ \text{eV}^2$ is allowed but yields a constraint $m_\chi  \lesssim 10^{-10}\, \text{eV}$ from eq. (\ref{eq:active_sterile_mass_diff}). 

For a complex $\phi$, the scenario $\alpha_{22',33'} \lesssim 10^{-2}$ is allowed, since we can have $E\dot{\Phi} \gg m_\chi^2$. It can be shown \cite{Sen:2023uga}  that $\dot{\Phi}$ can take a value between $0$ and $m_\phi$. For $\dot{\Phi} \sim m_\phi$ and $\alpha_{22',33'} \lesssim 10^{-2}$, one would get $m_\chi/m_\phi < 10^6$. The scenario $\Delta m_{22',33'}^2 \lesssim 10^{-12}\ \text{eV}^2$ continues to give the constraint of $m_\chi \lesssim 10^{-10}\, \text{eV}$ from eq. (\ref{eq:active_sterile_mass_diff}).

We now analyze the diagonalization of the effective Hamiltonian in the presence of matter inside the Sun, in addition to the uniform DM background. The forward scattering of neutrinos on the electrons, protons and neutrons inside the Sun yields
\begin{equation}
\mathbb{H_\odot} = (\mathbb{H} + \mathbb{V}_\odot),
\end{equation}
where $\mathbb{H}$ is given by eq. (\ref{eq:hamltonian_H}), and 
\begin{equation}
    \mathbb{V}_\odot = \begin{pmatrix}
        V_{\text{CC}}+V_{\text{NC}} & 0 & 0 & 0 & 0\\
        0 & V_{\text{NC}} & 0 & 0 & 0\\
        0 & 0 & V_{\text{NC}} & 0 & 0\\
        0 & 0 & 0 & 0 & 0\\
        0 & 0 & 0 & 0 & 0
    \end{pmatrix}, 
\end{equation}
with $V_{\text{CC}} = \sqrt{2} G_F N_e$ and $V_{\text{NC}} = -\frac{1}{\sqrt{2}} G_F N_n$. Here $G_F,\ N_e\ \text{and}\ N_n$ are the Fermi constant, solar electron density and the solar neutron density, respectively. For our numerical calculations, we take the matter density profile inside the Sun from \cite{Bahcall:2000nu}. 

Clearly, re-phasing $\lvert\chi_1\rangle$ and $\lvert \chi_2 \rangle$ to obtain eq. (\ref{eq:tilde_basis}) keeps $\mathbb{V}_\odot$ invariant as the sterile neutrinos do not interact with matter, i.e. the last two diagonal elements of $\mathbb{V}_\odot$ are zero. Thus, we can go to the ``tilde basis" as in the earlier subsection, where the Hamiltonian becomes $\widetilde{\mathbb{H}}_\odot = \widetilde{\mathbb{H}} + \mathbb{V}_\odot$. In this basis, the neutrino states evolve as
\begin{equation}\label{eq:prop_orig_two_flav}
    i\frac{d}{d t}\lvert\widetilde{\nu}_\alpha\rangle = \widetilde{\mathbb{H}}_\odot\lvert\widetilde{\nu}_\alpha\rangle = \left(\mathbb{P}\cdot\mathbb{H}_D\cdot\mathbb{P^\dagger} + \mathbb{V}_\odot\right)\lvert\widetilde{\nu}_\alpha\rangle.
\end{equation}
We denote the mixing matrix that diagonalizes $\widetilde{\mathbb{H}}_\odot$ as $\mathbb{P}_\odot$ and parametrize it in the same way as in eq. (\ref{eq:parameterization}).

Unless specified, in the calculations hereafter, we take the effective neutrino mass and mixing parameters in the presence of uniform DM background to be consistent with the current neutrino oscillation data \cite{Esteban:2024eli}. That is, we take $\Delta m^2_{21} = 7.5\times10^{-5}\, \text{eV}^2$, $\Delta m^2_{31} = 2.5\times10^{-3}\, \text{eV}^2$, $\theta_{12} = 34.4^\circ$, $\theta_{23} = 49.1^\circ$, $\theta_{13} = 8.54^\circ$ and $\delta_{CP}=0$. This corresponds to $m_{\text{sol}} \approx 8.7 \times 10^{-3}$ eV and $m_{\text{atm}} \approx 0.05$ eV. In the small active-sterile mixing scenario, we take $m_\phi = 10^{-9}\, \text{eV}$, $m_\chi^2 = 10^{-7}\, \text{eV}^2$, $\dot{\Phi} = m_\phi$, and $\theta_{2'3'} = 25^\circ$ in the uniform DM background, which set the other relevant quantities to $E_R = 50\,  \text{eV}$, $\alpha_{22'} = 0.008^\circ$ and $\alpha_{33'} = 0.04^\circ$. In the small active-sterile mass-squared difference scenario, we take $m_\phi = 10^{-22}\, \text{eV}$, $m_\chi^2 = 10^{-20}\, \text{eV}^2$, $\dot{\Phi}=0$, and $\theta_{2'3'} = 25^\circ$ in the uniform DM background, which set the other relevant quantities to $E_R = 50\,  \text{eV}$, $\Delta m^2_{22'} = 1.7 \times 10^{-12}\, \text{eV}^2$ and $\Delta m^2_{33'} = 10^{-11}\, \text{eV}^2$. We consider the entire local DM density to be composed of the field $\phi$, and thus set $\rho_\phi =  0.3\ \text{GeV\,cm}^{-3}$ in the solar system.

\subsubsection{Small active-sterile mixing angles} \label{subsubsec:small_active_sterile_mixing}

    For small $\alpha_{22'}$ and $\alpha_{33'}$, the matrices $R_{22'}$ and $R_{33'}$ are nearly identity matrices whose commutators with $\mathbb{H}_D$ are suppressed by $\alpha_{22'}$ and $\alpha_{33'}$ respectively. Further, for analytic simplicity one may take $R_{13} \approx \mathbb{I}$ since $\theta_{13}$ has been measured to be small. Also note that $R_{2'3'}$ commutes with $R_{12}$. Thus,
    \begin{equation}
        \mathbb{P}\cdot\mathbb{H}_D\cdot\mathbb{P^\dagger} \approx R_{23}\cdot R_{2'3'}\cdot R_{12} \cdot\mathbb{H}_D\cdot R_{12}^\dagger \cdot R_{2'3'}^\dagger \cdot R_{23}^\dagger\, .
    \end{equation}
 We take the rotations $R_{23}$ and $R_{2'3'}$ to be approximately time-independent. Further, note that $\mathbb{V}_\odot$ remains invariant under the rotations $R_{23}$ and $R_{2'3'}$, i.e. $R_{2'3'}^\dagger\cdot R_{23}^\dagger \cdot \mathbb{V}_\odot \cdot R_{23} \cdot R_{2'3'} = \mathbb{V}_\odot$. Then in the basis
 \begin{equation}
     \lvert\widehat{\nu}_\alpha\rangle \equiv R_{2'3'}^\dagger \cdot R_{23}^\dagger \lvert\widetilde{\nu}_\alpha\rangle\,,
 \end{equation}
 we obtain the evolution equations as
\begin{equation}
    i\frac{d}{d t}\lvert\widehat{\nu}_\alpha\rangle = \left(R_{12}\cdot\mathbb{H}_D\cdot R_{12}^\dagger + \mathbb{V}_\odot\right)\lvert\widehat{\nu}_\alpha\rangle\, .\label{eq:two_flavour}
\end{equation}
 Thus, we have reduced the propagation of refractive solar neutrinos to a two-flavor mixing scenario with  oscillations taking place dominantly between $\lvert\nu_e\rangle$ and $\left(\text{cos}\,\theta_{23}\,\lvert\nu_\mu\rangle - \text{sin}\,\theta_{23}\,\lvert\nu_\tau\rangle\right)$.
 
The unitary matrix $\mathbb{P}$ parametrized in eq. (\ref{eq:parameterization}) in terms of 6 mixing angles is not the most general $5\times5$ unitary matrix (which should include 10 mixing angles). Still, $\mathbb{P}_\odot$ with possibly time-dependent $\theta_{12}$ is sufficient to approximately diagonalize not only $\widetilde{\mathbb{H}}$ but also $\widetilde{\mathbb{H}}_\odot$. This is because, if the other 4 angles are taken to be zero in the uniform DM background, they are not subsequently generated during the evolution [eq. (\ref{eq:two_flavour})] inside the Sun. 

\begin{figure}[t!]
    \centering
    \includegraphics[width=0.375\linewidth]{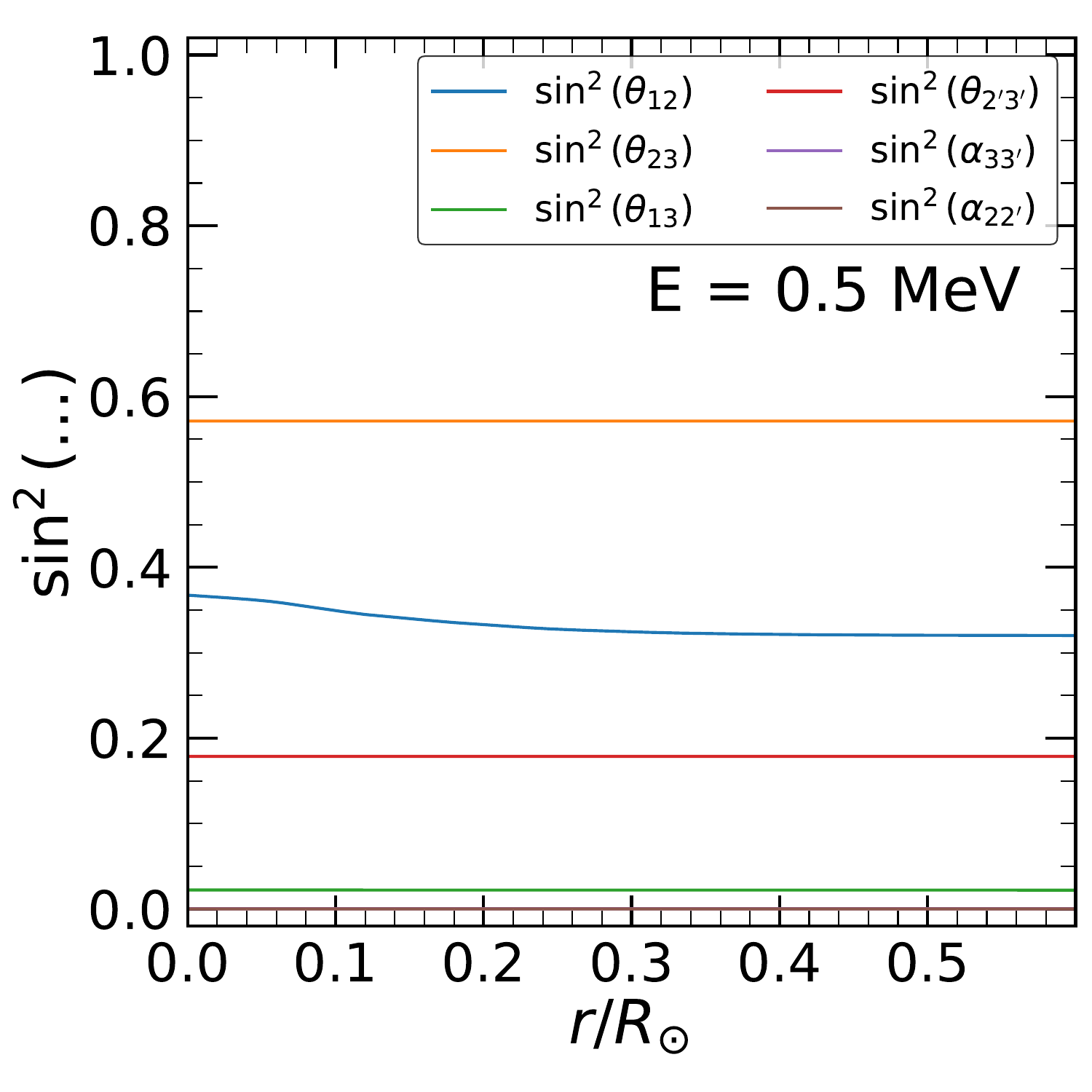}
    \includegraphics[width=0.375\linewidth]{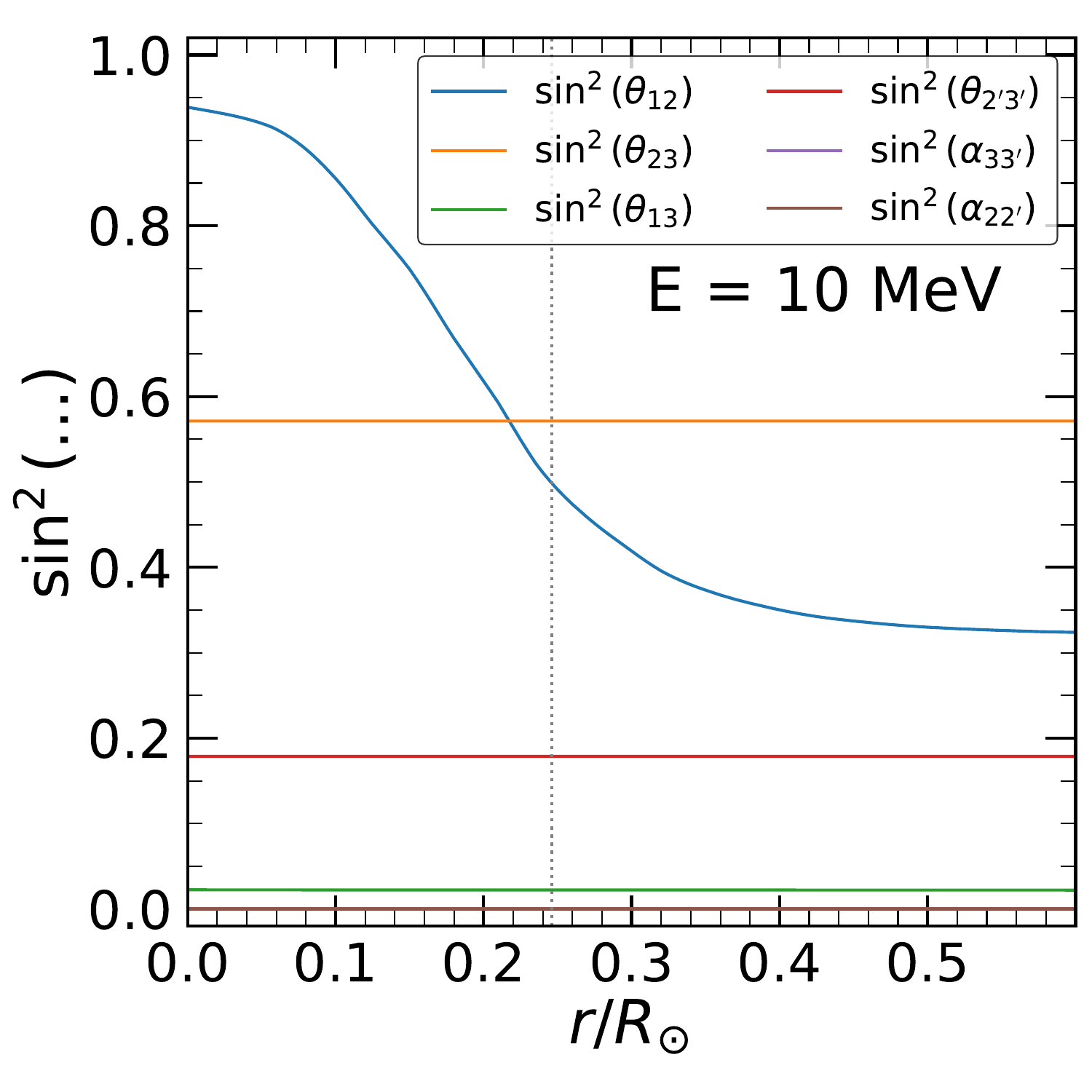}
    \includegraphics[width=0.395\linewidth]{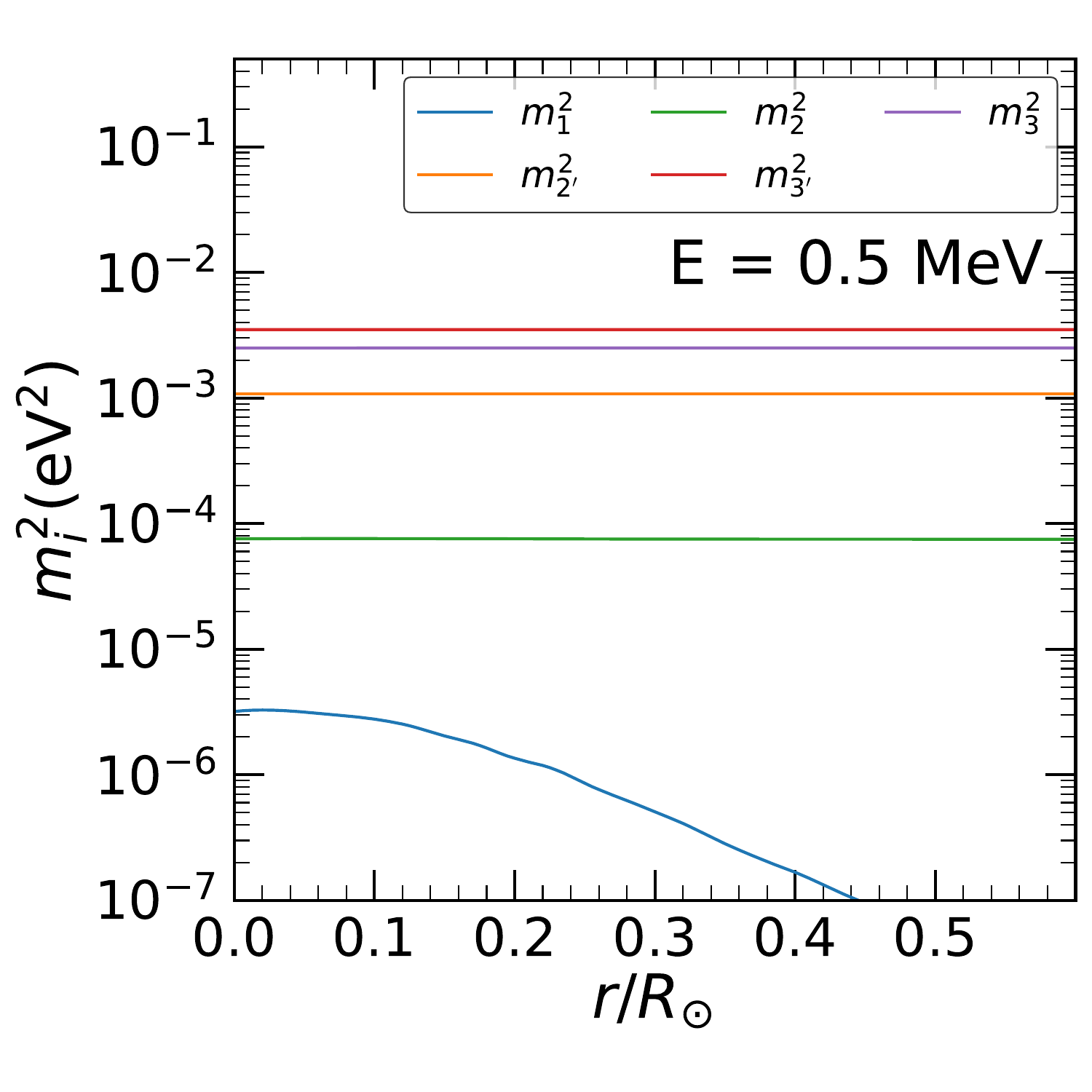}
    \includegraphics[width=0.395\linewidth]{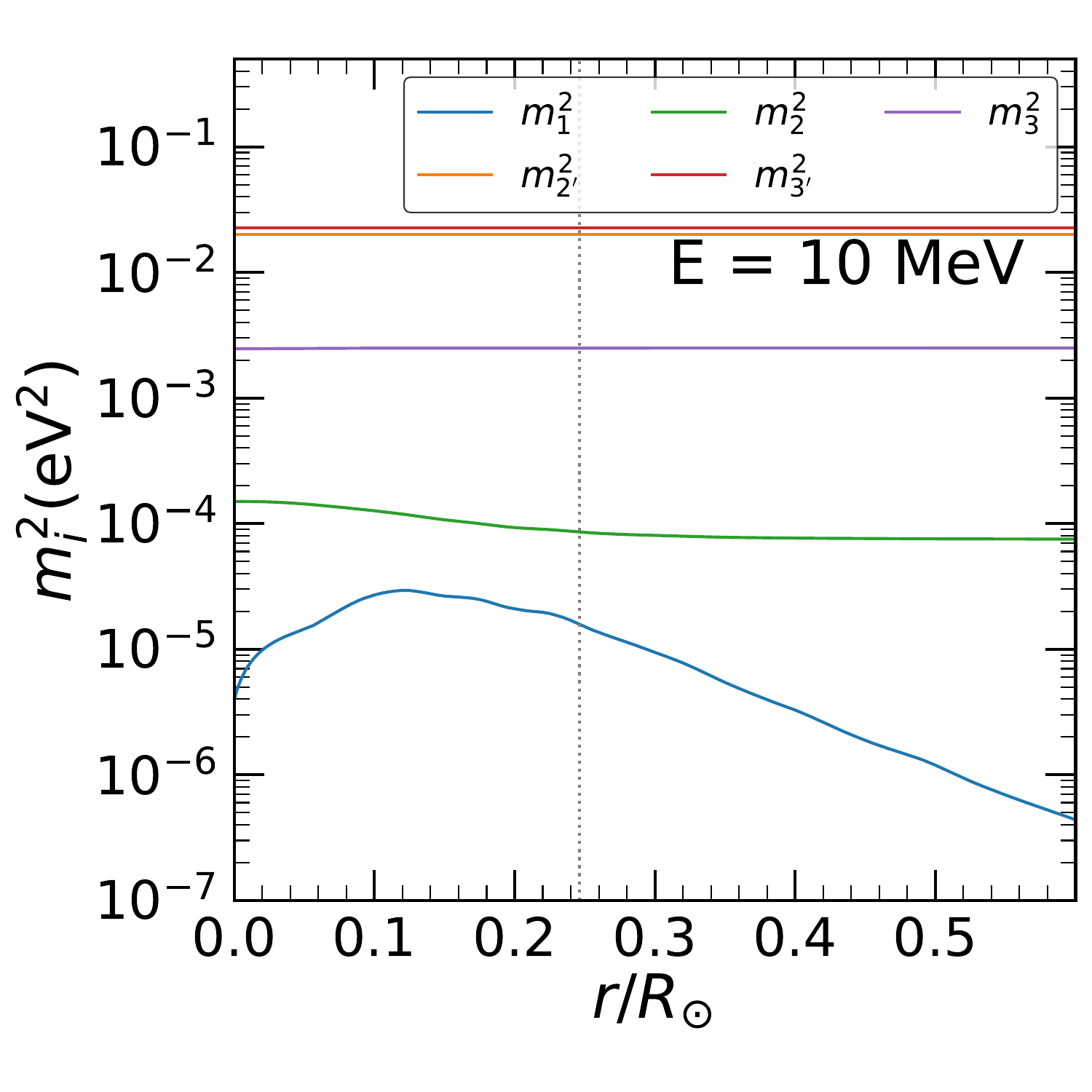}
    \caption{Mixing angles (upper panels) and mass-squared values (lower panels) in the 5-neutrino system for $E =0.5 \ \text{MeV (left panels) and}\ E =10\ \text{MeV (right panels)}$ when $m_\phi = 10^{-9}$ eV, $m_\chi^2 = 10^{-7}\, \text{eV}^2$ and $\dot{\Phi}=m_\phi$. The choice of parameters ensures small active-sterile mixing angles. We take $s_{12}^2 = 0.32$ in the uniform DM background. The vertical dotted line in the right panels represents the radial distance at which the MSW resonance occurs for $E = 10$ MeV. It is evident that in this scenario, the neutrino evolution can be treated effectively as a two-flavor problem.}\label{fig:mixing_angle_no_halo}
\end{figure}

  In Fig. \ref{fig:mixing_angle_no_halo}, we show the values of mixing angles and mass-squares as functions of radial distance from the center of the Sun, for two energies. From the figure, we observe the following:
  \begin{itemize}
      \item All the angles except $\theta_{12}$ and all the masses except $m_1^2$ and $m_2^2$ are approximately constant inside the Sun. This validates our assumptions made for reaching eq. (\ref{eq:two_flavour}) that reduce the situation to a two-flavor mixing scenario.

      \item At high energies (here $E = 10\, \text{MeV}$), the angle $\theta_{12}$ shows the MSW resonance ($\sin^2 \theta_{12} = 0.5$) around $r \approx 0.246\, R_\odot$. The value of $\Delta m_{21}^2$ also reaches a minimum at this distance, as expected in a 2-flavor scenario. At low energies (here $E = 0.5\, \text{MeV}$), there is no MSW resonance.

      \item The masses $m_{2'}^2$ and $m_{3'}^2$ change with energy whereas $m_{2}^2$ and $m_{3}^2$ remain constant with energy in the uniform DM background. This behavior can be well-understood as the energy dependence of $\overline{m^2}_{\!22'}$ in eq. (\ref{eq:avg_mass}) exactly cancels the energy dependence of $\Delta m_{22'}^2/2$ in eq. (\ref{eq:active_sterile_mass_diff}) when $m_\chi^2 \to 0$. Same is the case for $\overline{m^2}_{\!33'}$ and $\Delta m_{33'}^2/2$.

  \end{itemize}
    This simplified picture of oscillations in the presence of matter effects, as discussed in section \ref{subsec:texture}, brings out the hidden power of our parametrization, in particular of the ordering of the unitary matrices in eq. (\ref{eq:parameterization}).

\subsubsection{Small active-sterile mass-squared differences}

When $\Delta m_{22'}^2\ll\overline{m^2}_{\!22'}$ in the uniform DM background, then $m_{a_{1}} \approx  m_{\text{sol}}$ (see section \ref{subsec:texture}). We have shown that $m_\chi \lesssim 10^{-10}\, \text{eV}$ when $\Delta m_{22'}^2\ll 10^{-12}\, \text{eV}^2$ in section \ref{subsec:two_flavor_red}. Using these, eq. (\ref{eq:vacuum_mix_angles}) gives 
\begin{equation}
    \tan 2 \alpha_{22'} = \frac{2\, m_{\text{sol}}\ m_\chi}{m_\chi^2 + 2 E\, \dot{\Phi}}  \geq \frac{m_{\text{sol}}\ m_\chi}{(E_R+  E)\,m_\phi} = \frac{2\,m_{\text{sol}}}{m_\chi\left(1+ \frac{E}{E_R}\right)} \gg 1
\end{equation}
for the solar neutrino energy range ($E \lesssim 20\, \text{MeV}$). Here we have used $10\, \text{eV} \leq E_R \leq 0.1\, \text{MeV}$ and $\dot{\Phi} \leq m_\phi$ \cite{Sen:2023uga}. Similar arguments lead to $\tan 2 \alpha_{33'} \gg 1$. Thus, $\alpha_{22',33'} \approx \pi/4$ in the uniform DM background.

While analyzing the neutrino propagation inside the Sun, we may treat $\mathbb{V}_\odot$ as a perturbation on top of $\widetilde{\mathbb{H}}$ as in eq. (\ref{eq:prop_orig_two_flav}). When $2E\, V_{\text{CC,NC}} \ll \Delta m^2_{22',33'}$, active-sterile mixing angles remain $\approx \pi/4$ as argued above. However, when $\Delta m_{22'}^2\ll 2E\, V_{\text{CC,NC}}\ll\overline{m^2}_{\!22',33'}$, the two sets of eigenvalues of the unperturbed Hamiltonian are very close, and the non-degenerate perturbation series converges very slowly. (This is exactly the situation very close to the surface of the Sun, near $r\sim 0.95\, R_\odot$ where matter density falls sharply and neutrino propagation may not be adiabatic.) To overcome this, we use the nearly-degenerate perturbation theory. 

The Hamiltonian $\widetilde{\mathbb{H}}_\odot$ in eq. (\ref{eq:prop_orig_two_flav}) may be written as
\begin{equation}
    \widetilde{\mathbb{H}}_\odot= \mathbb{P}\cdot\mathbb{H}_D\cdot\mathbb{P^\dagger} + \mathbb{V}_\odot \equiv \widetilde{\mathbb{H}}^{(0)} + \mathbb{H}_{\text{pert}}\,, \label{eq:H_tot}
\end{equation}
where
\begin{equation}
    \widetilde{\mathbb{H}}^{(0)} = \frac{1}{2E}\, \mathbb{P}\cdot \text{diag}\left\{0\,,m_2^2\,,m_3^2\,,m_2^2\,,m_3^2\right\}\cdot\mathbb{P}^\dagger
\end{equation} 
and
\begin{equation}\label{eq:H_pert}
    \mathbb{H}_{\text{pert}} = \mathbb{V}_\odot + \frac{1}{2E}\,\mathbb{P}\cdot\text{diag}\left\{0,0,0,\Delta m_{22'}^2,\Delta m_{33'}^2\right\}\cdot\mathbb{P^\dagger}\, .
\end{equation}
Thus, the unperturbed system corresponds to one non-degenerate and two pairs of degenerate eigenvalues. Since $\Delta m_{22',33'}^2 \ll \overline{m^2}_{\!22',33'}$, we may neglect the second term in eq. (\ref{eq:H_pert}) and $\mathbb{H}_{\text{pert}} \approx \mathbb{V}_\odot$. Then, for the nearly-degenerate perturbation theory, since $\mathbb{P}^\dagger \cdot \widetilde{\mathbb{H}}^{(0)}\cdot\mathbb{P}$ is already diagonal, we need to choose the unperturbed basis such that the $\{22'\}$ and $\{33'\}$ blocks of $\mathbb{P}^\dagger \cdot \mathbb{V}_\odot\cdot\mathbb{P}$ are diagonal. In the ``tilde" basis used in eq. (\ref{eq:H_tot}), these blocks are
\begin{equation}
    w_{22'}\begin{pmatrix}
        1&1\\
        1&1
    \end{pmatrix}\ \ \text{and}\ \ 
    w_{33'}\begin{pmatrix}
        1&1\\
        1&1
    \end{pmatrix}\,, \label{eq:H_22'_33'}
\end{equation}
respectively, where
\begin{eqnarray}
    w_{22'} &=& \frac{1}{8}\left(4\, V_{\text{CC}}+3\, V_{\text{NC}} + V_{\text{NC}} \cos 2\theta_{12} - 2\, V_{\text{NC}} \cos 2\theta_{13} \sin^2\theta_{12}\right)\,,\\ \nonumber
    w_{33'} &=& \frac{1}{4}\left(2\, V_{\text{CC}} + V_{\text{NC}} + V_{\text{NC}} \cos 2\theta_{13}\right)\,.
\end{eqnarray}
Here, we have used $\alpha_{22',33'} = \pi/4$ in $\mathbb{P}$. The blocks in eq. (\ref{eq:H_22'_33'}) can be diagonalized by additional rotations $R_{22'}(\pi/4)$ and $R_{33'}(\pi/4)$, respectively. The net unitary matrix $\mathbb{P}^{(0)}$ diagonalizing $\widetilde{\mathbb{H}}^{(0)}$ is then
\begin{eqnarray}
    \mathbb{P}^{(0)} &=& \mathbb{P}\cdot R_{22'}(\pi/4)\cdot R_{33'}(\pi/4)\\ \nonumber
    &=& R_{23}(\theta_{23})\cdot R_{13}(\theta_{13},\delta_{\text{CP}}) \cdot R_{12}({\theta_{12}}) \cdot R_{2'3'}(\theta_{2'3'})\cdot R_{22'}\left(\pi/2\right)\cdot R_{33'}\left(\pi/2\right)\,,
\end{eqnarray}
where we have used the fact that $R_{22'}$ and $R_{33'}$ commute with each other. The matrix $\mathbb{P}^{(0)}$ that we derived approximately diagonalizes the complete $\widetilde{\mathbb{H}}_\odot$ when $\Delta m_{22'}^2\ll 2E\, V_{\text{CC,NC}}\ll\overline{m^2}_{\!22',33'}$. Note that we could have chosen to diagonalize the above blocks by $R_{22'}(-\pi/4)$ and $R_{33'}(-\pi/4)$. However, our convention maintains the ordering $m_{2,3}^2 < m_{2',3'}^2$ even inside the Sun, so that the continuity of levels in the level crossing diagrams remains intact.

The rotations $R_{22',33'}(\pi/2)$ correspond to 
the interchange of the order of the eigenvalues of 
$\{2 \leftrightarrow 2'\}$ and $\{3 \leftrightarrow 3'\}$. Therefore, the MSW level crossing inside the Sun now happens between $\nu_1\leftrightarrow\nu_{2'}$ instead of between $\nu_1\leftrightarrow\nu_{2}$. However the argument in section \ref{subsubsec:small_active_sterile_mixing}, reducing the situation to two-flavour mixing remains unchanged and hence,  the only angle undergoing significant change is $\theta_{12}$.

\begin{figure}[t!]
    \centering
    \includegraphics[width=0.375\linewidth]{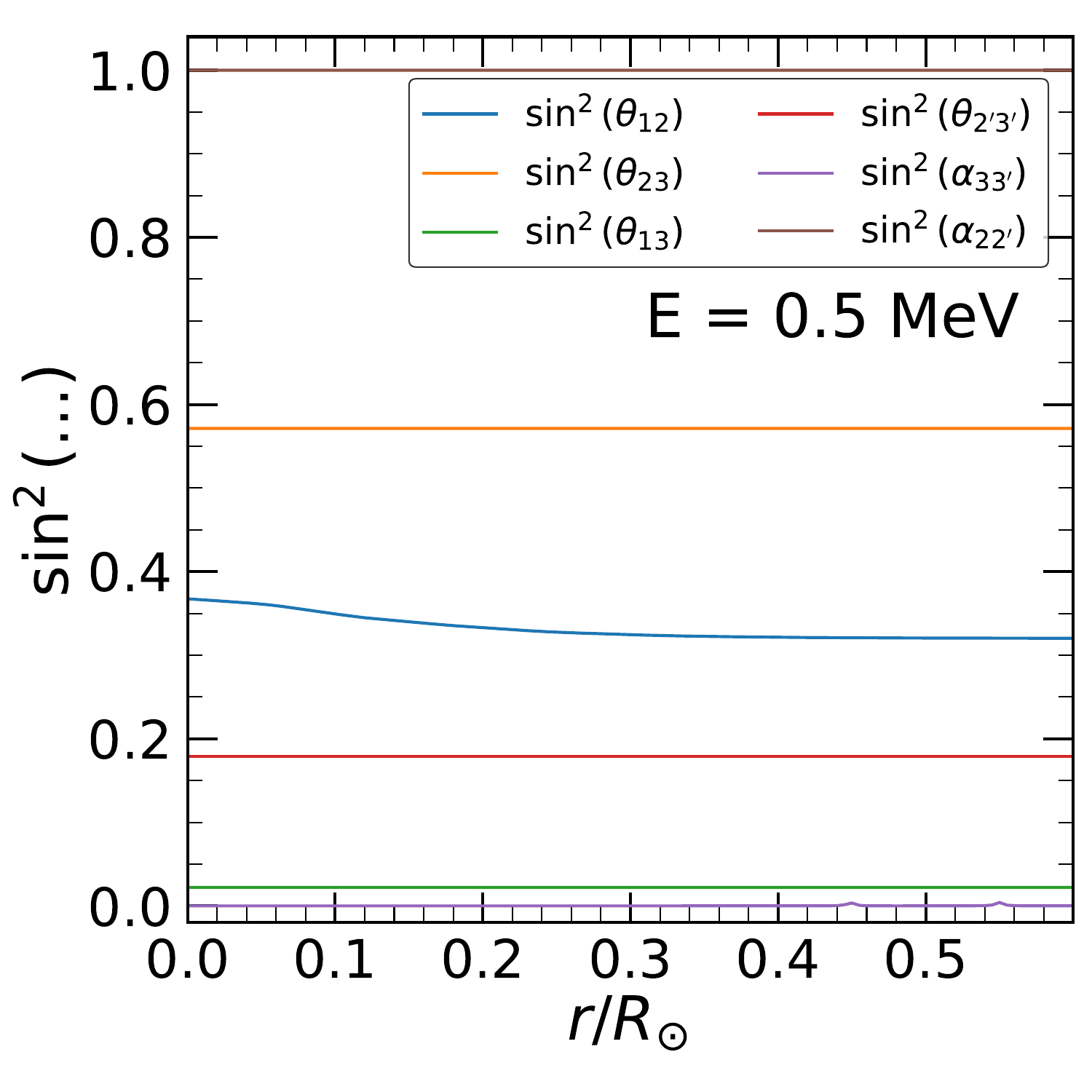}
    \includegraphics[width=0.375\linewidth]{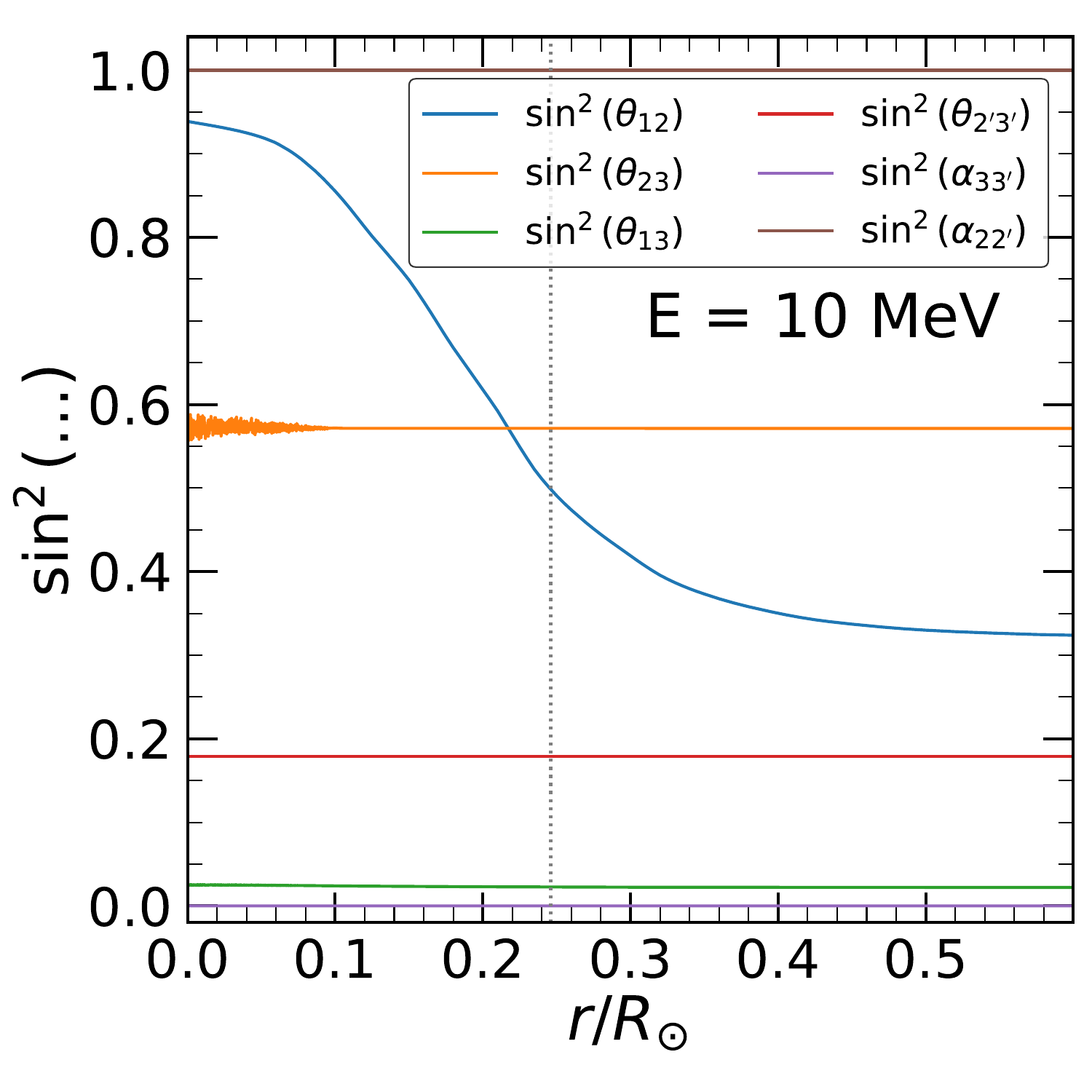}
    \includegraphics[width=0.395\linewidth]{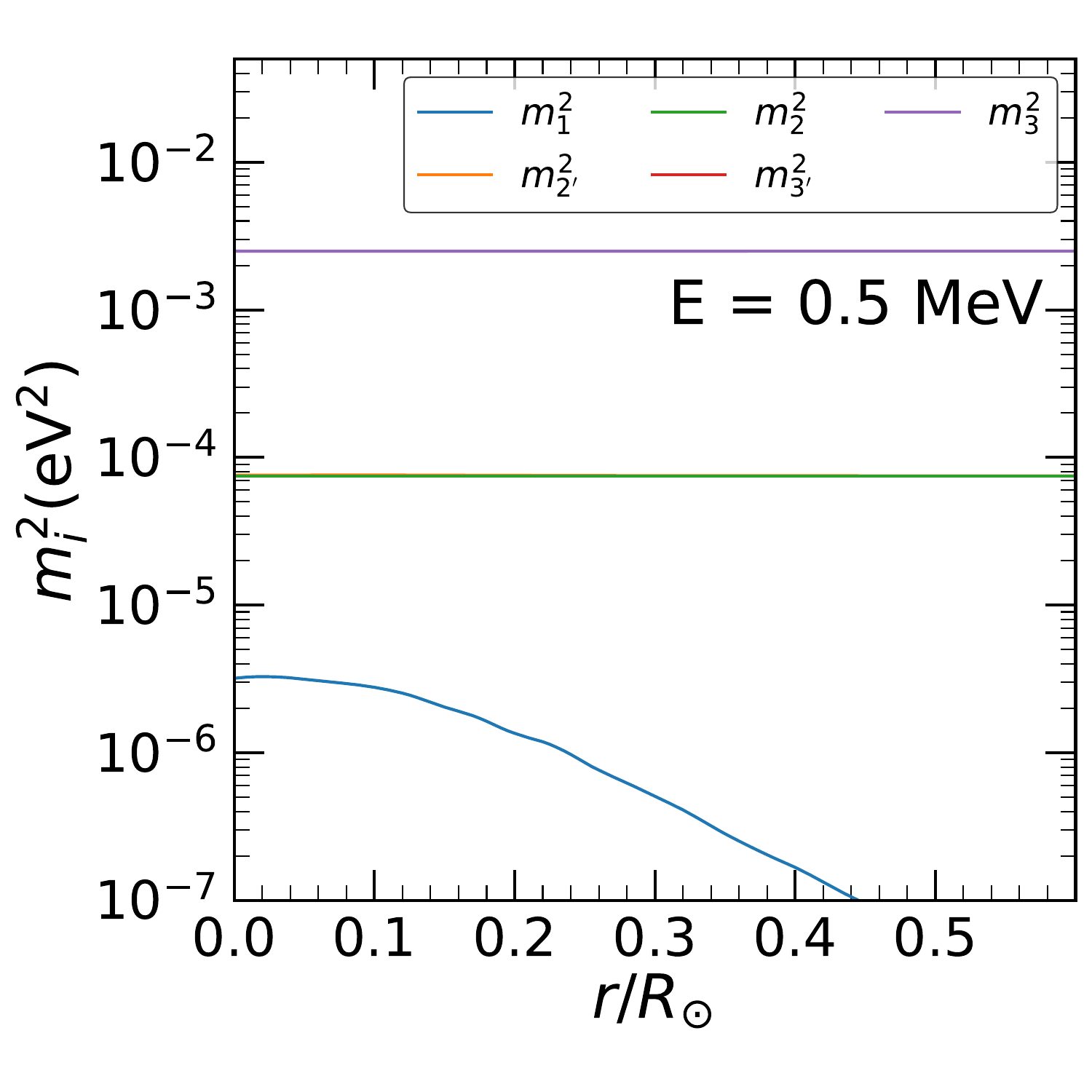}
    \includegraphics[width=0.395\linewidth]{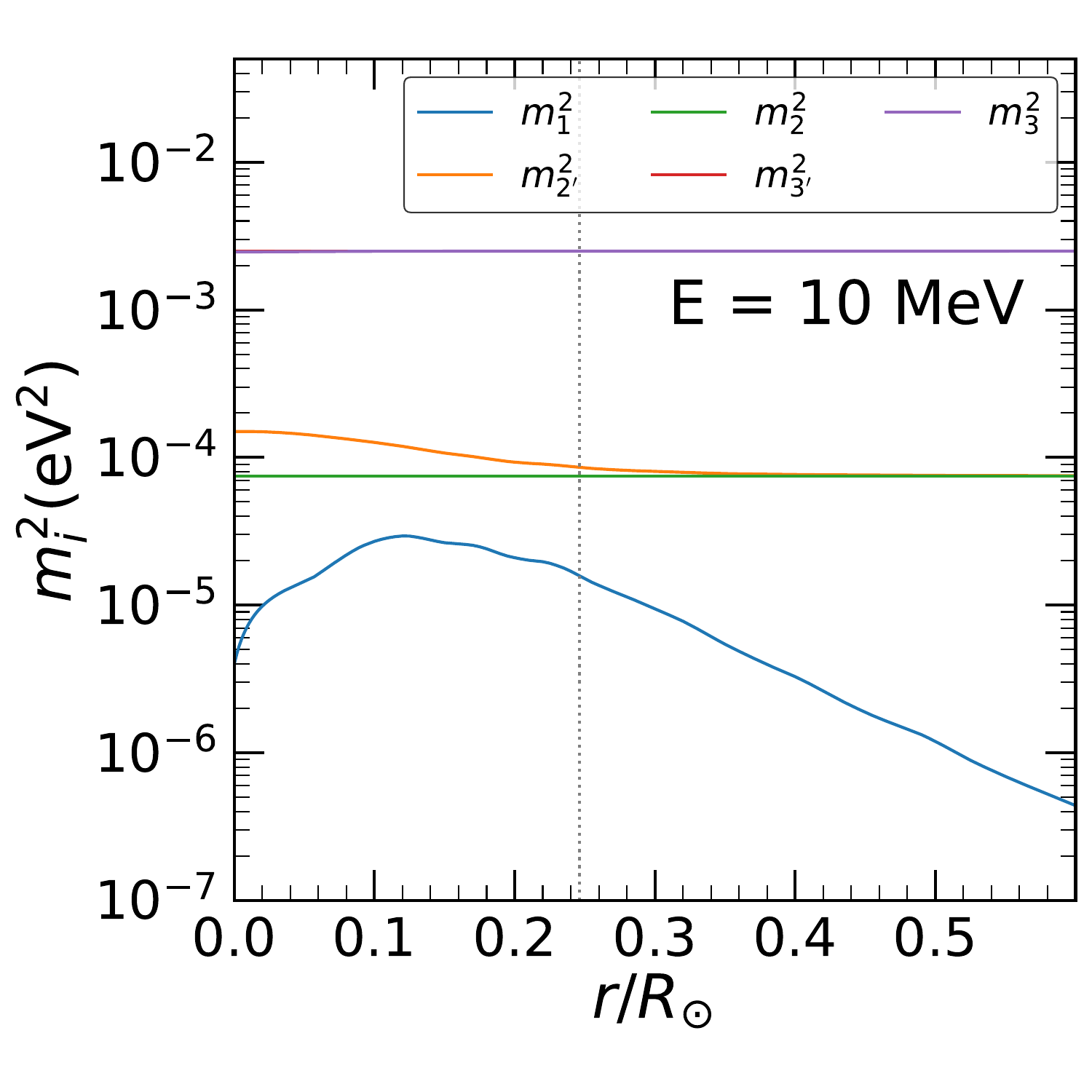}
    \caption{Mixing angles (upper panels) and mass-squared values (lower panels) in the 5-neutrino system for $E =0.5 \ \text{MeV (left panels) and}\ E =10\ \text{MeV (right panels)}$ when $m_\phi = 10^{-22}$ eV, $m_\chi^2 = 10^{-20}\, \text{eV}^2$ and $\dot{\Phi}=0$. The choice of parameters ensures small active-sterile mass-squared difference. We take $s_{12}^2 = 0.32$ in the uniform DM background. The vertical dotted line in the right panels represents the radial distance at which the MSW resonance occurs for $E = 10$ MeV. It is evident that in this scenario, the neutrino evolution can be treated effectively as a two-flavor problem.}\label{fig:mixing_angle_no_halo_small_mass_sq}
\end{figure}

The above analytical insights may be verified numerically from Fig. \ref{fig:mixing_angle_no_halo_small_mass_sq}. Here the following observations can be made.
\begin{itemize}
      \item All the angles except $\theta_{12}$ and all masses except $m_{1}^2$ and $m_{2'}^2$ are approximately constant inside the Sun.
      
      \item At high energies (here $E = 10\, \text{MeV}$), the angle $\theta_{12}$ shows the MSW resonance ($\sin^2 \theta_{12} = 0.5$) around $r \approx 0.246\, R_\odot$. The value of $\Delta m_{2'1}^2$ also reaches a minimum at this distance. At low energies (here $E = 0.5\, \text{MeV}$), there is no resonance mimicking the standard MSW scenario.

      \item Both $\alpha_{22'}$ and $\alpha_{33'}$ remain $\pi/2$ inside the Sun as explained earlier.

      \item All the masses remain essentially unchanged with energy in the uniform DM background. This is because $\Delta m_{22'}^2 \ll \overline{m^2}_{\!22'}$ and $\overline{m^2}_{\!33'}$ and $\Delta m_{33'}^2$ in this scenario.

  \end{itemize}

\section{Solar Neutrinos and ``Dark" Matter Effects}\label{sec:dark_matter_effects}

When the DM density is uniform in space, the discussion of section \ref{sec:matter_effects} implies that the solar neutrino survival probability remains the same as in the standard MSW framework. In this section, we consider the possibility that there is a DM halo around (or inside) the Sun produced due to the gravitational capture of $\phi$ by the Sun. In this scenario, we explore in detail the effects of this solar DM halo on the survival probability of solar neutrinos.

DM halos around/inside the Sun have been studied in literature earlier \cite{Budker:2023sex, Banerjee:2019xuy, Banerjee:2019epw}. As discussed in \cite{Budker:2023sex, Banerjee:2019xuy, Banerjee:2019epw}, such DM halos may have typical radii $R_\star$ greater or smaller than $R_\odot$. In these two scenarios, which we denote by X and Y, respectively, the density profiles are given by
\begin{eqnarray}
    X: \quad \rho_\phi &=& \rho_0\ \text{Exp} (-2r/R_\star)\ \ \text{where}\ \ \rho_0 = \frac{M_\star}{\pi\,  R_\star^3}\,,\ \ R_\star = \frac{M_\text{Pl}^2}{m_\phi^2} \frac{1}{M_\odot}\,,\\
    Y: \quad \rho_\phi &=& \rho_0\ \text{Exp}(-r^2/R_\star^2)\ \  \text{where}\ \ \rho_0 = \frac{M_\star}{\pi^{3/2}\,  R_\star^3}\,, \ \ R_\star= \left(\frac{M_\text{Pl}^2}{m_\phi^2} \frac{R_\odot^3}{M_\odot}\right)^{1/4}, \label{eq:halo_density}
\end{eqnarray}
where $\rho_0$ is the DM density at the center of the  halo, $M_\text{Pl}$ is the Planck scale and $M_{\odot(\star)}$ denote the solar (halo) mass. Note that in both these scenarios, $M_\star \ll M_\odot$.

Thus, a solar DM halo can be characterized by two quantities, namely, the mass of the particle $m_\phi$ which it is composed of, setting the radius $R_\star$ of the halo, and the total mass of the halo $M_\star$, setting the DM overdensity at its center. We would like to emphasize that, though the analysis employs particular forms of the halo density profile, the qualitative features would be valid for other profiles as well.

\subsection{Propagation of solar neutrinos through the solar DM halo}\label{subsec:dark_matter_effects}

When neutrinos propagate through the Sun with a DM halo, they feel two effects due to the medium: (i) the usual charged and neutral current interactions of neutrinos with the solar matter, and (ii) the change in the refractive mass matrix due to the space-dependent DM density inside the halo. The Hamiltonian for propagation can be expressed as $\widetilde{\mathbb{H}}^\text{halo}_\odot = \widetilde{\mathbb{H}}+\mathbb{V}_\odot$, where now $\widetilde{\mathbb{H}}$ (see eq. (\ref{eq:H_tilde_Fg})) is not a constant matrix but has a dependence on the distance from the center of the Sun, $r$, due to the space-dependent DM density. Let us parametrize the overdensity due to the halo with 
\begin{equation}
\xi(r) \equiv \sqrt{\frac{\rho_\phi(r)}{\rho_\infty}} = \frac{F(r)}{F_\infty},    
\end{equation}
where $F(r) = \lvert \langle \phi(r)\rangle_{\text{coh}}\rvert$ as defined in section \ref{subsec:texture} and $F_\infty$ is the uniform value of $F$ outside the halo. Deep inside the DM halo, $\xi \gg 1$ and we may use the approximation $\widetilde{\mathbb{H}}^\text{halo}_\odot \simeq \widetilde{\mathbb{H}}$ so that
\begin{equation}
    m_1 \approx 0\,,\ \ m_{2,2'}(r) \approx m_{a_{1}}(r) \approx \xi(r)\ m_{\text{sol}}\,,\ \ m_{3,3'}(r) \approx m_{a_{2}}(r) \approx \xi(r)\ m_{\text{atm}}\,.
\end{equation}

 The unitary matrices $[U]$ and $[V]$ involved in the SVD of $[g]_{3\times2}$ (see eq. (\ref{eq:SVD})) do not change. Hence, the mixing angles, $\theta_{12}\,, \theta_{23}\,, \theta_{13}\,, \theta_{2'3'}$ are not affected by the DM halo and the solar matter in this region. On the other hand, the two active-sterile mixing angles 
\begin{equation}\label{eq:alpha22p_DMhalo_exp}
 \alpha_{22'} \simeq \frac{1}{2} \text{tan}^{-1}\left(\frac{2\xi(r)\,m_{\text{sol}}\,m_\chi}{m_\chi^2 + 2 E \dot{\Phi}}\right)\ \  \text{and}\ \  \alpha_{33'} \simeq \frac{1}{2} \text{tan}^{-1}\left(\frac{2\xi(r)\,m_{\text{atm}}\,m_\chi}{m_\chi^2 + 2 E \dot{\Phi}}\right)   
\end{equation}
do not remain small any more, even though they may be small in the absence of any DM halo. Moreover, $\Delta m_{22',33'}^2 \simeq 2 m_{a_{1,2}}\, m_\chi \simeq 2\xi(r)\,  m_{\text{sol}}\, m_\chi$ also becomes large. The other 4 angles, which may be required to parametrize a general $5\times5$ unitary matrix, remain zero as the Hamiltonian is now approximately $\widetilde{\mathbb{H}}$ (with an $r$-dependence of $F$), for which only the 6 angles are sufficient following the discussion in section \ref{subsec:texture}. However, unlike the situation in section \ref{subsubsec:small_active_sterile_mixing}, we see that the commutators of $R_{22'}$ and $R_{33'}$ with $\mathbb{H}_D$ are not negligible. This would result in active-sterile oscillations inside the DM halo. We will call this region ($\xi \gg 1$) as the ``region of halo domination" (RHD) in the subsequent sections of the paper.

With the approximation $\theta_{13} \approx 0$, an electron neutrino produced at $r=r_P$ inside the Sun, when represented in the instantaneous mass eigenbasis, corresponds to
\begin{equation}
    \lvert\nu_e^P(r_P)\rangle = \begin{pmatrix}
        c_{12}^P\\
        c_{22'}^P s_{12}^P\\
        0\\
        s_{22'}^P s_{12}^P\\
        0
    \end{pmatrix}\,,
\end{equation}
where $c_{ij}$ and $s_{ij}$, respectively, denote the cosines and sines of angles $\theta_{ij}$ or $\alpha_{ij}$, and the superscript $P$ indicates the angle at production. After adiabatic radial propagation\footnote{We have checked numerically that adiabaticity holds in the range of parameters considered.}, the state becomes
\begin{equation}
    \lvert\nu_e^P(r)\rangle = \begin{pmatrix}
        c_{12}^P\ e^{-i\phi_1(r)}\\
        c_{22'}^P s_{12}^P\ e^{-i\phi_2(r)}\\
        0\\
        s_{22'}^P s_{12}^P e^{-i\phi_{2'}(r)}\\
        0
    \end{pmatrix}\,,
\end{equation}
where 
\begin{equation}
    \phi_i(r) = \int_{r_P}^{r}\frac{m_i^2(r')}{2E} dr'.
\end{equation} 
The probability that this state is detected as electron neutrino $\lvert \nu_e^D \rangle$ at the Earth is
\begin{equation}
    P_{ee} = \lvert\langle \nu_e^D(r_{E})\lvert\nu_e^P(r_P)\rangle\rvert^2\,,
\end{equation}
where $r_E$ is the radius of Earth's orbit around the Sun. Since the mass eigenstates effectively decohere, $P_{ee}$ becomes\footnote{Without neglecting $\theta_{13}$, we get
\begin{equation}
P_{ee} = \left(c_{13}^P c_{13}\right)^2\left[\left(c_{12}^P c_{12}\right)^2 + \left(s_{12}^P s_{12}\right)^2\left\{\left(c_{22'}^P c_{22'}\right)^2+\left(s_{22'}^P s_{22'}\right)^2\right\}\right]+\left(s_{13}^P s_{13}\right)^2\left[\left(c_{33'}^P c_{33'}\right)^2+\left(s_{33'}^P s_{33'}\right)^2\right]   \nonumber 
\end{equation}
Here, $s_{13}^P \simeq s_{13}$ as $\theta_{13}$ remains almost constant within the Sun even in the presence of a DM halo.}
\begin{equation}\label{eq:surv_prob}
    P_{ee} = \left(c_{12}^P c_{12}\right)^2 + \left(s_{12}^P s_{12}\right)^2\left[\left(c_{22'}^P c_{22'}\right)^2+\left(s_{22'}^P s_{22'}\right)^2\right] + \mathcal{O}\left(s_{13}^2\right), 
\end{equation}
where the sines and cosines without a superscript $P$ are for angles at the site of detection. When the neutrino is produced outside the RHD, eq. (\ref{eq:surv_prob}) is still valid with the production angles now governed by the standard solar matter. 

We see that the survival probability $P_{ee}$ inside the RHD, given by eq. (\ref{eq:surv_prob}), differs from the standard MSW expression, $P_{ee}^{\text{MSW}}$, which is given by
\begin{equation}\label{eq:surv_prob_MSW}
    P_{ee}^{\text{MSW}} = \left(c_{12}^P c_{12}\right)^2 + \left(s_{12}^P s_{12}\right)^2.
\end{equation} 
Only in the limit when both $\alpha_{22'}^P \ll 1$ as well as $\alpha_{22'} \ll 1$, do we retrieve eq. (\ref{eq:surv_prob_MSW}) from eq. (\ref{eq:surv_prob}). 

If the DM halo engulfs the Sun, i.e. $R_\star > R_\odot$, then the production angles for both low energy $(\sim 0.1\ \text{MeV})$ as well as the high energy $(\sim 10\ \text{MeV})$ solar neutrinos will be the same, resulting in the same survival probability at all energies. This is not favored by the present experimental observations of the solar neutrino spectrum which suggests $P_{ee} \sim 0.53$ at low energy and $P_{ee} \sim 0.32$ around $5 - 10\ \text{MeV}$. Thus, we will always consider a DM halo confined within the Sun ($R_\star < R_\odot$) in the later parts of the paper. 

As the solar neutrinos are mostly produced within the core of the Sun, we further have $R_\star \lesssim 0.3\, R_\odot$. For such values of $R_\star$, eq. (\ref{eq:halo_density}) implies $m_\phi \gtrsim 10^{-12}$ eV. This, in turn, also puts bounds on the value of $m_\chi$ to be $\gtrsim 10^{-5}$ eV given that $E_R \gtrsim 10\ \text{eV}$. As discussed in section \ref{subsec:two_flavor_red}, for the real scalar field $\phi$ to lead to a small active-sterile mass difference avoiding active-sterile oscillations of solar neutrinos, we need $m_\chi \lesssim 10^{-10}$ eV. Therefore, we can have viable DM halos only if $\phi$ is a complex scalar field. In this work, we therefore choose a complex $\phi$ and focus on the possibility of small active-sterile mixing angles that was discussed in section \ref{subsec:two_flavor_red}.

\subsection{Mixing angles and level crossings in the presence of a solar DM halo}\label{subsec:mixing_angle_evolution}

\begin{figure}[t!]
    \centering
    \includegraphics[width=0.375\linewidth]{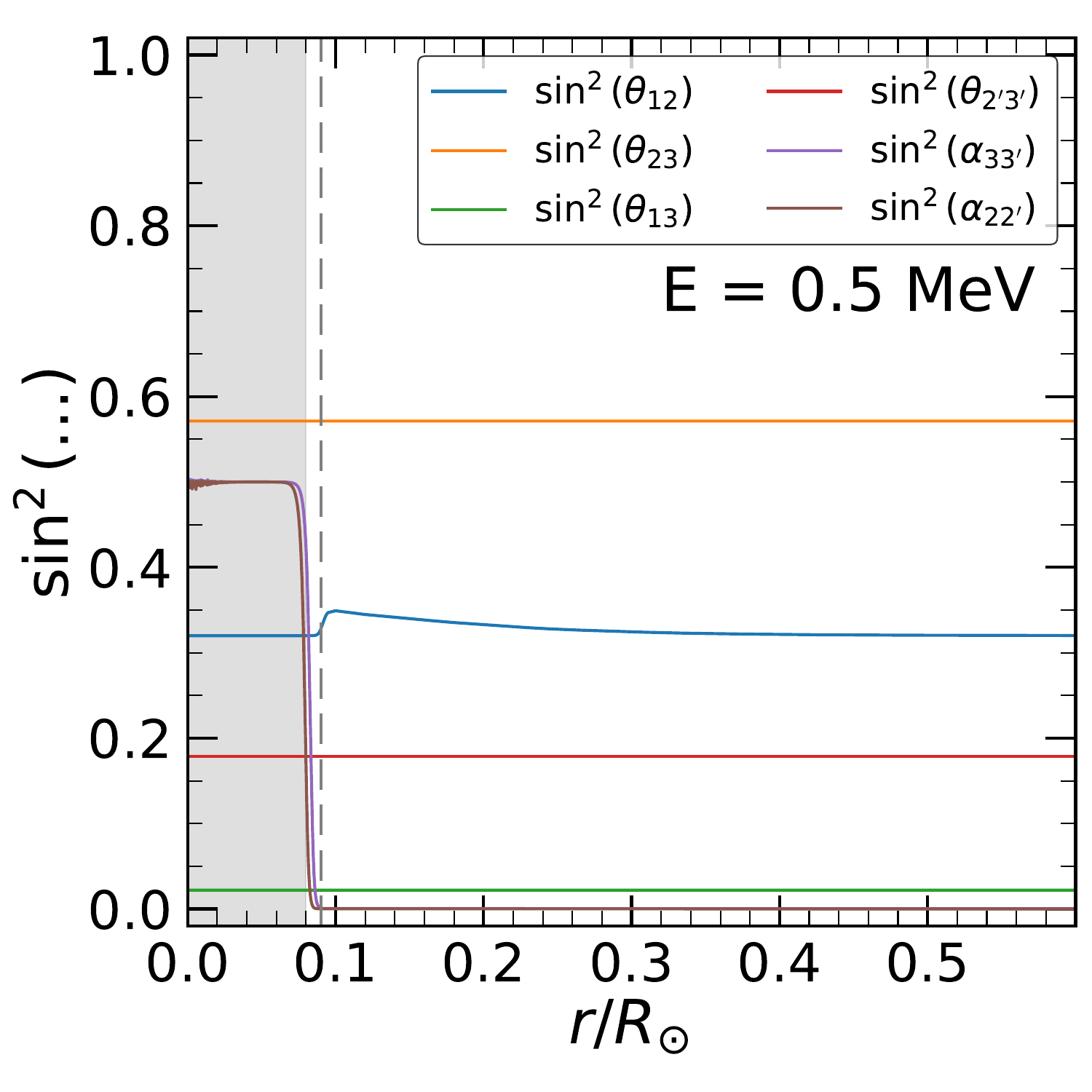}
    \includegraphics[width=0.375\linewidth]{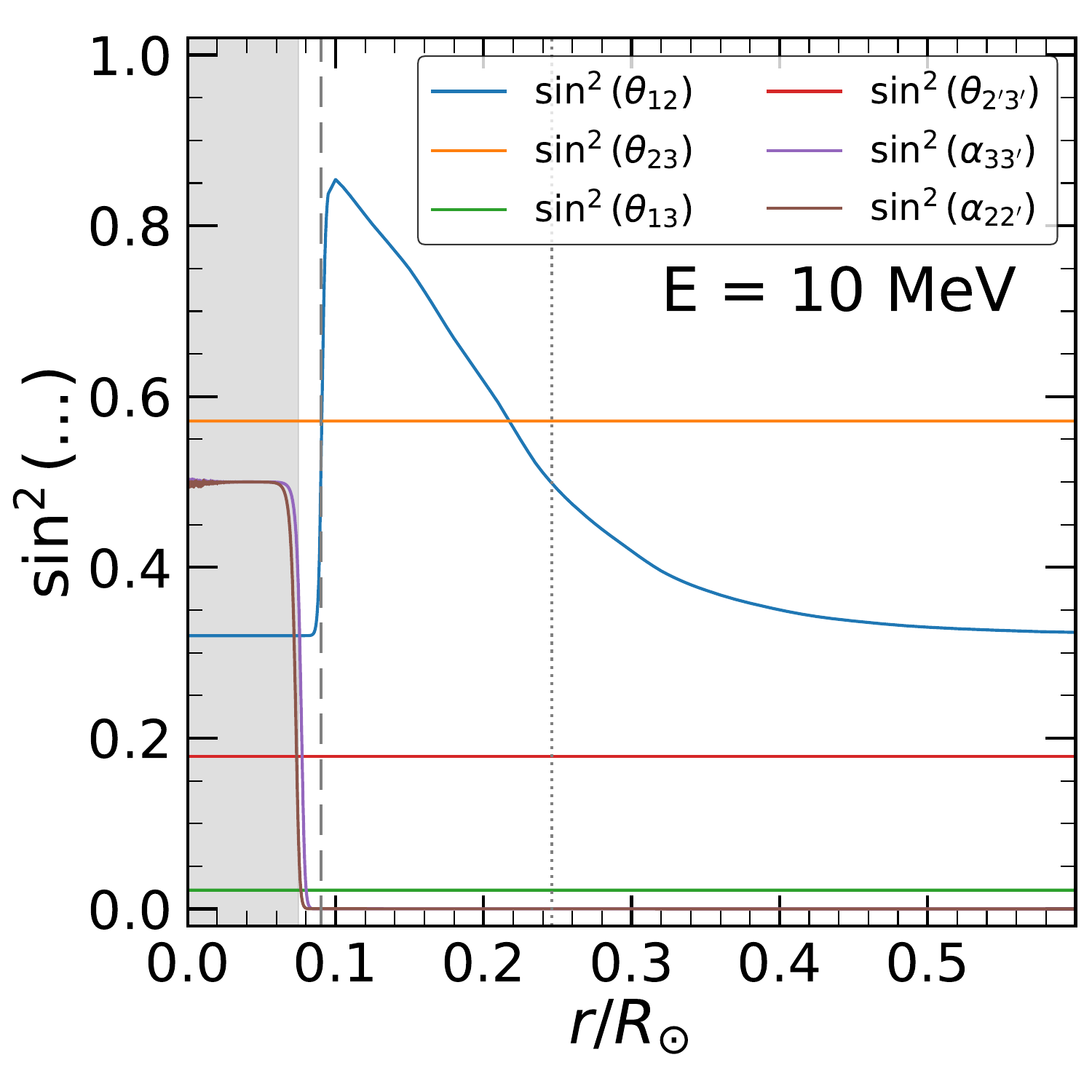}
    \includegraphics[width=0.395\linewidth]{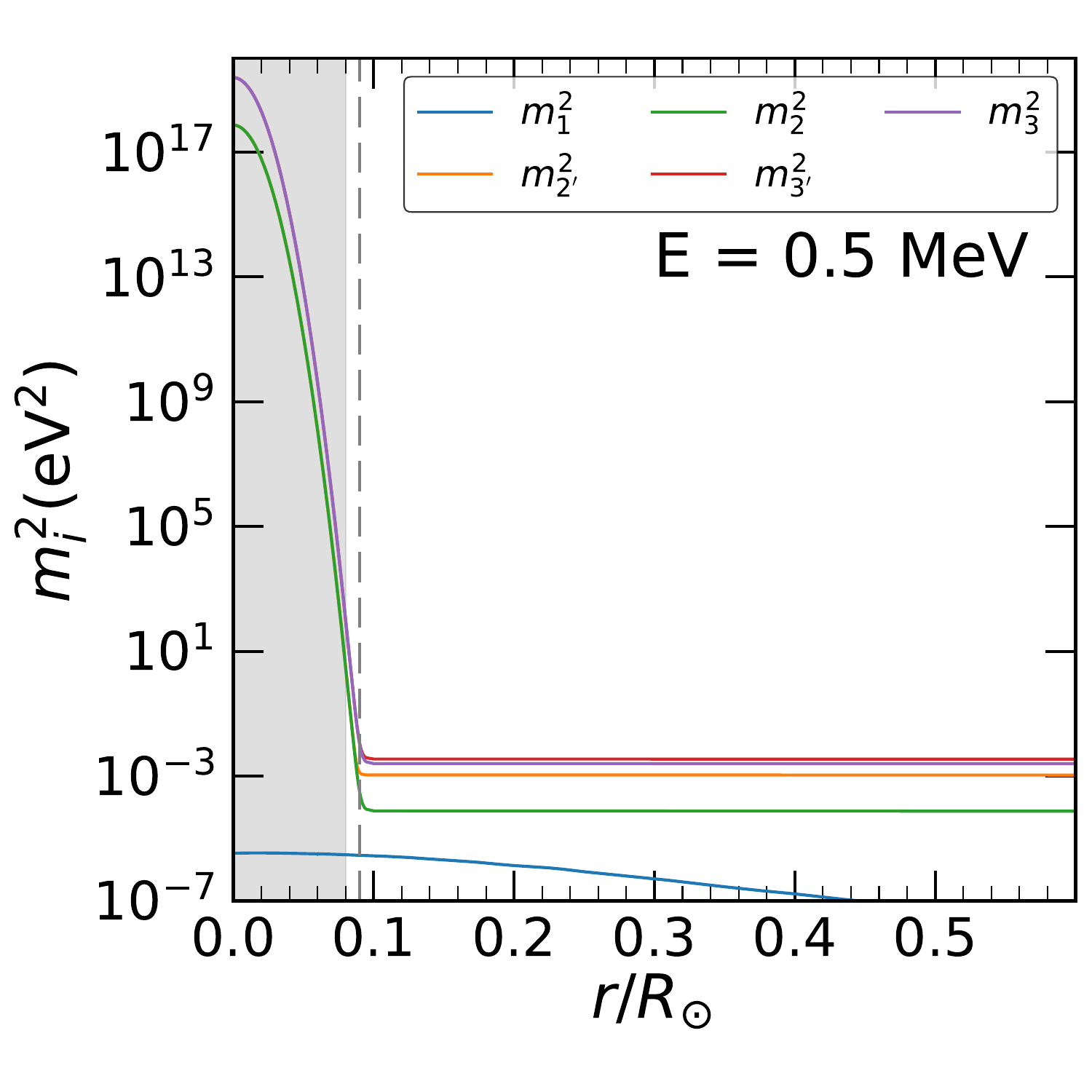}
    \includegraphics[width=0.395\linewidth]{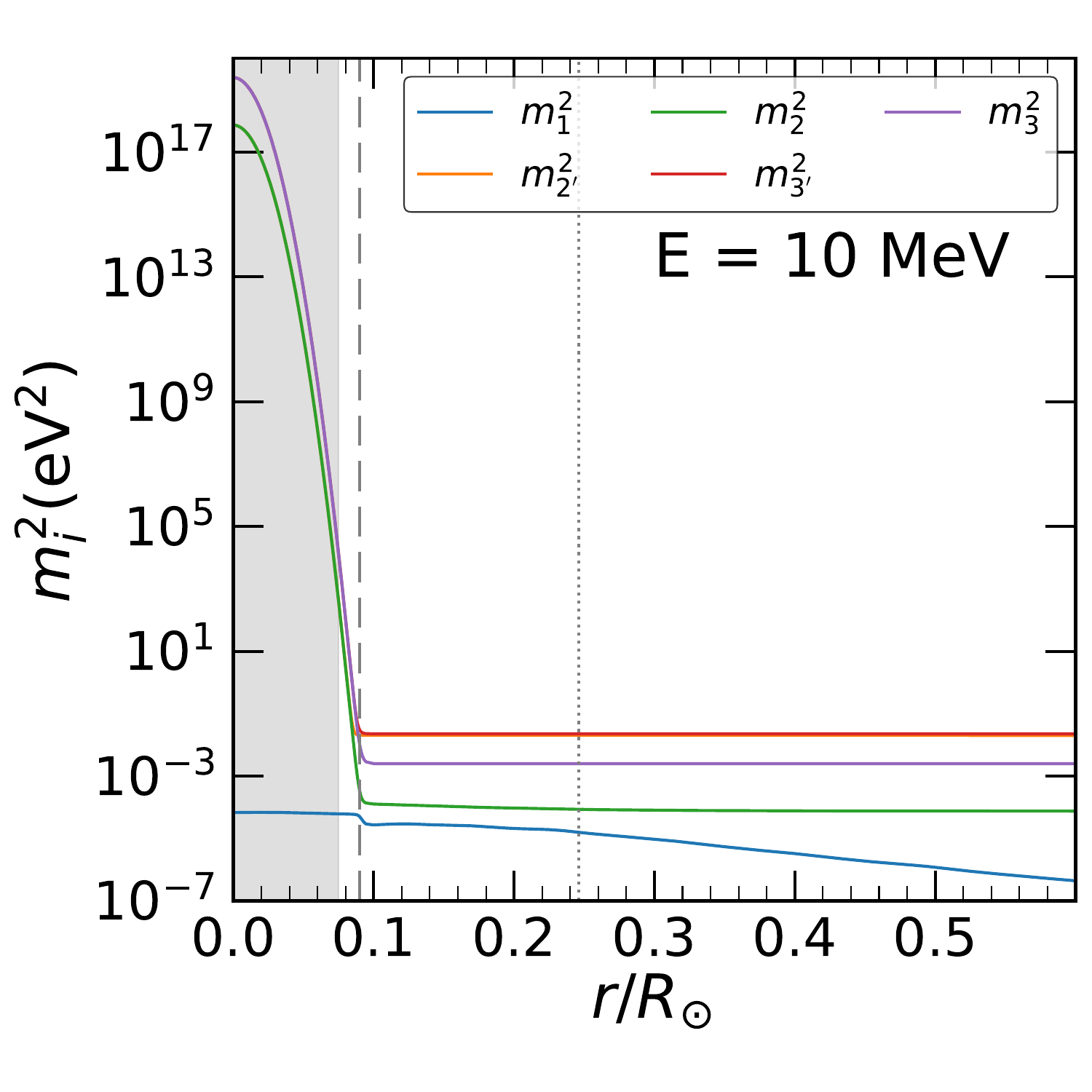}
    \caption{Mixing angles (upper panels) and mass-squared values (lower panels) in the 5-neutrino system for $E =0.5 \ \text{MeV (left panels) and}\ E =10\ \text{MeV (right panels)}$ when $m_\phi = 10^{-9}$ eV, $m_\chi^2 = 10^{-7}\, \text{eV}^2$ and $\dot{\Phi}=m_\phi$,  $M_\star =10^{-8}\,M_\odot$. We take $s_{12}^2 = 0.32$ in the uniform DM background. The vertical dashed line indicates the boundary of RHD ($r = r_0$) and the shaded region represents the RHD core ($r \leq r_1$). Notice the sharp core-periphery transition. Beyond RHD, the plots mimic the standard MSW evolution.}
    \label{fig:mixing_angles_large_mhalo}
\end{figure}
In this subsection, we look into the mixing angles and level crossings for different energy ranges in the presence of a DM halo within the Sun. In Fig. \ref{fig:mixing_angles_large_mhalo}, we present the numerical results for the mixing angles (top panel) and $m^2$ values (bottom panel) for a low energy, i.e. $0.5$ MeV (left panel) and a high energy, i.e. $10$ MeV (right panel) solar neutrino for $M_\star = 10^{-8}\, M_\odot$ and $m_\phi = 10^{-9}$ eV. We estimate the RHD boundary, $r_0$, by solving $\Delta m_{21}^2(r_0) \simeq 2E\,V_{\text{CC}}(r_0)$ and show it with a vertical dashed line. 
\iffalse
\begin{equation}\label{eq:RHD_boundary}
    r_0 = \frac{1}{4 m_\phi} \left( 
M_{\text{Pl}} \sqrt{\frac{R_\odot}{M_\odot}} \, \alpha + 
\sqrt{ 
\frac{ 
M_{\text{Pl}} R_\odot 
\left( 
M_{\text{Pl}} \alpha^2 - 
8 m_\phi \sqrt{M_\odot R_\odot} 
\log \left( 
\frac{ 
E_\nu M_{\text{Pl}}^3 \pi^3 R_\odot^{9/2} V_{\text{CC}}^0 \rho_{\text{local}}^2 
}{ 
M_\star^2 m_{\odot}^2 M_\odot^{3/2} m_\phi^3 
} 
\right) 
\right) 
}{M_\odot} 
} 
\right)
\end{equation}\
where $V_{\text{CC}}(r) = V_{\text{CC}}^0\ \text{Exp}\left[-\alpha \frac{r^2}{R_\odot^2}\right]$\fi
From the figure, we observe the following
\begin{itemize}
    \item As expected, within RHD, $\theta_{12}$ is equal to its value in the uniform DM background. 
    \item A finite but very small value of $m_1^2$ within the RHD can be attributed to the perturbation $\mathbb{V}_\odot$ due to the solar matter. Outside the RHD, it follows its corresponding value with standard MSW matter effects.
    \item The $m_{2,2'}^2$ and $m_{3,3'}^2$ values within RHD are downward opening parabolas when plotted on the log-scale, as expected since $\xi \sim \text{Exp}(-r^2/2R_\star^2)$. 
    \item The value of $\theta_{12}$ sharply rises from its DM background value to its standard MSW value at the RHD boundary.
    \item Outside RHD, the plots mimic the standard MSW evolution with the usual MSW-resonance (shown by a vertical dotted line) for higher energies. 
    \item Within RHD, there are two distinct regions where $\alpha_{22'} \approx \pi/4$ in one region and $\alpha_{22'} \ll 1$ in the other. This follows from eq. (\ref{eq:alpha22p_DMhalo_exp}) where $\alpha_{22'}(r)$ is almost flat, $\approx \pi/4$, for $\xi \gg 1$, and sharply descends to its value in uniform DM background, $\alpha_{22'}$, as $\xi(r)$ decreases.
\end{itemize}
For simplicity of analysis, we divide the RHD into two: one where $\xi(r) \,m_{\text{sol}}\, m_\chi \gg E\, m_{\phi}$, (i.e. $\alpha_{22'} \sim \pi/4$) and the other where $\xi(r)\, m_{\text{sol}}\, m_\chi \lesssim E\, m_{\phi}$ (i.e. $\alpha_{22'} \ll 1$). We refer to the former as the ``RHD core" and the latter as ``RHD periphery". The RHD core radius, $r_1$, may be obtained by solving $\xi(r_1)\, m_{\text{sol}}\, m_\chi \simeq E\, m_{\phi}$. In Fig. \ref{fig:mixing_angles_large_mhalo}, we show the RHD core as a gray shaded region. Note that the RHD core and RHD periphery regions depend on the neutrino energy.

\begin{figure}[t!]
    \centering
    \includegraphics[width=0.375\linewidth]{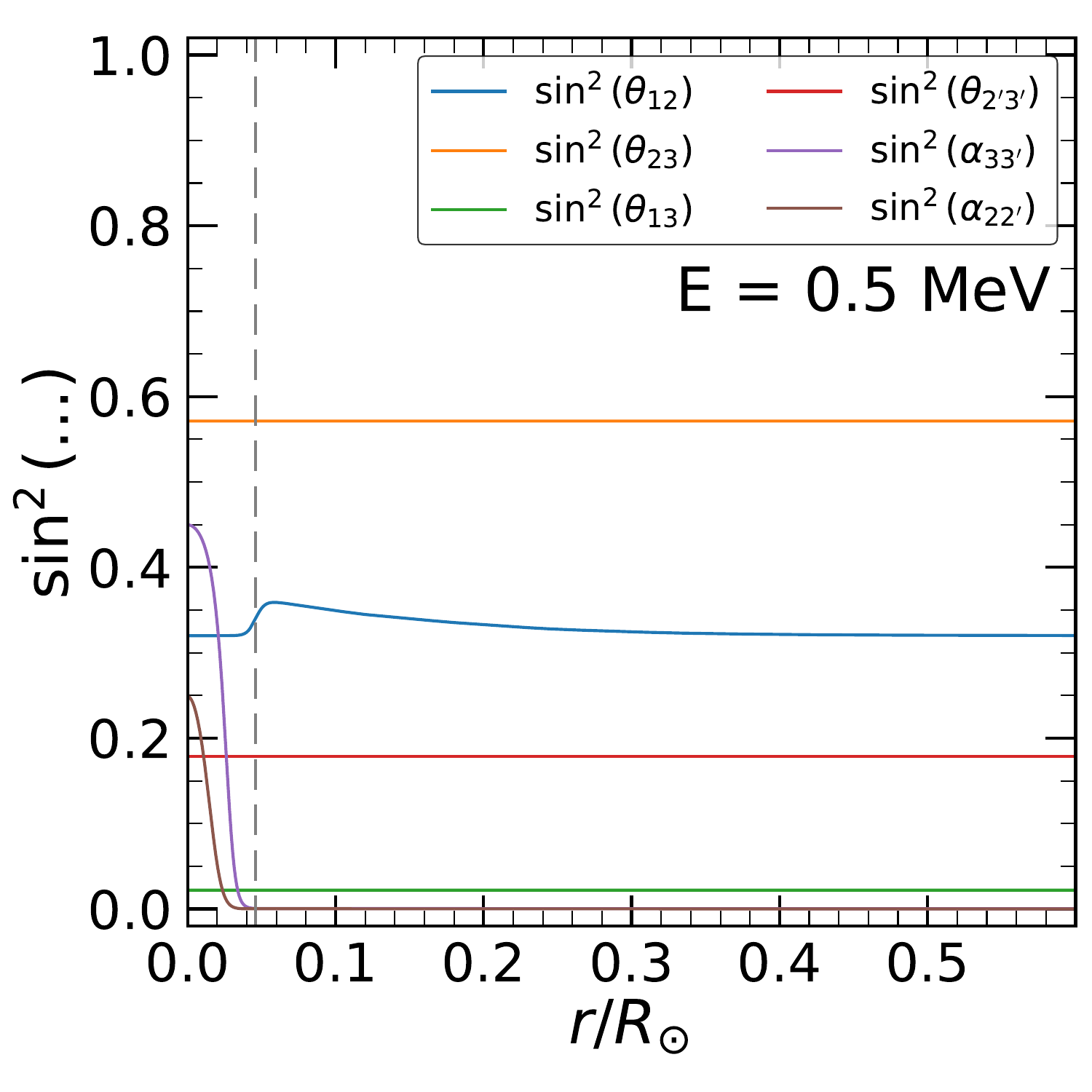}
    \includegraphics[width=0.375\linewidth]{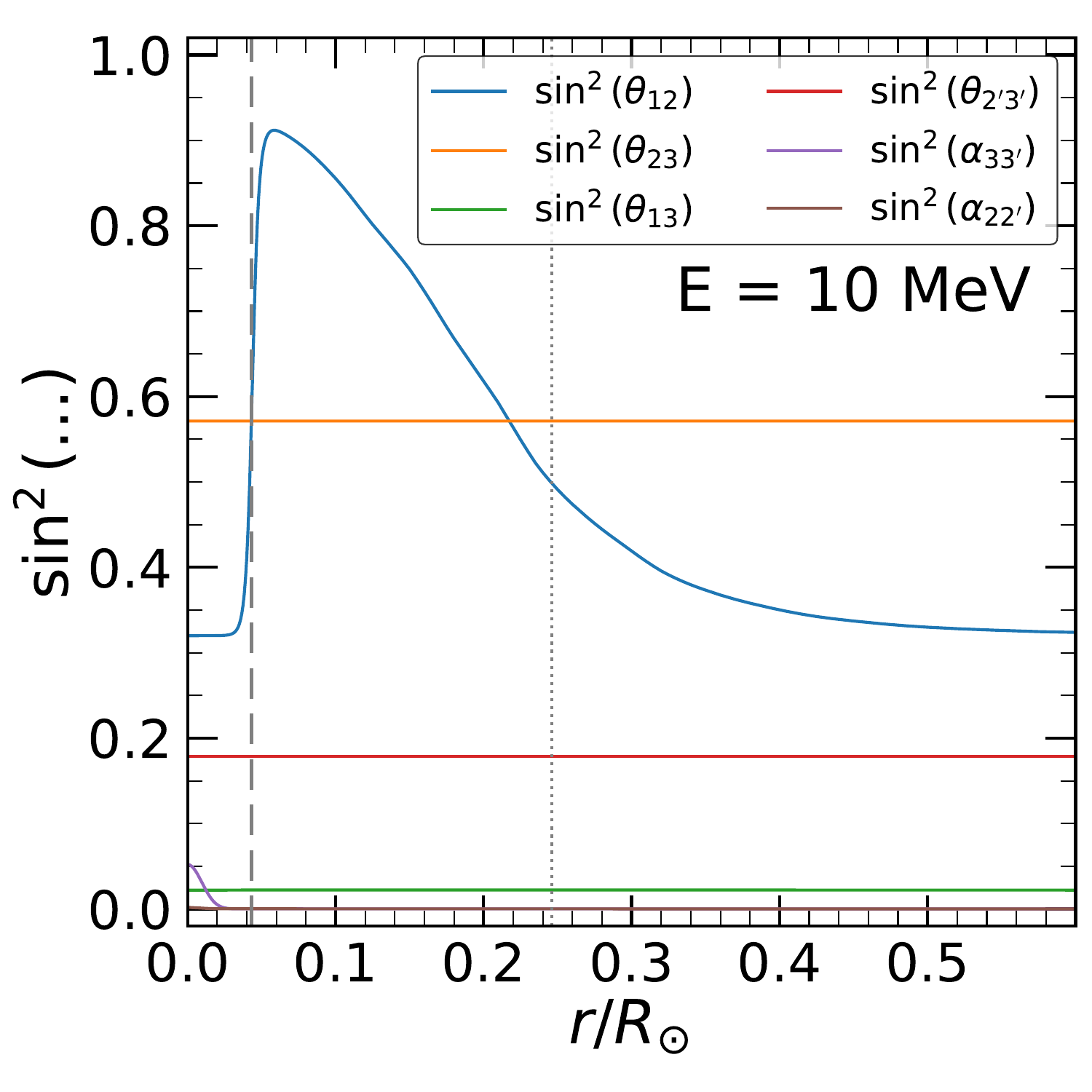}
    \includegraphics[width=0.395\linewidth]{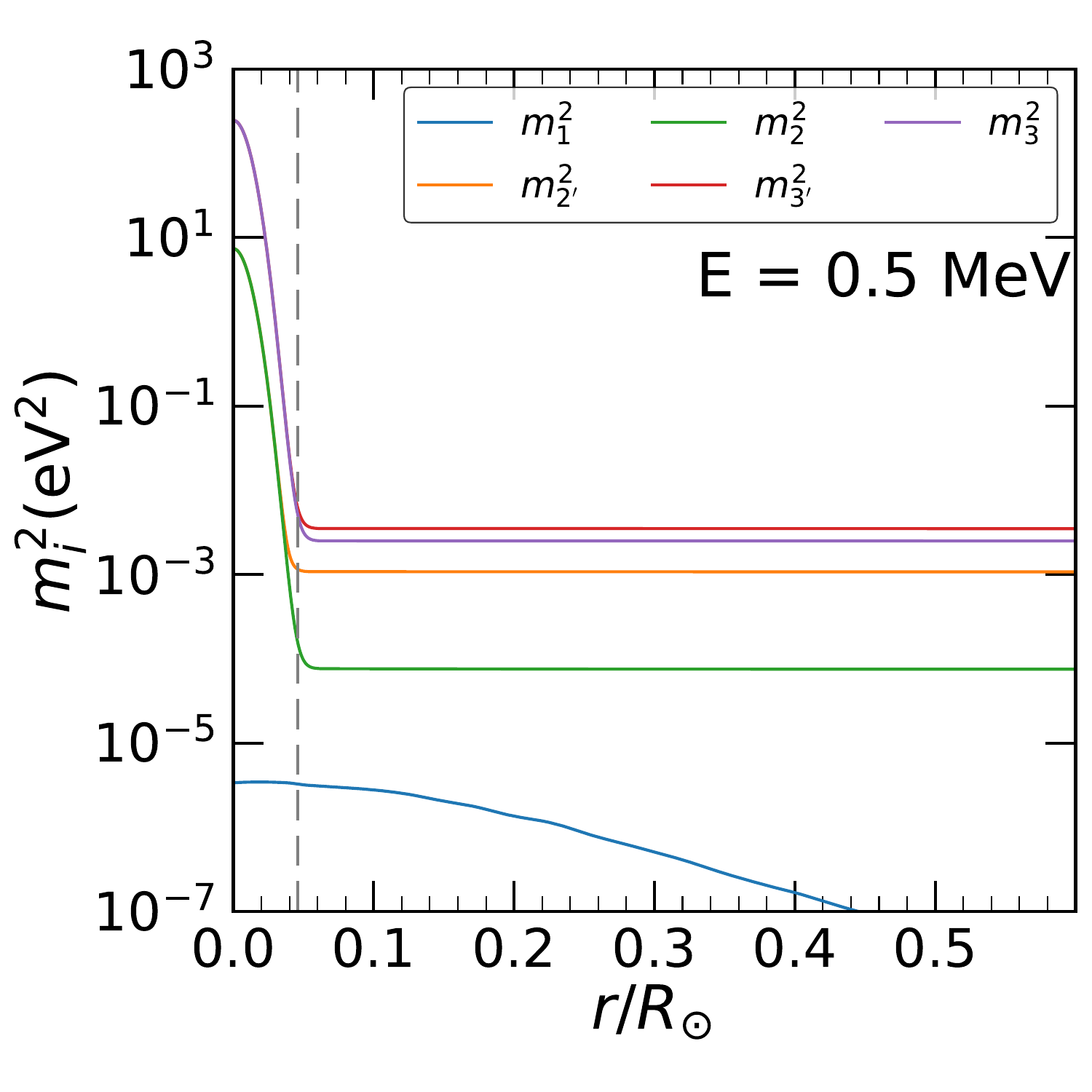}
    \includegraphics[width=0.395\linewidth]{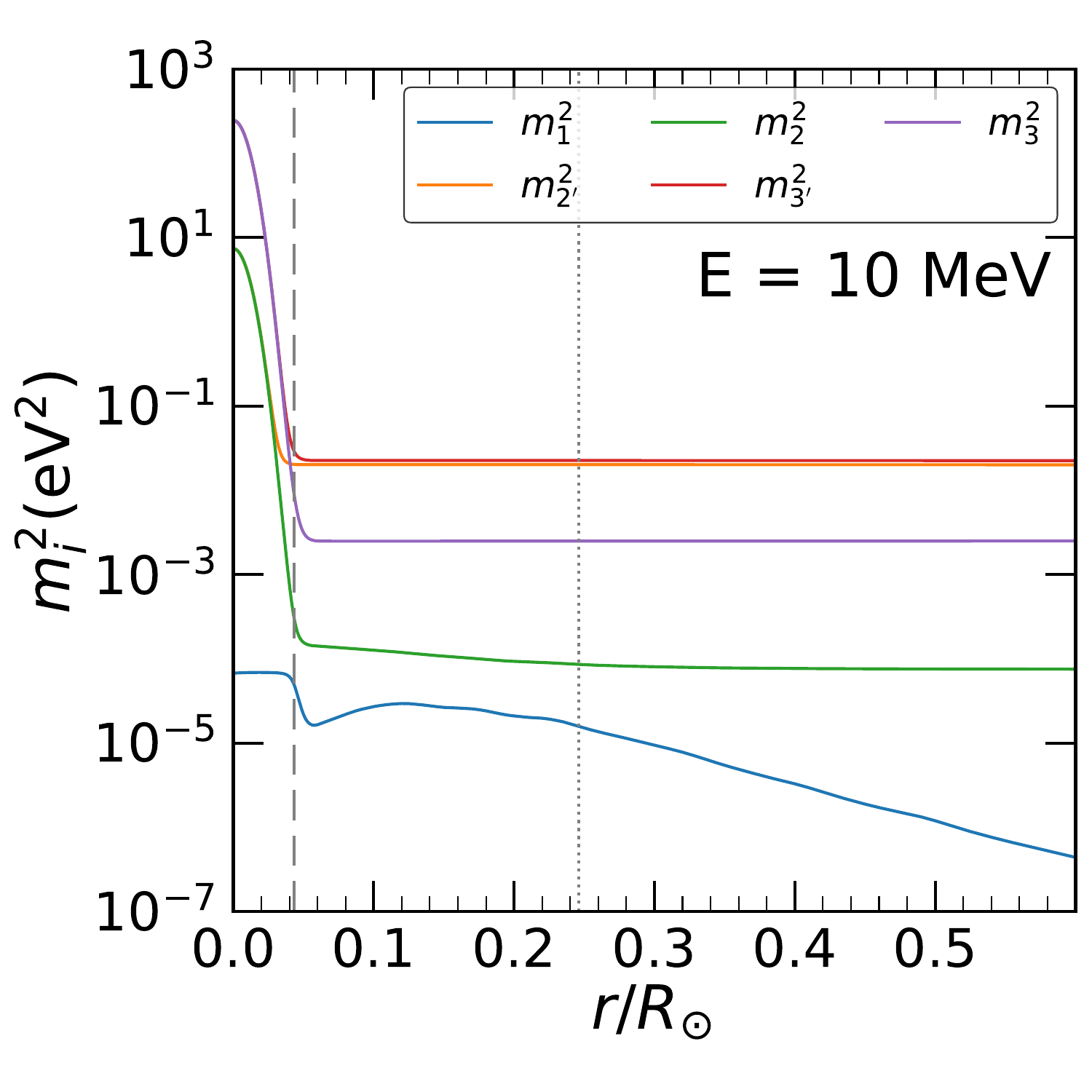}
    \caption{Mixing angles (upper panels) and mass-squared values (lower panels) in the 5-neutrino system for $E =0.5 \ \text{MeV (left panels) and}\ E =10\ \text{MeV (right panels)}$ when $m_\phi = 10^{-9}$ eV, $m_\chi^2 = 10^{-7}\, \text{eV}^2$, $\dot{\Phi}=m_\phi$ and  $M_\star =10^{-25}\,M_\odot$. We take $s_{12}^2 = 0.32$ in the uniform DM background. The vertical dashed line indicates the boundary of RHD ($r = r_0$). Within RHD, there is no sharp core-periphery transition. Beyond RHD, the plots mimic the standard MSW evolution.}
    \label{fig:mixing_angles_small_mhalo}
\end{figure}

In Fig. \ref{fig:mixing_angles_small_mhalo}, we present the numerical results for the mixing angles and $m^2$ values in the presence of a DM halo when the fraction of DM inside the Sun is very small: we take $M_\star = 10^{-25}\, M_\odot$ and $m_\phi = 10^{-9}$ eV. We note that, although the RHD and the non-RHD regions are distinguishable, there exists no clear separation between the RHD core and RHD periphery. This is because for the chosen value of $M_\star$, the inverse tangent function for $\alpha_{22'}$ in eq. \ref{eq:alpha22p_DMhalo_exp} does not have a very steep slope. Thus, for small halo masses, we may not be able to precisely define an RHD core and a periphery. This observation will be used later in sections \ref{subsec:surv_prob_comp}, \ref{subsec:flux_avg_spectrum_light_LMA} and \ref{sec:Dark_LMA}.

\subsection{Survival probability in the presence of a solar DM halo}\label{subsec:surv_prob_comp}

From eq. (\ref{eq:surv_prob}), it is clear that the survival probability of an electron neutrino depends not only on its energy but also on the conditions --- the densities of normal matter and DM --- at the production point. In this section, we explore the dependence of survival probability on the production region. Since solar neutrinos from different production channels (pp, $^7$Be, $^{13}$N, $^{15}$O, $^{17}$F, pep, $^{8}$B, hep) are typically produced at different radial distances, this consideration will be crucial in our analysis.

Since  $\alpha_{22'} \ll 1$ in the uniform DM background (and hence at the detection point), the term $\left(s_{22'}^P s_{22'}\right)^2$ in eq. (\ref{eq:surv_prob}) is negligible, and also $c_{22'} \approx 1$. Thus,
\begin{equation}
    P_{ee} \approx \left(c_{12}^P c_{12}\right)^2 + \left(s_{12}^P s_{12}\right)^2\left(c_{22'}^P\right)^2\,.
\end{equation}
For a neutrino with a given energy, the survival probability depends on its production region, as described below.

\begin{itemize}
    \item If the neutrino is produced inside the RHD core, then $c_{12}^P \approx c_{12}$ and $s_{12}^P \approx s_{12}$. Also $\alpha_{22'}^P \approx \pi/4$. Therefore, the survival probability becomes 
\begin{equation}\label{eq:surv_prob_approx_1}
    P_{ee,\,\text{core}} \approx c_{12}^4 + \frac{s_{12}^4}{2}\,.
\end{equation} 
The individual values of $P_{e\mu}$ and $P_{e\tau}$ within the core depend on $\theta_{23}$ and the values of $P_{e\chi_1}$ and $P_{e\chi_2}$ within the core depend on $\theta_{2'3'}$. However, their pairwise sums are independent of these angles. We define

\begin{equation}
    P_{e\alpha,\,\text{core}} \equiv P_{e\mu,\,\text{core}} + P_{e\tau,\,\text{core}} \approx \frac{3}{2}\, s_{12}^2\, c_{12}^2\,,
\end{equation}
\begin{equation}
    P_{e\chi,\,\text{core}} \equiv P_{e\chi_1,\,\text{core}} + P_{e\chi_2,\,\text{core}} \approx \frac{s_{12}^2}{2}\, ,
\end{equation}
as the probabilities of $\nu_e$ produced inside the core converting to active and sterile neutrinos, respectively.

    \item If the neutrino is produced within the RHD periphery, we have 
\begin{equation}\label{eq:surv_prob_approx_2}
    P_{ee,\,\text{peri}} \approx c_{12}^4 + s_{12}^4 \left(c_{22'}^P\right)^2\,,
\end{equation} 
where for a sharp RHD core-periphery transition, $\left(c_{22'}^P\right)^2 \approx 1$. If the transition is not sharp enough, $\left(c_{22'}^P\right)^2$ could range from $0.85-1$. The other two probabilities are
\begin{equation}
    P_{e\alpha,\,\text{peri}} \approx \left(1+\left(c_{22'}^P\right)^2\right)s_{12}^2c_{12}^2 \approx  \left(1-P_{ee,\,\text{peri}}\right)\,, \quad P_{e\chi,\,\text{peri}} \approx (s_{22'}^P)^2 s_{12}^2 \approx 0\,,
\end{equation}
where the second approximation in each equation is valid when there is a sharp RHD core-periphery transition.
    \item If the neutrino is produced outside the RHD, then the probabilities become
\begin{equation}\label{eq:surv_prob_approx_3}
    P_{ee,\,\text{out}} \approx c_{12,\odot}^2\ c_{12}^2+s^{2}_{12,\odot}\ s_{12}^2\,,\quad P_{e\alpha,\,\text{out}} \approx 1-P_{ee,\,\text{out}}\,, \quad P_{e\chi,\,\text{out}} \approx 0\,,
\end{equation}
where the subscript $\odot$ indicates the production angle if there was only the uniform DM background and we get back the `standard' MSW survival probability expression. 
\end{itemize}

\begin{figure}[t!]
    \centering
    \includegraphics[width=0.4\linewidth]{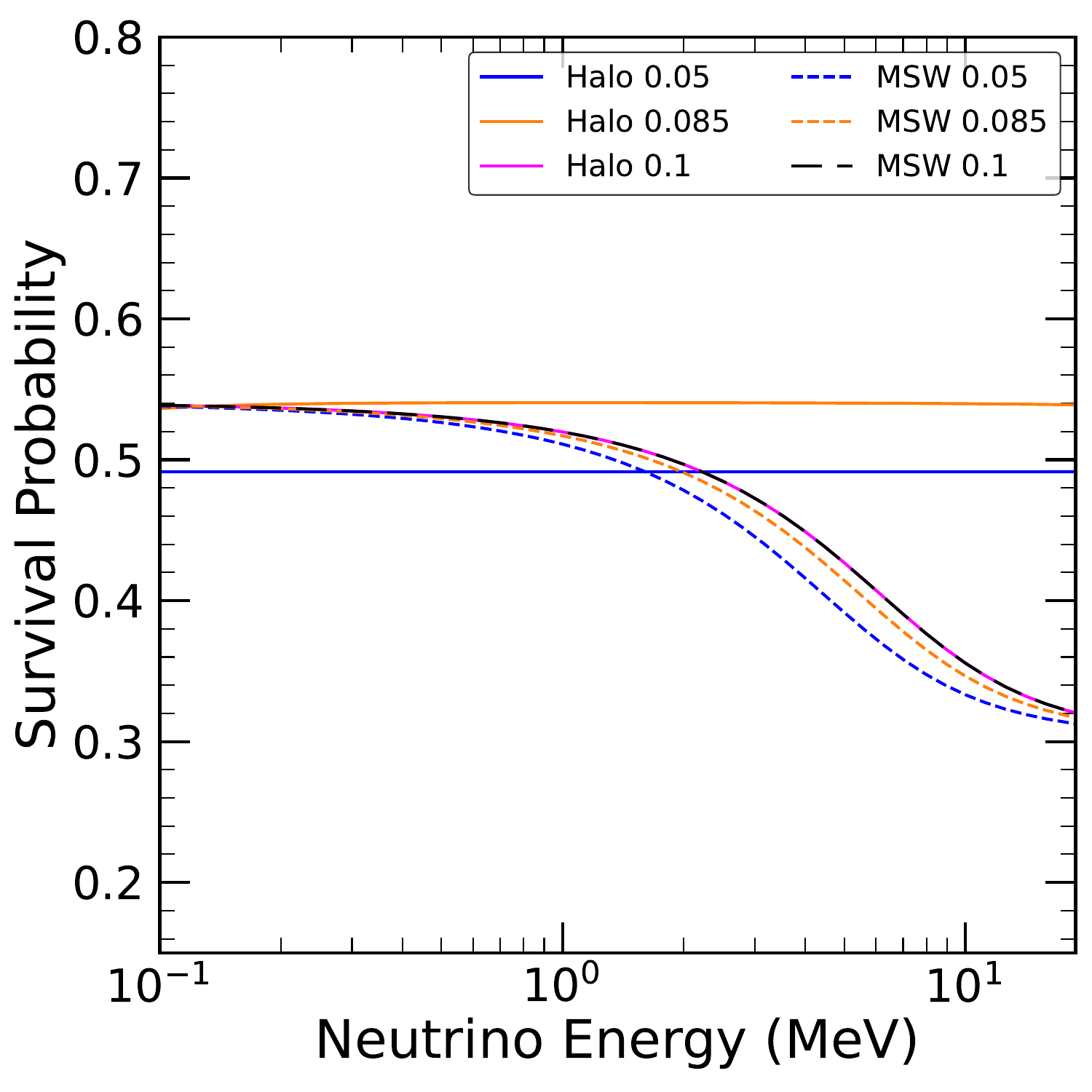}
    \includegraphics[width=0.4\linewidth]{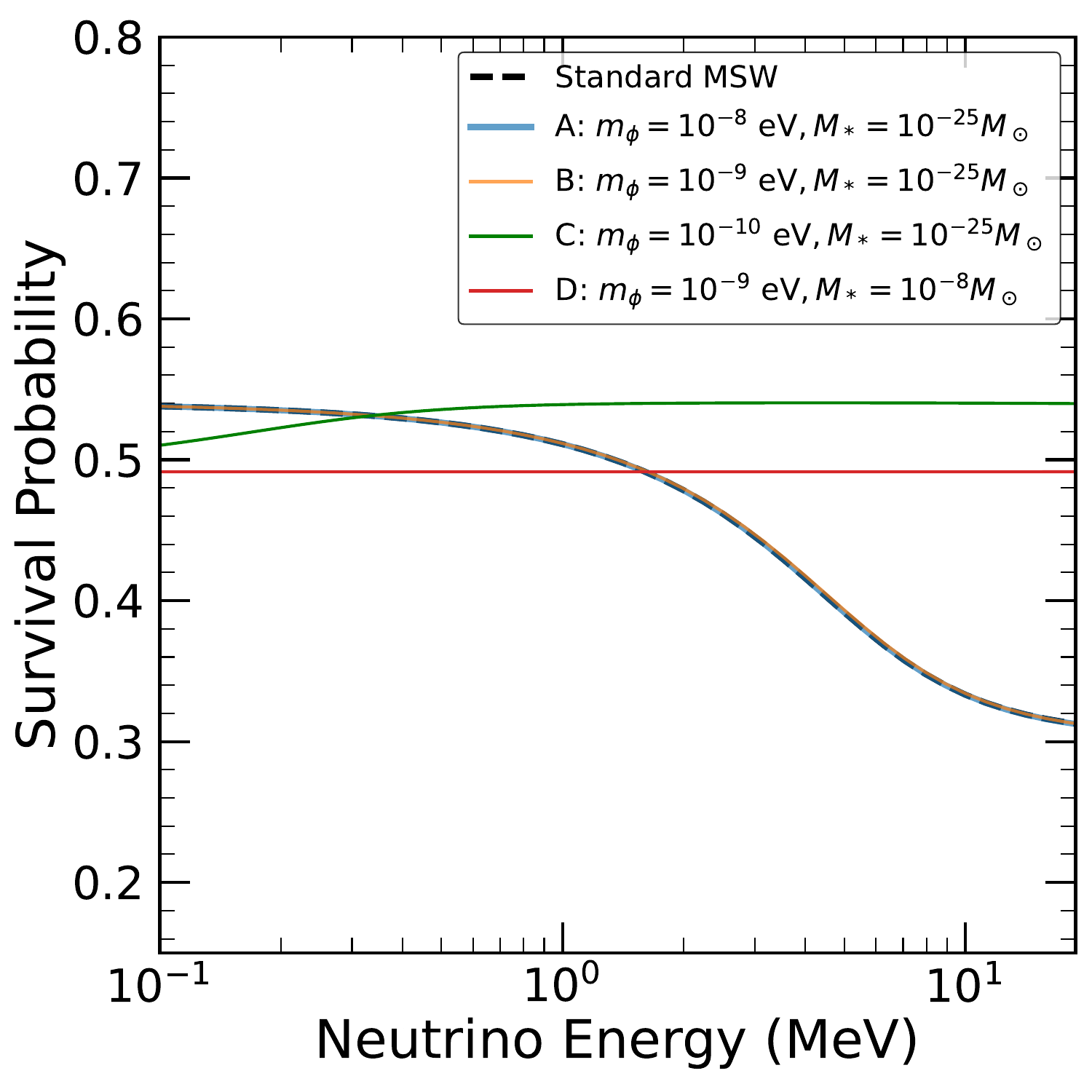}
    \caption{Left panel: Survival probability of an electron neutrino produced at different radial distances $r=0.05\, R_\odot$, $r=0.085\, R_\odot$ and $r=0.1\, R_\odot$, for a DM halo with $m_\phi = 10^{-9}$ eV, $m_\chi^2 = 10^{-7}\, \text{eV}^2$ and $\dot{\Phi}=m_\phi$,  $M_\star =10^{-8}\,M_\odot$. The three radial distances are chosen to correspond to regions within the RHD core, the RHD periphery and outside the RHD, respectively. Right panel: Survival probability of an electron neutrino produced at $r=0.05\, R_\odot$ for different halo parameters. We have taken $m_\chi^2 = 2\, E_R\, m_\phi$ with $E_R=50$ eV. The value of $s_{12}^2$ has been taken to be 0.32 for all scenarios, for illustration.}
    \label{fig:indv_surv_prob}
\end{figure}

In the left panel of Fig. \ref{fig:indv_surv_prob}, we show the survival probability of a neutrino which is produced at different distances from the center of the Sun for a particular choice of halo parameters as in Fig. \ref{fig:mixing_angles_large_mhalo}. We have considered three radii of production regions, namely, $0.05\, R_\odot$, $0.085\, R_\odot$ and $0.1\, R_\odot$, which lie within the RHD core, within the RHD periphery and outside RHD, respectively. For comparison with the standard MSW case, we choose $s^2_{12} = 0.32$ for all cases. We observe the following.
\begin{itemize}
    \item In the standard MSW case, the dependence on the production region is quite weak. The probabilities in all three cases --- MSW 0.05, MSW 0.085 and MSW 0.1 --- are almost identical. The probability changes from $\sim 0.56$ at low energies ($E\approx 0.1\ \text{MeV}$) to $\sim 0.32$ at high energies ($E\approx 10$ MeV).
    \item As expected, when the neutrino is produced outside the RHD (Halo 0.1), its survival probability mimics the corresponding standard MSW probability (MSW 0.1).
    \item The probabilities when the neutrino is produced inside the RHD core (Halo 0.05) and within the RHD periphery (Halo 0.085) differ significantly from the MSW solutions. The probability spectra are almost flat, which may be inferred from eq. (\ref{eq:surv_prob_approx_1}) and eq. (\ref{eq:surv_prob_approx_2}).
    These probability spectra will not be able to satisfy the experimental constraints at high energy for this set of halo parameters. 
\end{itemize}
 
In the right panel of Fig. \ref{fig:indv_surv_prob}, we show the dependence of the survival probability of a neutrino produced at $0.05\, R_\odot$, for different benchmark halo parameters. We observe that when the halo mass and halo radius are sufficiently small (i.e. scenario A: $r_0 = 0.016\, R_\odot$ or scenario B: $r_0 = 0.046\, R_\odot$), the neutrino is produced outside the RHD, and the probability spectrum is almost the same as the standard MSW case. On the other hand, when the neutrino comes from within RHD (for scenario C: $r_0=0.128\, R_\odot$ and scenario D: $r_0=0.092\, R_\odot$), we get almost flat probability spectrum as expected from eq. (\ref{eq:surv_prob_approx_1}) and eq. (\ref{eq:surv_prob_approx_2}). Note that the small energy dependence in scenario C comes from the lack of a sharp core-periphery transition for the chosen halo parameters. The strong dependence of the nature of survival probability spectra on the halo parameters indicates that solar neutrino data can be used to tightly constrain the halo parameters. 

The neutrino production inside the Sun, occurring in different regions for different production channels, offers us the opportunity to test the predictions of the model comprising refractive masses of neutrinos. This will be explored later in section \ref{subsec:flux_avg_spectrum_light_LMA}. However, before exploring the effect of refractive neutrino masses and halo on the solar neutrino oscillation probabilities in more detail, we need to carefully interpret the data from solar neutrino experiments. We perform this task in the next section.

\subsection{Reinterpretation of SK and SNO data}

Although the small active–sterile mixing angles $\alpha_{22'}$ and $\alpha_{33'}$ suppress active–sterile oscillations in a uniform DM background, these angles are $\approx \pi/4$ within the RHD core (see eq. (\ref{eq:alpha22p_DMhalo_exp})), giving rise to a significant probability for such oscillations. Thus, the SK and SNO datasets, normally analyzed under the standard three active-neutrino framework, require reinterpretation for a consistent comparison with the predictions of the refractive neutrino mass scenario in the presence of DM halo. 

SK detects solar neutrinos via the process of elastic scattering (ES) on electrons, which is sensitive to all the three active neutrino flavors, albeit with different strengths. The ES cross-section of electron neutrinos is about six times that of muon or tau neutrinos. The SK observations are often presented as ``data/MC", i.e. the observed ES event rate divided by the rate expected without oscillations. The electron neutrino survival probability can be inferred from data/MC via
\begin{equation}\label{eq:RSK_defn}
    R_{\text{SK}} = P_{ee} + r_\sigma\, P_{e\alpha}\,,
\end{equation}
where $R_{\text{SK}}$ denotes the data/MC as measured by SK, $r_\sigma \equiv \sigma(\nu_{\mu/\tau}\,e)/\sigma(\nu_{e}\,e) \approx 0.16$ is the ratio of cross-sections, and $P_{e\alpha}$ denotes the total conversion probability of an electron neutrino to active (muon and tau) neutrinos. In the standard scenario, 
\begin{equation}\label{eq:RSK_std_exp}
    R_{\text{SK}} = P_{ee}^\text{std} + r_\sigma\, (\,1-P_{ee}^\text{std}\,)\,.
\end{equation}
In the scenario with refractive masses and halo, 
\begin{equation}\label{eq:RSK_refr_exp}
    R_{\text{SK}} = P_{ee}^{\text{halo}} + r_\sigma\, (\,1-P_{ee}^{\text{halo}}\, - P_{e\chi}^{\text{halo}}\,)\,,
\end{equation}
where $P_{ee}^{\text{halo}}$ and $P_{e\chi}^{\text{halo}}$ are the electron neutrino survival probability and the total electron-to-sterile neutrino conversion probability, respectively.

The SNO experiment has measurements of three different processes, sensitive to three different combinations of the electron neutrino flux $\phi_e$ and the total muon and tau neutrino flux $\phi_{\mu\tau}$: (i) The charged-current (CC) process sensitive to $\phi_{CC} = \phi_e$, (ii) the neutral-current (NC) process sensitive to $\phi_{NC} = \phi_e + \phi_{\mu\tau}$, and (iii) The elastic-scattering (ES) process sensitive to $\phi_{ES} = \phi_e + r_\sigma\, \phi_{\mu\tau}$. From these three flux measurements, usually two ratios $R_{\text{SNO}}^{\text{CC/NC}}$ and $R_{\text{SNO}}^{\text{CC/ES}}$ are determined, which are related to the neutrino oscillation probabilities via
\begin{equation}\label{eq:RSNO_defn}
    R_{\text{SNO}}^{\text{CC/NC}} = \frac{\phi_{\text{CC}}}{\phi_{\text{NC}}} = \frac{P_{ee}}{P_{ee} + P_{e\alpha}}\ \ \text{and}\ \ R_{\text{SNO}}^{\text{CC/ES}} = \frac{\phi_{\text{CC}}}{\phi_{\text{ES}}} = \frac{P_{ee}}{P_{ee} + r_\sigma\, P_{e\alpha}}\,.
\end{equation}
In the standard scenario,
\begin{equation}
    R_{\text{SNO}}^{\text{CC/NC}} = P_{ee}^{\text{std}}\ \ \text{and}\ \ R_{\text{SNO}}^{\text{CC/ES}} = \frac{P_{ee}^{\text{std}}}{P_{ee}^{\text{std}} + r_\sigma\, (1-P_{ee}^{\text{std}})}\,.
\end{equation}
In the scenario with refractive masses and halo,
\begin{equation}\label{eq:RSNO_refr_exp}
    R_{\text{SNO}}^{\text{CC/NC}} = \frac{P_{ee}^\text{halo}}{1-P_{e\chi}^\text{halo}}\ \ \text{and}\ \ R_{\text{SNO}}^{\text{CC/ES}} = \frac{P_{ee}^\text{halo}}{P_{ee}^\text{halo} + r_\sigma\, (1-P_{ee}^\text{halo}-P_{e\chi}^\text{halo})}\,.
\end{equation}
Thus, the values of $P_{ee}^\text{halo}$ inferred from the measurements depend on the corresponding values of $P_{e\chi}^\text{halo}$. Note that in the standard scenario, $P_{ee}^{\text{std}} + P_{e\alpha}^{\text{std}} = 1$, a constraint which is not valid with oscillations to sterile neutrinos. Therefore, instead of only one independent quantity $P_{ee}^{\text{std}}$ parameterizing the measured quantities in eqs. (\ref{eq:RSK_defn}) and (\ref{eq:RSNO_defn}) in the standard scenario, now we have two independent parameters, namely, $P_{ee}^{\text{halo}}$ and $P_{e\chi}^{\text{halo}}$.

\begin{figure}[t!]
    \centering
    \includegraphics[width=0.65\linewidth]{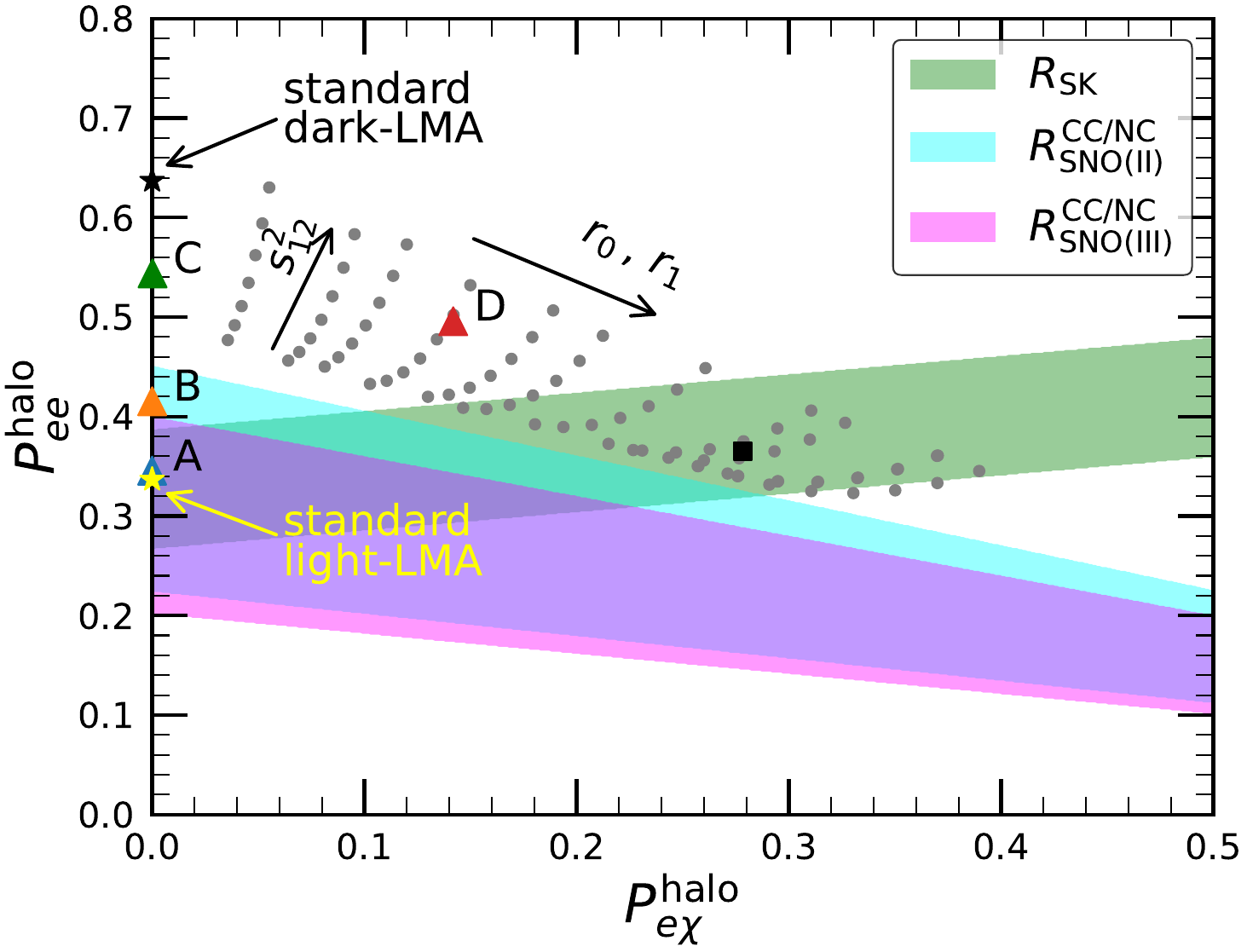}
    \caption{Reinterpretation of the SK and SNO data in terms of $P^{\text{halo}}_{e\chi}$ and $P^{\text{halo}}_{ee}$. The green, cyan and magenta bands indicate the values of these parameters allowed by the ratios $R_{\text{SK}}$, $R_{\text{SNO}}^{\text{CC/NC}}$ for SNO phase-II and $R_{\text{SNO}}^{\text{CC/NC}}$ for SNO phase-III, respectively, at the $3\sigma$ level. The colored triangles show ($\langle P_{e\chi}\rangle$, $\langle P_{ee}\rangle$) for $E=10$ MeV for the scenarios A, B, C and D in Fig. \ref{fig:indv_surv_prob}, whereas  the yellow (black) star represents the standard light-LMA (dark-LMA) MSW solution. The gray dots represent ($\langle P_{e\chi}\rangle$, $\langle P_{ee}\rangle$) at $E=10$ MeV for different halo parameters when $s_{12}^2$ is in the dark-LMA region allowed by KamLAND at $3\sigma$ level. The black square represents the value of ($\langle P_{e\chi}\rangle$, $\langle P_{ee}\rangle$) at $E = 10$ MeV for $s_{12}^2 = 0.68$ and $m_\phi = 10^{-9}$ eV, $m_\chi^2 = 10^{-7}\, \text{eV}^2$, $\dot{\Phi}=m_\phi$, $M_\star = 10^{-8}\, M_\odot$.} \label{fig:SK_SNO_reinterpret}
\end{figure}
In Fig. \ref{fig:SK_SNO_reinterpret}, we show the $3\sigma$-allowed values of $P_{ee}^{\text{halo}}$ and $P_{e\chi}^\text{halo}$, given the measurements of $R_{\text{SK}}$ \cite{Super-Kamiokande:2023jbt} and $R_{\text{SNO}}^{\text{CC/NC}}$ \cite{SNO:2005oxr,SNO:2011ajh}. We do not show the $R_{\text{SNO}}^{\text{CC/ES}}$ in the figure as the SNO data is predominantly sensitive to CC and NC interactions (because of a smaller $\nu$-$e$ cross-section), and anyway the quantity $\phi_{ES}$ is determined at a higher accuracy by SK. The standard scenario corresponds to the y-axis in the figure, i.e. $P_{e\chi} = 0$. Note that the $P_{ee}^{\text{halo}}$ and $P_{e\chi}^{\text{halo}}$ bands shown in this figure are independent of any halo parameters, and only depend on the measurements.

From the figure, we observe that for a given range of $R_{\text{SK}}$, higher values of $P_{ee}^{\text{halo}}$ are allowed for higher values of $P_{e\chi}^{\text{halo}}$. On the other hand, for a given range of $R_{\text{SNO}}^{\text{CC/NC}}$, lower values of $P_{ee}^{\text{halo}}$ are allowed for higher values of $P_{e\chi}^{\text{halo}}$. These features can be understood from eq. (\ref{eq:RSK_refr_exp}) and (\ref{eq:RSNO_refr_exp}). Further, in order to satisfy both the SNO and SK constraints at the $3 \sigma$ level, $P_{e\chi}^{\text{halo}} \lesssim 0.3$. 

Figure \ref{fig:SK_SNO_reinterpret} also indicates the $P_{ee}^{\text{halo}}$ values in the benchmark scenarios A, B, C, D corresponding to different halo parameters. These probabilities have been calculated using the flux-averaging as described in the next section. Scenarios C and D seem to be highly disfavored by the $^8$B neutrino data both at SK and SNO.

\subsection{Flux-averaged probability spectrum} \label{subsec:flux_avg_spectrum_light_LMA}

\begin{figure}[t!]
    \centering
    \includegraphics[width=\linewidth]{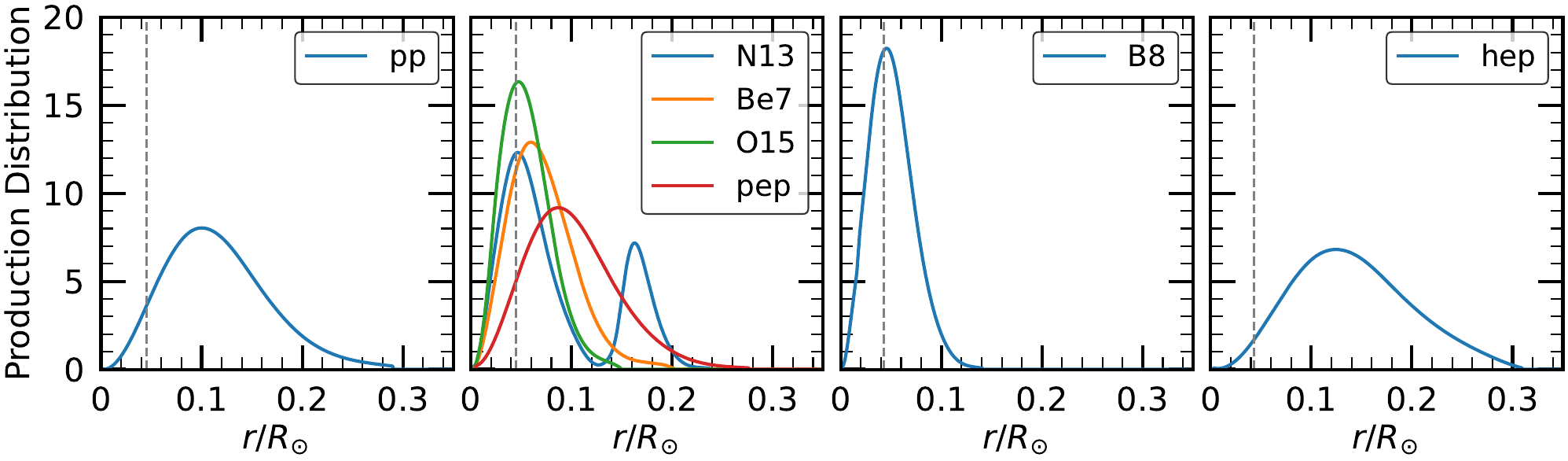}
    \caption{Production distributions $\eta_i(r)$ for low energy (pp), intermediate energy ($^{13}$N, $^{7}$Be, $^{15}$O, pep), high energy ($^{8}$B) and very high energy (hep) solar neutrinos \cite{Vinyoles:2016djt}. The vertical dashed lines indicate the RHD boundary for $m_\phi = 10^{-9}$ eV, $m_\chi^2 = 10^{-7}\, \text{eV}^2$  and $M_\star = 10^{-25}\, M_\odot$ (benchmark scenario B).}
    \label{fig:light_LMA_prod_dist}
\end{figure}

According to the standard solar model (SSM), solar neutrinos are produced via different pp-chain and CNO-cycle reactions inside the Sun. Different production channels produce neutrinos in different regions and at different energies. In Fig. \ref{fig:light_LMA_prod_dist}, we show the normalized neutrino production rate, $\eta_i(r)$, for each channel $i$ at a distance $r$ from the center of the Sun \cite{Vinyoles:2016djt}. This illustrates the different production regions for different neutrino channels. We also overlay the RHD boundaries for the scenario $m_\phi = 10^{-9}$ eV and $M_\star = 10^{-25} M_\odot$ (scenario B). Since the RHD boundaries have an energy dependence, albeit very mild, for illustrating the RHD boundaries in the figure, we use $E = 0.5$ MeV for the two left panels (low and intermediate energies) and $E = 10$ MeV for the two right panels (high and very high energies). In this scenario, the core-periphery transition is not sharp.

Following the discussion in section \ref{subsec:surv_prob_comp}, the survival probability depends strongly on whether the production point is within RHD core, in the RHD periphery or outside RHD. Clearly, from Fig. \ref{fig:light_LMA_prod_dist}, the fractions of neutrinos produced inside RHD are different for different channels. Therefore, we expect that even for neutrino with a given energy, the survival probability will depend significantly on its production channel.  At a solar neutrino detector at the Earth, we do not observe the individual survival probability spectra but a flux-averaged effective survival probability spectrum. 

For a particular production channel $i$, the survival probability depends on the exact production point within the production region. The survival probability for a neutrino with given energy, averaged over the production region for the channel $i$, may be approximated as
\begin{equation}\label{eq:flux_avg_prob}
    P_{ee}^{(i)} = f_{\text{core}}^{(i)} \left(c_{12}^4+\frac{s^{4}_{12}}{2}\right) + f_{\text{peri}}^{(i)} \left(c_{12}^4+s^{4}_{12}\right) + f_{\text{out}}^{(i)} \left(c_{12,\odot}^2\ c_{12}^2+s^{2}_{12,\odot}\ s_{12}^2\right), 
\end{equation}
where $f_{\text{core}}^{(i)}\,, f_{\text{peri}}^{(i)}\,\text{and}\ f_{\text{out}}^{(i)}$ are the fractions of fluxes for the channel $i$ coming from the RHD core, the RHD periphery and outside the RHD, respectively. These fractions are given by
\begin{equation}
    f_{\text{core}}^{(i)} = \int_0^{r_1} \eta_i(r)\, dr\,,\ \ f_{\text{peri}}^{(i)} = \int_{r_1}^{r_0} \eta_i(r)\, dr \ \  \text{and}\ \
    f_{\text{out}}^{(i)} = 1-f_{\text{core}}^{(i)}-f_{\text{peri}}^{(i)}\,.
\end{equation}

Finally, the net flux-averaged survival probability becomes
\begin{equation}
    \langle P_{ee}(E) \rangle = \frac{\sum_i \left(\frac{d\mathcal{F}^{(i)}}{dE}\right) P_{ee}^{(i)}(E)}{\sum_i \left(\frac{d\mathcal{F}^{(i)}}{dE}\right)},
\end{equation}
where $\frac{d\mathcal{F}^{(i)}}{dE}$ is the differential neutrino flux in channel $i$ with energy $E$ \cite{Vinyoles:2016djt,Bahcall:2005va}. In the flux-averaged survival probability spectrum $\langle P_{ee}(E)\rangle$, whenever the dominant production channel for the neutrino flux changes, we expect to see discontinuities. This transition usually happens near the end-point energies of a dominant channel where the flux from that channel sharply goes to zero. Thus, the end-point energies of various production channels leave their signatures as kinks in the flux-averaged probability spectrum. 

\begin{figure}[t!]
    \centering
    \includegraphics[width=0.75\linewidth]{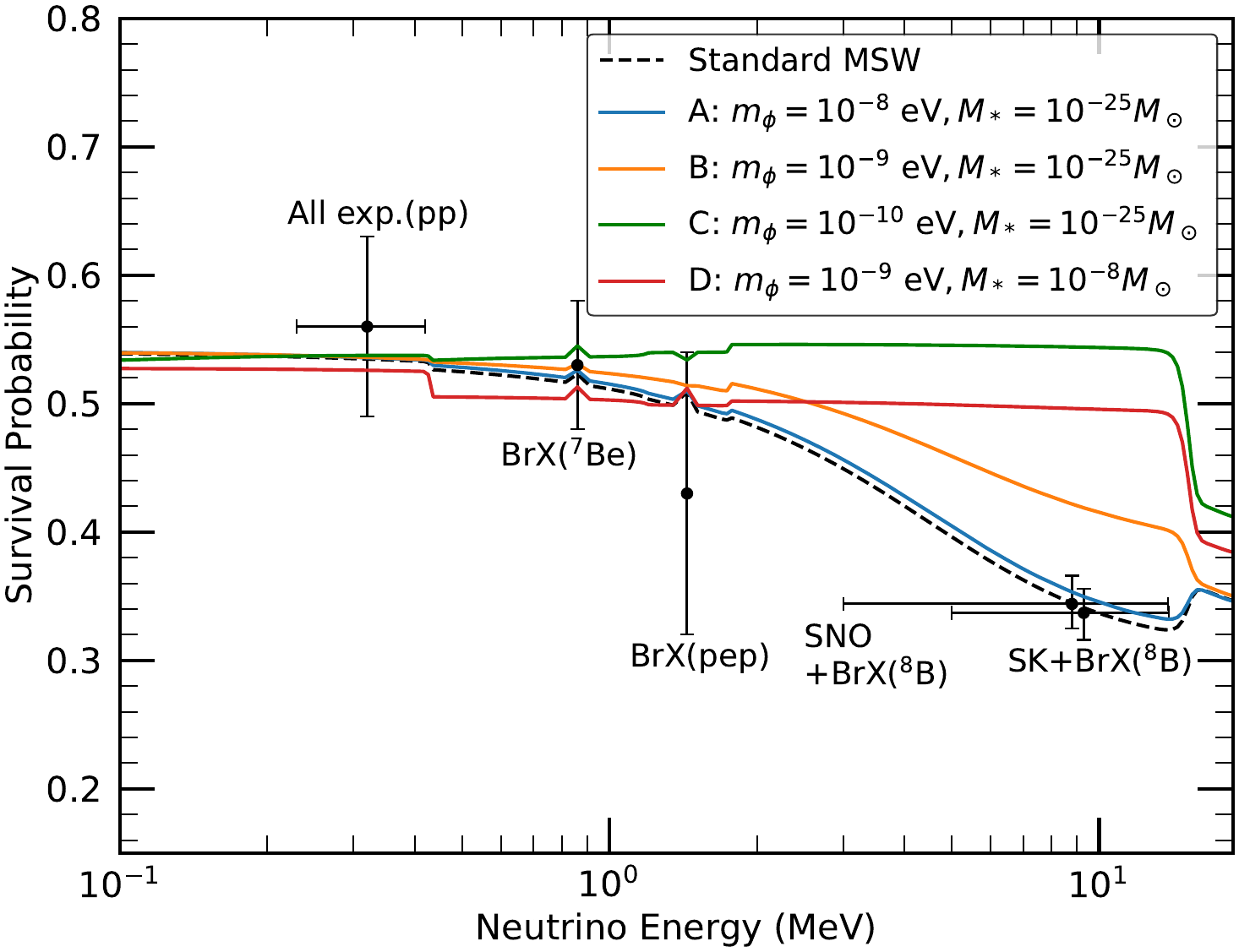}
    \caption{Flux-averaged probability spectrum $\langle P_{ee}(E)\rangle$ for the scenarios A, B, C, D in Fig. \ref{fig:indv_surv_prob}. The black points with error bars indicate the measurements by various experiments assuming a standard 3-flavor framework (`BrX' denotes the measurements made by the Borexino experiment). For our case, these data need to be reinterpreted from Fig. \ref{fig:SK_SNO_reinterpret} to compare with our predictions. The kinks due to the $^7$Be and pep neutrinos have been exaggerated for better visibility.} \label{fig:light_LMA_flux_avg}
\end{figure}

In Fig. \ref{fig:light_LMA_flux_avg}, we show $\langle P_{ee}(E)\rangle$ for the same set of benchmark values of $m_\phi$ and $M_\star$ as in the right panel of Fig. \ref{fig:indv_surv_prob}. Due to neutrinos coming from multiple channels, $\langle P_{ee}(E)\rangle$ can be very different from the $P_{ee}$ spectrum when neutrinos are assumed to be produced at a fixed distance from the center of the Sun. For instance, the $P_{ee}(E)$ spectrum in the benchmark scenario B ($m_\phi = 10^{-9}$ eV and $M_\star = 10^{-25} M_\odot$) matches very well with the standard MSW case when all neutrinos are produced at the same radial distance, as can be seen in the left panel of Fig. \ref{fig:indv_surv_prob}. However, $\langle P_{ee}(E)\rangle$ for the same DM halo does not match with its MSW counterpart, as can be seen in Fig. \ref{fig:light_LMA_flux_avg}. This may be traced to a large fraction of $^8$B neutrinos being produced inside the RHD.

We can identify several distinct features in the $\langle P_{ee}(E)\rangle$ spectra in Fig. \ref{fig:light_LMA_flux_avg}.
\begin{itemize}
    \item Low energy region ($E \lesssim 0.4$ MeV): At these energies, the dominant contribution to neutrino flux is from the $\text{pp}$ neutrinos. The survival probabilities for $\text{pp}$ neutrinos are observed to be almost the same for standard MSW and in the presence of a DM halo. This is because a very small fraction of $\text{pp}$ neutrinos are produced within the RHD. The small differences in some scenarios are due to slightly different values for $P_{ee,\,\text{core}}$, $P_{ee,\,\text{peri}}$ and $P_{ee,\,\text{out}}$. 
    
    \item Intermediate energy region ($0.4\ \text{MeV} \lesssim E \lesssim 1.7$ MeV): In this energy range, the dominant contribution comes from $^{13}\text{N},\ ^7\text{Be},\ ^{15}\text{O}\ \text{and}\ \text{pep}$ neutrinos. There are also $^{17}\text{F}$ neutrinos that have some contribution at these energies, but their flux is smaller by two orders of magnitude compared to the fluxes of $^{13}\text{N}$ and $^{15}\text{O}$ neutrinos. Therefore, we ignore the effects of $^{17}\text{F}$ neutrinos. The figure shows two small kinks at $\sim 0.4\ \text{MeV}$ and $\sim 1.7$ MeV due to the endpoint energies of $\text{pp}$ and $^{15}\text{O}$ neutrinos, respectively. Beyond $0.4\ \text{MeV}$, the dominant contribution to the neutrino flux shifts from $\text{pp}$ to $^{13}\text{N}$ neutrinos. Similarly, at $1.7\ \text{MeV}$, $^8\text{B}$ neutrinos take over $^{15}\text{O}$ neutrinos. The kink for the endpoint energy of $^{13}\text{N}$ neutrinos around $1.2$ MeV is suppressed because there is not much hierarchy between the fluxes of $^{13}\text{N}$ and $^{15}\text{O}$ neutrinos. 
    
    \item High energy region ($1.7\ \text{MeV} \lesssim E \lesssim 15$ MeV): $^8\text{B}$ neutrinos have the most significant contribution in this range of the energy spectrum. Though there are $\text{hep}$ neutrinos also at these energies, their flux is about three orders of magnitude smaller than $^8\text{B}$ neutrinos, so we will comment about them in the next energy range. The upturn of the probability spectrum at $E\sim 1-10\ \text{MeV}$ can have different shapes depending on the DM halo parameters. This is because the fractions $f_{\text{core}}^{(i)}$ and $f_{\text{peri}}^{(i)}$ depend on the halo parameters. In the limit $f_{\text{core,peri}}^{(i)} \ll1$, the upturn mimics the standard MSW spectrum.
    
    \item Very high energy region ($15\ \text{MeV} \lesssim E \lesssim 18.7$ MeV): Around $15\ \text{MeV}$, which is the end-point energy for $^8\text{B}$ neutrinos, we observe a small upward/downward jump in the spectrum. At these energies, the neutrino flux comprises only $\text{hep}$ neutrinos. The production regions of $^8\text{B}$ and $\text{hep}$ are significantly different: the $^8\text{B}$ neutrinos are produced near the core of the Sun while the $\text{hep}$ are produced over a broader region. Due to this, their average $\theta_{12}^P$ values differ from each other, causing the difference in their survival probabilities\footnote{This jump was also recently discussed in \cite{Denton:2025cbo}, where they studied the prospects of utilizing this discontinuity to determine the solar density profile assuming standard MSW probability.}. We see that for scenarios A and B, the refractive solar neutrinos mimic the standard MSW probability at these energies. This happens because the values of $f_{\text{core,peri}}^{\text{hep}}$ are small for $\text{hep}$ neutrinos, due to their broad production region.
\end{itemize}
In near future, the $\langle P_{ee}(E)\rangle$ spectrum will be measured with higher precision. In particular, the DUNE experiment can make the first measurements of the yet-undiscovered $\text{hep}$ neutrinos \cite{Capozzi:2018dat, DUNE:2020ypp, Meighen-Berger:2024xbx, Denton:2025cbo}. The JUNO experiment will be able to measure the survival probability in the intermediate energy region \cite{JUNO:2023zty} and Hyper-Kamiokande will target the high energy region \cite{Hyper-Kamiokande:2022smq,Hyper-Kamiokande:2016srs}. These measurements will allow us to put strong constraints on the halo parameters, and consequently, on the model comprising refractive masses of neutrinos.

\section{Revival of the Dark-LMA Solution ?}\label{sec:Dark_LMA}

The neutrino mixing parameters $\theta_{12}$ and $\Delta m_{21}^2$ are determined primarily from solar neutrino and reactor neutrino experiments. While the solar neutrino data with the standard MSW mechanism favors the solution with $s^2_{12} < 0.5$ (the ``light-LMA" region), the reactor neutrinos also allow the ``mirror" solution with $s^2_{12} > 0.5$ \cite{deGouvea:2000pqg}. This solution is the reflection of the standard light-LMA solution about $s_{12}^2 = 0.5$; the degeneracy between these two solutions being broken by the matter effects inside the Sun. Recently, it has been pointed out \cite{Coloma:2016gei, Bakhti:2014pva, Miranda:2004nb} that this degeneracy can be restored by specific Non-Standard Interactions (NSIs) and solutions with $s_{12}^2 > 0.5$ may become allowed. Such solutions are referred collectively as `dark-LMA' solutions \cite{Denton:2022nol}. In this section, we investigate the possibility of reviving the dark-LMA solution for refractive neutrinos in the presence of a DM halo.

There are strong motivations for expecting that the dark-LMA solution can be revived in the context of the refractive neutrino mass model in presence of a solar DM halo.
When $\theta_{12}$ is in the dark-LMA region, $P_{ee,\,\text{core}}$ is very close to the standard light-LMA MSW survival probability, i.e. 
\begin{equation}
    \left(\cos^4\theta_{12}^{\text{dark}} + \frac{\sin^4\theta_{12}^{\text{dark}}}{2}\right) \simeq \sin^2\theta_{12}^{\text{light}}\,,
\end{equation}
 where $ \theta_{12}^{\text{dark}} = \pi/2 - \theta_{12}^{\text{light}}$. Both sides of the equation become exactly equal when $\sin^2{\theta_{12}^{\text{light}}} = 1/3$, which is very close to the best fit for the standard light-LMA solution. Due to this interesting degeneracy, the present solar neutrino detectors may not be able to resolve whether the $P_{ee}$ for $^8\text{B}$ neutrinos at high energies originates from the dark-LMA solution with the RHD core or the standard light-LMA solution. Moreover, if the low-energy pp neutrinos are produced within the RHD periphery, their survival probability $P_{ee,\,\text{peri}}$ is invariant under $\theta_{12} \leftrightarrow \pi/2 - \theta_{12}$ when $c_{22'}^P \approx 1$ (see eq. \ref{eq:surv_prob_approx_2}). Also, the matter effects on these low-energy neutrinos are small, and therefore if they are produced outside the RHD, then their survival probability $P_{ee,\, \text{out}}$ is also invariant under $\theta_{12} \leftrightarrow \pi/2 - \theta_{12}$.

Thus, the dark-LMA solution can be consistent with the present solar neutrino measurements if the DM halo has the following two desirable characteristics.
\begin{enumerate}
    \item The RHD boundary should be around $r_0 \gtrsim 0.09-0.1\, R_\odot$. This is required to ensure that almost all of $\,^8\text{B}$ neutrinos come from the RHD, since $P_{ee}^{\text{halo}} \approx 0.3-0.4$ from Fig. \ref{fig:SK_SNO_reinterpret} and $P_{ee,\, \text{core}} \approx 0.33$. The fraction coming from the RHD periphery ($P_{ee,\, \text{peri}} \approx 0.56$) and outside ($P_{ee,\, \text{out}} \approx 0.68$) should be small. 
    \item  Both pp and $^7\text{Be}$ neutrinos should be produced outside the RHD core, i.e. $r_1 \lesssim 0.07-0.08\, R_\odot$. In addition, the core-periphery transition should be sharp, so that $c_{22'}^P \approx 1$. 
\end{enumerate}

\begin{figure}[t!]
    \centering
    \includegraphics[width=\linewidth]{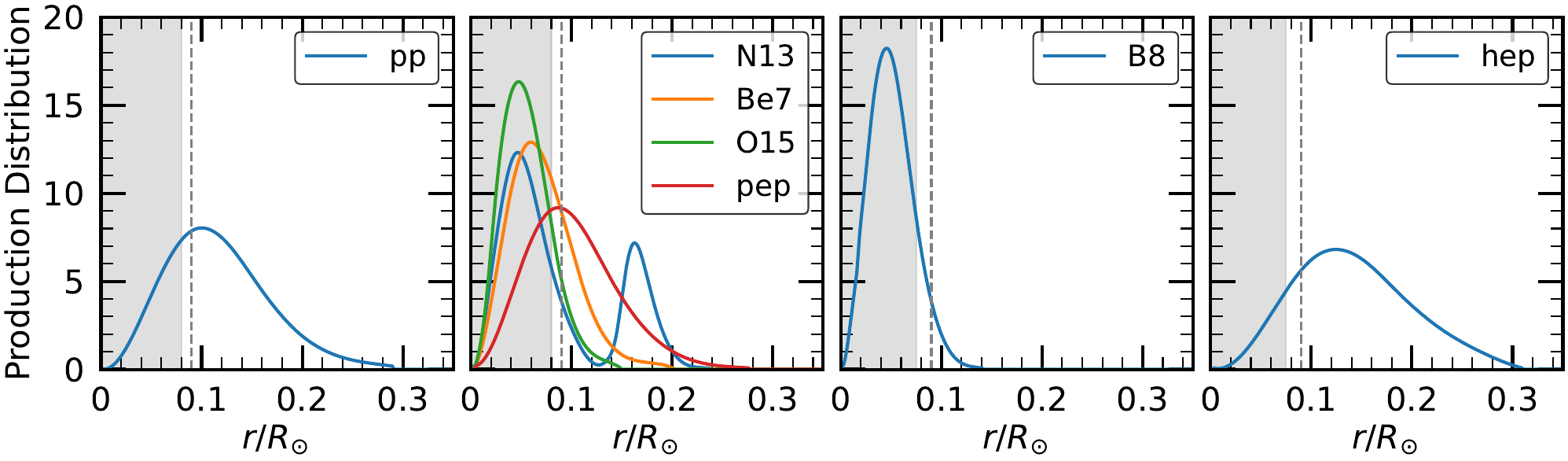}
    \caption{Production distributions $\eta_i(r)$ for neutrinos from different production channels. The shaded region represents the RHD core and the vertical dashed line represents the boundary of the RHD periphery for  $m_\phi = 10^{-9}$ eV, $m_\chi^2 = 10^{-7}\, \text{eV}^2$  and $M_\star = 10^{-8}\, M_\odot$.}
    \label{fig:dark_LMA_prod_dist}
\end{figure}

These two conditions guide us to a region of the ($m_\phi$--$M_\star$) parameter space near $m_\phi \approx 10^{-10}$--$10^{-9}$ eV and $M_\star \approx 10^{-16}$--$10^{-8}\, M_\odot$, where lower values of $m_\phi$ correlate with lower values of $M_\star$. In Fig. \ref{fig:dark_LMA_prod_dist}, we illustrate the production distributions $\eta_i(r)$ of the neutrino production channels along with the RHD core and RHD periphery boundaries, for $(m_\phi,M_\star) = (10^{-9}\, \text{eV}\,, 10^{-8}\, M_\odot)$. Table \ref{tab:dark_LMA_fractions} shows the corresponding fractions $f_{\text{core, peri, out}}^{(i)}$ of neutrinos from their respective  production channels. The table clearly demonstrates that the desirable conditions mentioned above are satisfied.

\begin{table}[h]
    \centering
    \begin{tabular}{|c|c|c|c|c|}
    \hline
         \textbf{Energy Range}& \textbf{Channel-}$\boldsymbol{i}$ & $\boldsymbol{f_{\text{core}}^{(i)}}$ & $\boldsymbol{f_{\text{peri}}^{(i)}}$ & $\boldsymbol{f_{\text{out}}^{(i)}}$ \\
         \hline
         \hline
         \multirow{1}{*}{$\lesssim 0.4$ MeV}
          & $\text{pp}$ & 0.25 & 0.08 & 0.67\\
         \hline
         \hline
         \multirow{5}{*}{$0.4\ \text{MeV} \lesssim E \lesssim 1.7$ MeV}
          & $^{13}\text{N}$ & 0.62 & 0.05 & 0.33\\
          \cline{2-5}
          & $^{15}\text{O}$ & 0.84 & 0.07 & 0.09\\
          \cline{2-5}
          & $^7\text{Be}$ & 0.66 & 0.1 & 0.24\\
          \cline{2-5}
          & $\text{pep}$ & 0.34 & 0.09 & 0.57\\
          \hline
        \hline
         \multirow{1}{*}{$1.7\ \text{MeV} \lesssim E \lesssim 15$ MeV}
          & $^8\text{B}$ & 0.86 & 0.09 & 0.05\\
         \hline
         \hline
         \multirow{1}{*}{$15\ \text{MeV} \lesssim E \lesssim 18.7$ MeV}
          & $\text{hep}$ & 0.12 & 0.08 & 0.8\\
         \hline
    \end{tabular}
    \caption{Fractions of neutrino flux for different channels from the RHD core, the RHD periphery and outside RHD, when $m_\phi = 10^{-9}$ eV, $m_\chi^2 = 10^{-7}\, \text{eV}^2$  and $M_\star = 10^{-8} M_\odot$.}
    \label{tab:dark_LMA_fractions}
\end{table}

\begin{figure}[t!]
    \centering
    \includegraphics[width=0.375\linewidth]{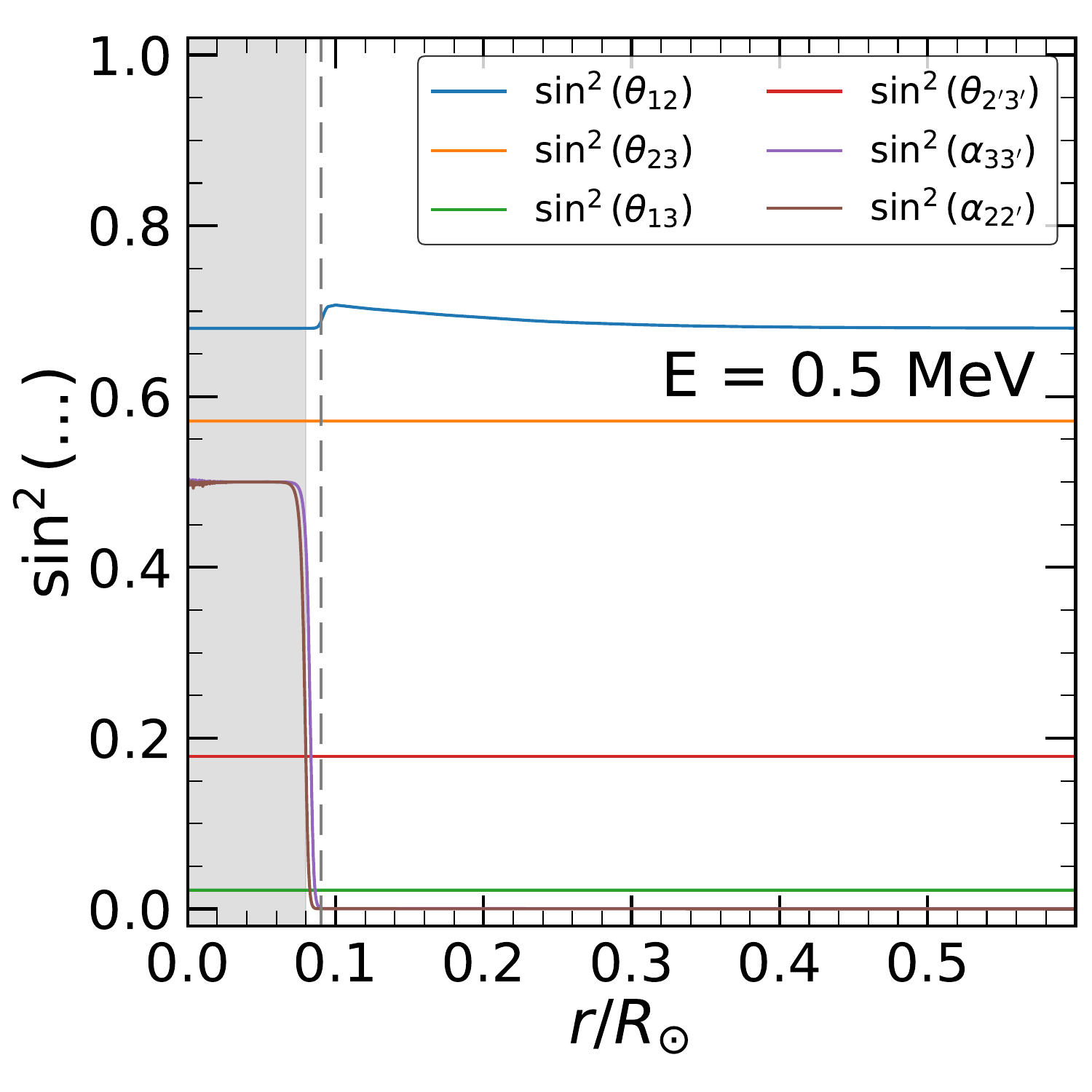}
    \includegraphics[width=0.375\linewidth]{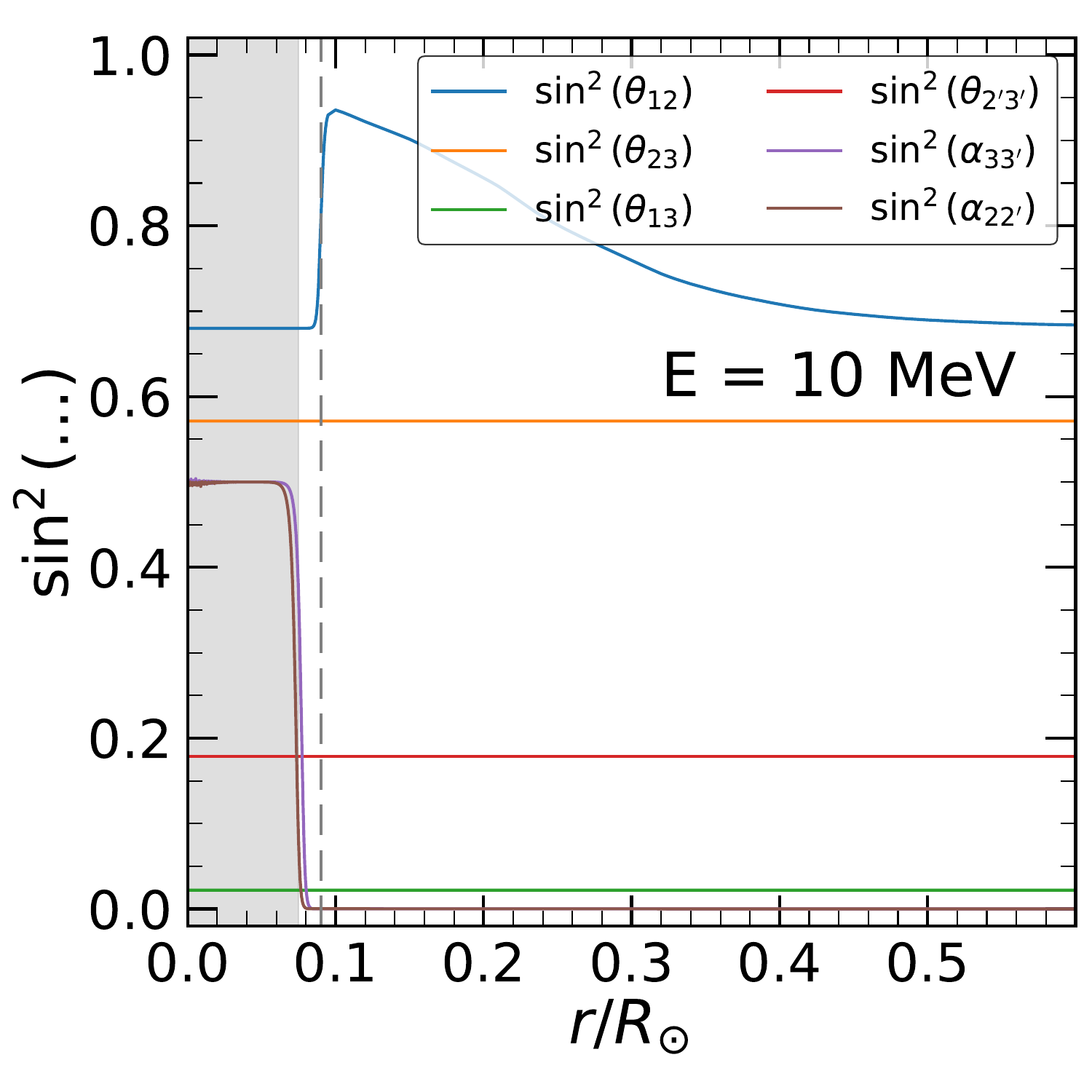}
    \includegraphics[width=0.395\linewidth]{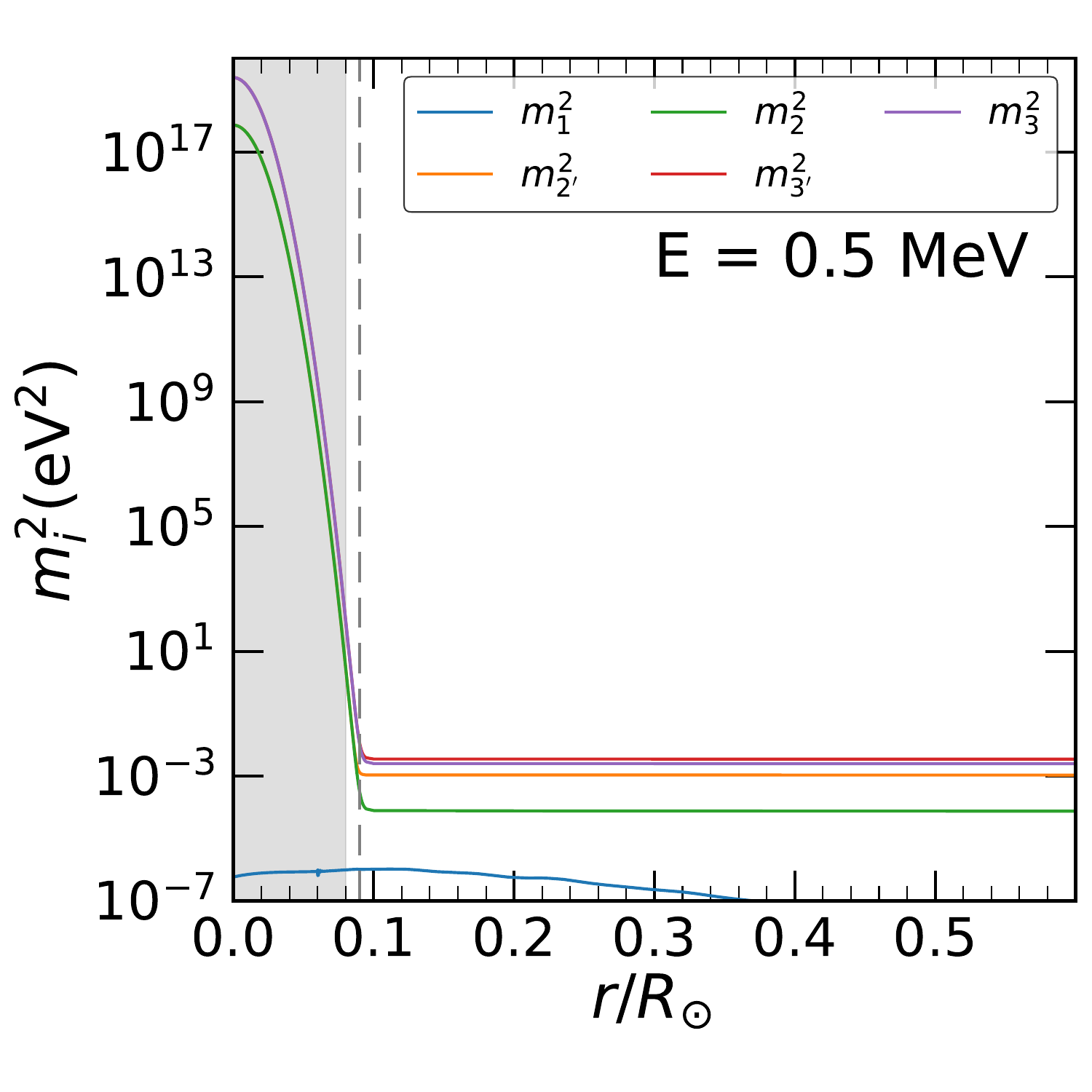}
    \includegraphics[width=0.395\linewidth]{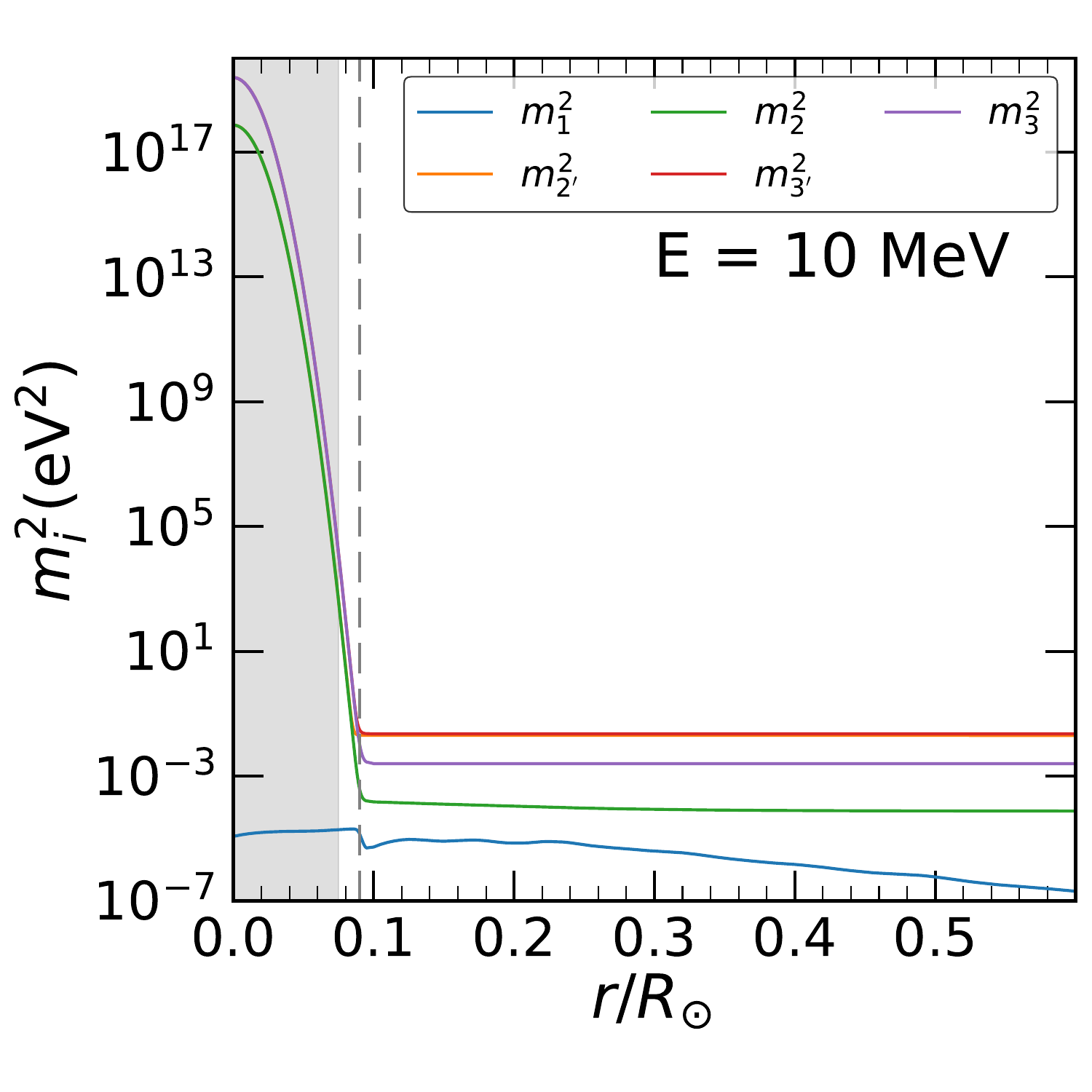}
    \caption{Mixing angles (upper panels) and mass-squared values (lower panels) in the 5-neutrino system for $E =0.5 \ \text{MeV (left panels) and}\ E =10\ \text{MeV (right panels)}$ when $m_\phi = 10^{-9}$ eV, $m_\chi^2 = 10^{-7}\, \text{eV}^2$ and $\dot{\Phi}=m_\phi$,  $M_\star =10^{-8}\,M_\odot$. We take $s_{12}^2 = 0.68$ (dark-LMA) in the uniform DM background. The vertical dashed line indicates the boundary of RHD ($r = r_0$) and the shaded region represents the RHD core ($r \leq r_1$).}\label{fig:dark_LMA_mixing_angles}
\end{figure}

With this DM halo, we can predict several features of $\langle P_{ee} \rangle$ in different energy ranges. For illustration, we start by showing the mixing angles and $m^2$'s as functions of $r$ for $s_{12}^2 =0.68$ in Fig. \ref{fig:dark_LMA_mixing_angles}. Note that there is no MSW resonance at low ($E=0.5\,\text{MeV}$) or high ($E=10\,\text{MeV}$) energy. The figure shows that the value of $s_{12}^2$ always stays above $0.5$. However, at high energies, there is a jump in the value of $s_{12}^2$ while crossing the RHD boundary. 

\begin{figure}[t!]
    \centering
    \includegraphics[width=0.75\linewidth]{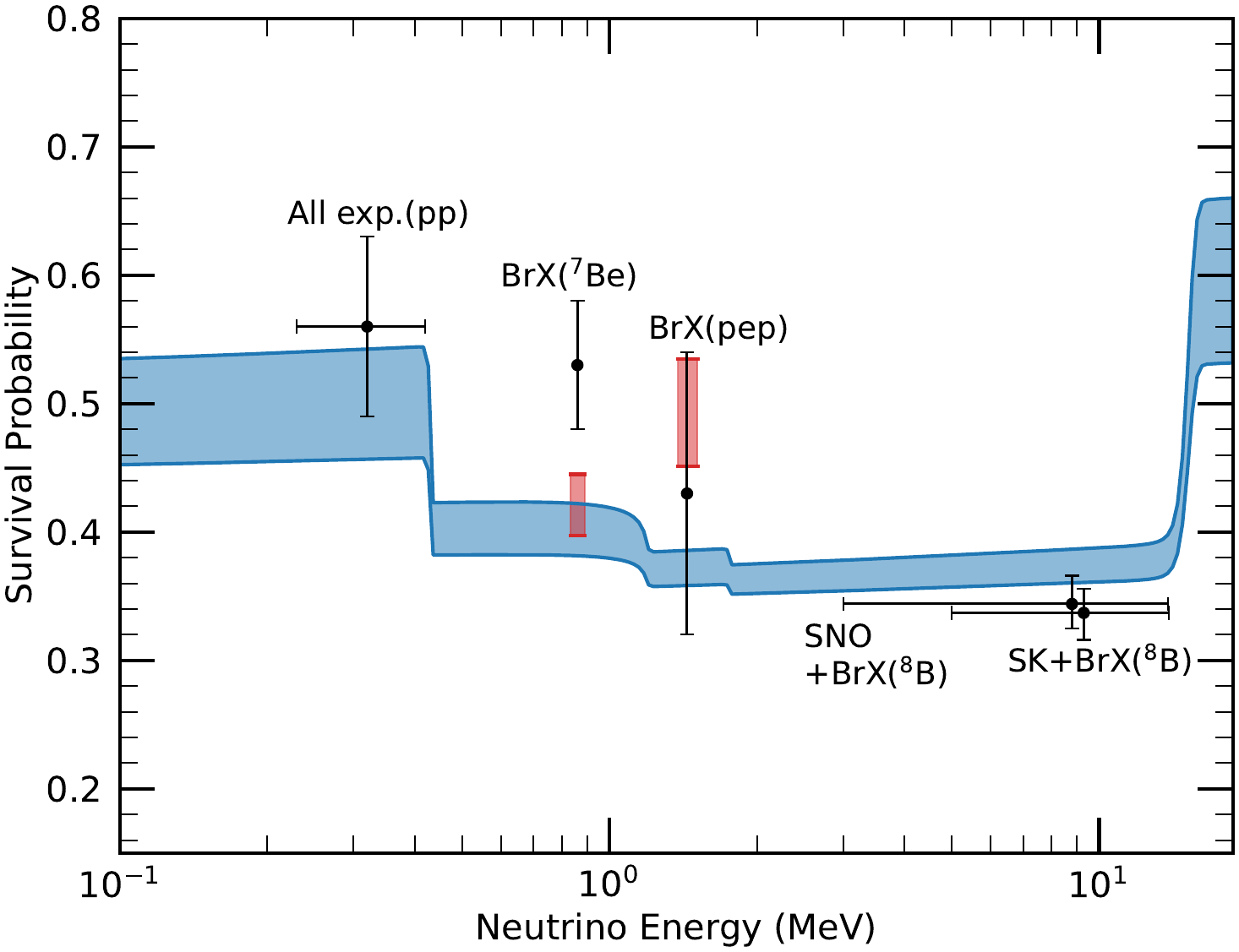}
    \caption{Flux-averaged probability spectrum $\langle P_{ee}(E)\rangle$ for $m_\phi = 10^{-9}$ eV, $m_\chi^2 = 10^{-7}\, \text{eV}^2$ and $\dot{\Phi}=m_\phi$,  $M_\star =10^{-8}\,M_\odot$. Here $\text{sin}^2(\theta_{12})$ is varied within $2\sigma$ allowed region of KamLAND dark-LMA solution. The black points with error bars indicate the measurements by various experiments assuming a standard 3-flavor framework. For our case, these data need to be reinterpreted as in Fig. \ref{fig:SK_SNO_reinterpret} for each value of $s_{12}^2$ to compare with our predictions. The predicted survival probabilities for $^7$Be and pep neutrinos have been shown in red, with their energy widths exaggerated for better visibility.}
    \label{fig:dark_LMA_surv_prob}
\end{figure}

In Fig. \ref{fig:dark_LMA_surv_prob}, we show the flux-averaged survival probability spectrum $\langle P_{ee}\rangle$, with $\theta_{12}$ varied within the dark-LMA region allowed at $2\sigma$ by the KamLAND experiment \cite{KamLAND:2008dgz}\footnote{Recently, the JUNO experiment has announced its first measurement of oscillation parameters \cite{JUNO:2025gmd}, which is the most precise measurement of $\theta_{12}$ so far. While the JUNO analysis focuses only on the light-LMA region, the dark-LMA solution would also be allowed, given the nature of the experiment. Further, since the central value of $\theta_{12}$ measured by the JUNO and KamLAND experiments are very similar, the arguments in this section would not change even if the JUNO results were added.}, i.e. $s_{12}^2 = (0.6-0.76)$.
The numerical values of $\langle P_{ee} \rangle$ in different energy ranges may be accounted for from simple analytic arguments as discussed below.   

\begin{itemize}

    \item Low energy region ($E \lesssim 0.4$ MeV): For $\text{pp}$ neutrinos, the production distribution is very broad. As these neutrinos have low energies, we may take $c_{12,\odot} \simeq c_{12}$ and $s_{12,\odot} \simeq s_{12}$, implying
    \begin{equation}
        \langle P_{ee} \rangle \approx f_{\text{core}}^{\text{pp}} \left(c_{12}^4+\frac{s_{12}^4}{2}\right) + (f_{\text{peri}}^{\text{pp}}+f_{\text{out}}^{\text{pp}}) \left(c_{12}^4+s_{12}^4\right).
    \end{equation}
    The pp neutrinos have an end-point energy of 0.42 MeV, till which point the above expression is valid. The survival probability changes in the intermediate energy region, giving rise to the kink as observed in the figure.
    
    \item Intermediate energy region ($0.4\ \text{MeV} \lesssim E \lesssim 1.7$ MeV): The $^{13}\text{N}$ neutrino production distribution has two peaks. The second peak at $r\approx 0.15\, R_\odot$ results in a significant fraction of neutrinos produced outside RHD. This causes the survival probability of $^{13}\text{N}$ neutrinos to have an energy dependence till its end-point energy ($\approx 1.2$ MeV). On the other hand, $^{15}\text{O}$ neutrinos have comparably narrow distribution of production points, with most of the flux coming from inside the RHD. Hence $\langle P_{ee} \rangle$ in the $1.2-1.5$ MeV energy range has a flat distribution as can be seen in Fig. \ref{fig:dark_LMA_surv_prob}.

    \item High energy region ($1.7\ \text{MeV} \lesssim E \lesssim 15$ MeV): For $^8\text{B}$ neutrinos, almost the entire flux comes from inside the RHD. Note that this also implies a flat survival probability spectrum for these neutrinos, since the energy dependence of $\langle P_{ee} \rangle$ is significant only for neutrinos produced outside the RHD.
    
    \item Very high energy region ($15\ \text{MeV} \lesssim E \lesssim 18.7$ MeV): The $\text{hep}$ neutrinos have a very broad probability distribution with $f_{\text{core}}^{\text{hep}} \approx 0.12$, $f_{\text{peri}}^{\text{hep}} \approx 0.08$ and $f_{\text{out}}^{\text{hep}} \approx 0.8$. Moreover $c_{12,\odot} \simeq 0$ and $s_{12,\odot} \simeq 1$. Thus, we have
    \begin{equation}
        \langle P_{ee}\rangle \approx 0.12 \left(c_{12}^4+\frac{s^{4}_{12}}{2}\right) + 0.08 \left(c_{12}^4+s^{4}_{12}\right) + 0.8\ s_{12}^2\ .
    \end{equation} 
    The last term leads to a higher value of $\langle P_{ee}\rangle$ in the dark-LMA scenario as compared to the light-LMA MSW solution. This jump of $\langle P_{ee} \rangle$ from the high energy to the very high energy region indeed could be a strong signature of the dark-LMA scenario with a DM halo inside the Sun.
\end{itemize}

\begin{figure}[t!]
    \centering
    \includegraphics[width=0.65\linewidth]{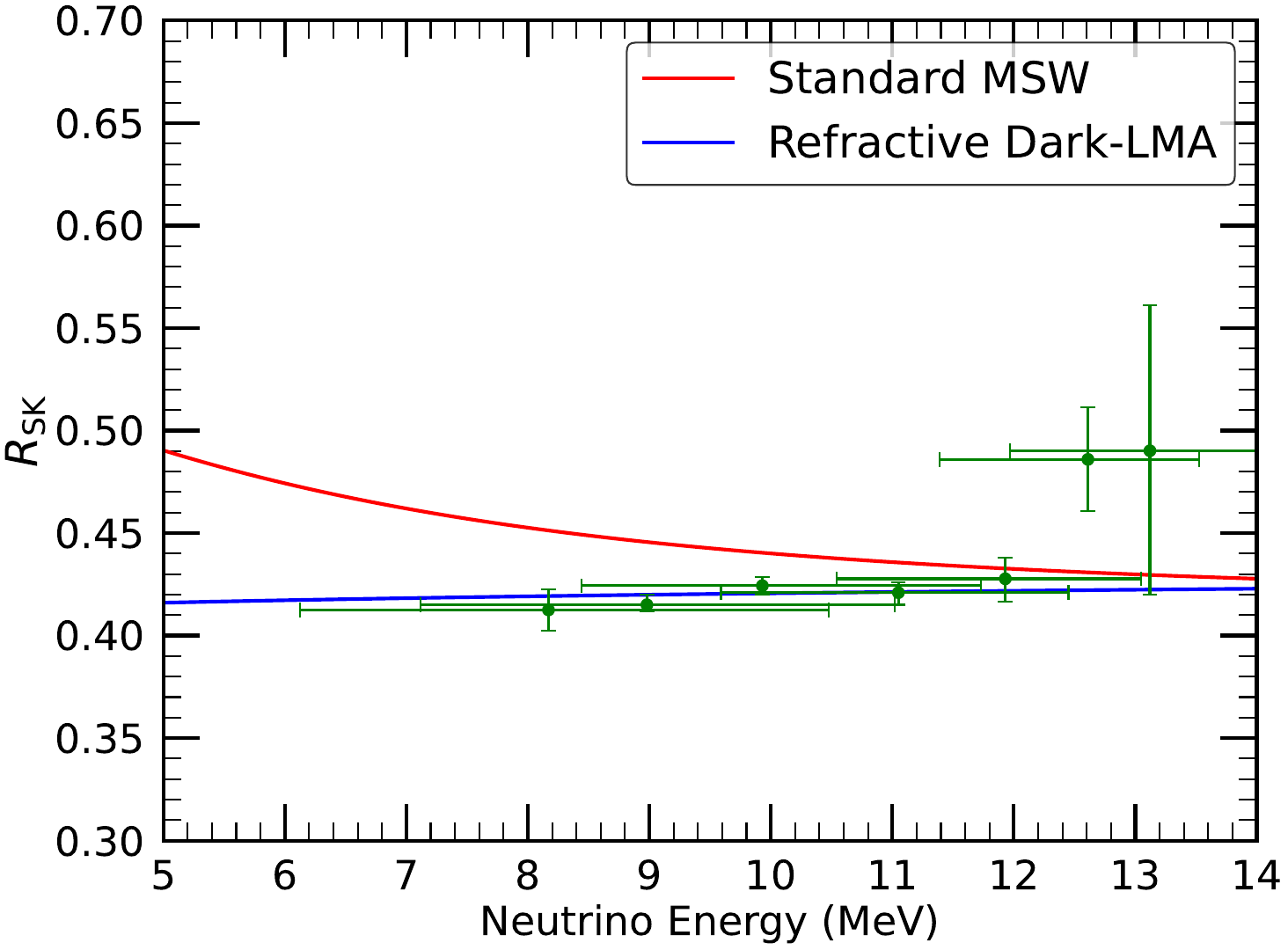}
    \caption{Comparison of $R_{\text{SK}}$ for standard MSW ($s_{12}^2 = 0.32$) and the prediction from the dark-LMA solution ($s_{12}^2 = 0.68$) for refractive neutrino masses in the presence of a DM halo with $m_\phi = 10^{-9}$ eV, $m_\chi^2 = 10^{-7}\, \text{eV}^2$ and $\dot{\Phi}=m_\phi$,  $M_\star =10^{-8}\,M_\odot$. The green points with error bars indicate the bin-wise $R_{\text{SK}}$ values calculated from the $P_{ee}$ values inferred \cite{Maltoni:2015kca} from the SK data.}
    \label{fig:dark_LMA_MSW_comp}
\end{figure}

The gray circles in Fig. \ref{fig:SK_SNO_reinterpret} represent the probabilities corresponding to different values of halo parameters $m_\phi \in [10^{-10}-10^{-9}]$ eV , $M_\star \in [10^{-20}-10^{-8}]\, M_\odot$ and $s^2_{12} \in [0.55,0.8]$. The range of $s_{12}^2$ corresponds to the $3\sigma$ region for the dark-LMA solution allowed by KamLAND \cite{KamLAND:2008dgz}. The positioning of these benchmark points in the plot is observed to follow a pattern based on the increasing values of $r_{0}$ and $r_1$ (which are in turn functions of $m_\phi$ and $M_\star$) as well as the increasing values $s^2_{12}$ as indicated by arrows in the figure. As $r_0$ and $r_1$ increase, the value of $P_{e\chi}^{\text{halo}}$ increases while $P_{ee}^{\text{halo}}$ decreases to saturate at a minimum of $\approx 0.33$ for all values of $s_{12}^2$. Note that these values fall outside the $3\sigma$ parameter space allowed by the CC/NC data at SNO. However, there are some regions that are perfectly consistent with $R_\text{SK}$ and lie within $3.5\sigma$ of $R_{\text{SNO}}^{\text{CC/NC}}$. The figure also shows a ``dark star" point, corresponding to the standard dark-LMA solution. Clearly, the solar DM halo helps in bringing the dark-LMA solution closer to the region allowed by the data. 

An important prediction of the dark-LMA scenario with the solar DM halo is the flatness of the $\langle P_{ee}\rangle$ spectrum in the $^8$B-neutrino region. This is a consequence of $f_{\text{out}}^{8\text{B}} \ll 1$, which suppresses the energy dependence arising from $P_{ee,\,\text{out}}$. This may be seen in Fig. \ref{fig:dark_LMA_MSW_comp} which presents the energy dependence of the standard MSW scenario and the dark-LMA scenario, while comparing them with the reconstructed bin-wise $R_{\text{SK}}$ data\footnote{We choose to present the results in terms of $R_{\text{SK}}$ rather than $\langle P_{ee}\rangle$ since the value of $P_{ee}$ inferred from the SK data depends on $P_{e\chi}$.} \cite{Maltoni:2015kca}. The flat spectrum of the dark-LMA scenario in this context is clearly very aligned to the experimental observations. Moreover, the high bin values in the last two bins may be indicative of a jump in the $\langle P_{ee} \rangle$ spectrum.

An envisaged experiment using a deuterated liquid scintillator may be sensitive to the NC events as well as to the features of the $\langle P_{ee} \rangle$ spectrum in the energy range $2-20$ MeV \cite{Chauhan:2021snf,DLS:2023xor}. Such an experiment could play a crucial role in identifying or ruling out the dark-LMA scenario in the context of refractive masses and solar DM halo. 

\section{Summary and Conclusions}\label{sec:conclusions}

In this work, we explore the possibility in which neutrinos acquire ``refractive masses" as a result of their propagation through a DM background. The particular model considered has an ultralight scalar $\phi$ that acts as DM and two sterile neutrinos $\chi_{1,2}$ in addition to the 3 active ones. We show that the $5\times 5$ Hamiltonian for the propagation of neutrinos can always be diagonalized using a unitary matrix $\mathbb{P}$ involving only 6 rotation angles and 1 complex phase. When $m_{\chi_1}=m_{\chi_2}$, the neutrino-DM coupling matrix $[g]_{3\times2}$ satisfying all experimental constraints is unique upto a rotation angle $\theta_{2'3'}$. The remaining 5 angles are the three active neutrino mixing angles $\theta_{12}$, $\theta_{23}$, $\theta_{13}$ and the two active-sterile mixing angles $\alpha_{22'}$ and $\alpha_{33'}$. 

To restrict active-sterile neutrino oscillations from the Sun to the Earth, we need to choose either small active-sterile mixing angles $\alpha_{22',33'} \lesssim 10^{-2}$, or small active-sterile mass squared differences $\Delta m_{22',33'}^2 \lesssim 10^{-12}\, \text{eV}^2$. In either case, our chosen parametrization for $\mathbb{P}$ allows us to analyze the neutrino propagation inside the Sun as an effective two-flavor problem. We support this analysis by numerically computing the mixing angles and level crossings inside the Sun. In the absence of any effect of the Sun on the background DM density, the solar neutrino survival probabilities in the refractive mass situation and the standard MSW framework are indistinguishable.

The gravity of the Sun and interactions of the DM may give rise to a overdense DM halo inside or around the Sun. 
We explore the effects of such a DM halo on the solar neutrinos with refractive masses. We find that DM halos should have a radius $\lesssim 0.3\, R_\odot$ to satisfy the observed $^8$B-neutrino survival probability. This constraint on the halo radius, and the constraints from cosmology, imply $m_\phi \gtrsim 10^{-10}$ eV and $m_\chi \gtrsim 10^{-5}$ eV. These magnitudes of masses cannot generate sufficiently small $\Delta m_{22',33'}^2$. Thus, we focus on the small $\alpha_{22',33'}$ scenario which can be achieved only with a complex $\phi$.

Inside the DM halo, $\alpha_{22',33'}$ may not remain small in spite of its small value in the uniform DM background. Depending on the effect of the DM overdensity on the mixing angles, we classify the region inside the DM halo into an RHD core and an RHD periphery. In both these regions, $\theta_{12}$ is equal to its uniform DM-background value. On the other hand, $\alpha_{22',33'} \approx \pi/4$ within the RHD core while $\alpha_{22',33'} \approx 0$ within the RHD periphery. Therefore, in the presence of a DM halo, the problem is no longer a two-flavor problem. This results in a survival probability which is different from the standard MSW framework. 

The solar neutrino data from SK and SNO are normally analyzed under the assumption of the standard three-neutrino framework, while a non-negligible $\alpha_{22'}$ inside the RHD core leads to a substantial active-sterile conversion $P_{e\chi}$. Thus, to consistently compare our predictions with experimental observations, the values of $P_{ee}$ and $P_{e\mu,e\tau}$ obtained from the 3 ratios $R_{\text{SK}}$, $R_{\text{SNO}}^{\text{CC/ES}}$ and $R_{\text{SNO}}^{\text{CC/ES}}$ need to be revised. We briefly outline the procedure for this reinterpretation and present allowed regions for $P_{ee}^{\text{halo}}$ and $P_{e\chi}^{\text{halo}}$. 
 
We derive simple analytic expressions for the survival probability $P_{ee}$ in the presence of a halo and demonstrate its dependence on the site of production. Since each production channel has a different radial and energy distribution, the quantity to compare with experiments is the flux-averaged survival probability $\langle P_{ee}(E)\rangle$. We calculate the $\langle P_{ee}(E)\rangle$ spectrum and show its features for different halo parameters $m_\phi$ and $M_\star$. We point out how end-point energies of various channels leave their imprints as kinks in the $\langle P_{ee}(E) \rangle$ spectrum. 

Finally, we examine the intriguing possibility of reviving the dark-LMA solution of the solar neutrino problem in the context of refractive neutrino masses with a solar DM halo. This is motivated by the interesting degeneracy between the $P_{ee}^{\text{halo}}$ in the RHD core for the dark-LMA scenario and the standard $P_{ee}^{\text{MSW}}$ for the light-LMA solution. With appropriately chosen halo, this degeneracy can yield $\langle P_{ee} \rangle \approx 0.5-0.6$ for $E \lesssim 0.4$ MeV and $\langle P_{ee} \rangle \approx 0.3-0.4$ for $2\, \text{MeV} \lesssim E \lesssim 15\, \text{MeV}$. Note that there is no MSW resonance at high energies for this case, unlike the standard light-LMA scenario. A simplified analysis suggests that this scenario can be easily consistent with the SK data, but it is consistent with the SNO data only at the $3.5-4\sigma$ level. Nevertheless, the presence of a solar DM halo can improve the viability of the dark-LMA solution significantly. We further identify several distinct features in the dark-LMA $\langle P_{ee} \rangle$ spectrum that make it attractive. Two notable features among them are: (i) a flat  $\langle P_{ee} \rangle$ spectrum in $2\, \text{MeV} \lesssim E \lesssim 15\, \text{MeV}$, and (ii) a jump in the value of $\langle P_{ee} \rangle$ at $E \approx 15\, \text{MeV}$. The data/MC spectrum of SK hints towards both these features. This raises the question whether the dark-LMA solution constitutes a plausible candidate for the solar neutrino problem. Further investigation in this direction through a detailed analysis of the re-interpreted experimental data is left for future work.

Upcoming solar neutrino experiments will determine the solar neutrino survival probability spectrum with significantly higher precision. While the intermediate energy CNO cycle neutrinos can be detected by JUNO, the high energy $^8$B and the very high energy hep neutrinos can be detected in large numbers at Hyper-Kamiokande and DUNE. Due to several unique features in the survival probability predictions for refractive neutrinos in the presence of the DM halo, one can constrain the DM halo for the light-LMA solution and test the dark-LMA solution. The open issues in the origins of neutrino masses and the DM interactions may find a common resolution in the solar neutrino data.

\section*{Acknowledgements}

We would like to thank Manibrata Sen for useful discussions. The work is supported by the Department of Atomic Energy, Government of India, under Project Identification Number RTI 4002. A.D. would like to acknowledge funding from the J. C. Bose Grant of the
Anusandhan National Research Foundation (ANRF), Government of India.

\bibliographystyle{JHEP}
\bibliography{biblio.bib}
\end{document}